# Digital Pathway Curation (DPC): a pipeline able to assess the reproducibility, consensus and accuracy in biomedical search retrieval by comparing Gemini, PubMed, and Scientific Reviewers

## Authors:

Flavio Lichtenstein[1*], Daniel Alexandre de Souza[2], Carlos Eduardo Madureira Trufen[5], Victor Wendel da Silva Gonçalves[3], Juliana de Paula Bernardes[3], Vinícius Miranda Baroni[3], Carlos DeOcesano-Pereira[1], Leonardo Fontoura Ormundo[8], Fabio Augusto Labre de Souza[3,4], Olga Celia Martinez Ibañez[6], Nancy Starobinas[6], Luciano Rodrigo Lopes[7], Aparecida Maria Fontes[3], Sonia Aparecida de Andrade[9], Ana Marisa Chudzinski-Tavassi[1,2]

1 - CENTD, Centre of Excellence in New Target Discovery, Butantan Institute, São Paulo, Brazil.
2 - Laboratory of Development and Innovation, Butantan Institute, São Paulo, SP, Brazil.
3 - Department of Genetics, Ribeirão Preto Medical School, University of São Paulo, Ribeirão Preto 14049-900, SP, Brazil
4- Department of Medical Imaging, Hematology, and Clinical Oncology, Ribeirão Preto Medical School, University of São Paulo, Ribeirão Preto 14049-900, SP, Brazil
5 - Immunochemistry Laboratory, Prevor, Liège, Belgium.
6 - Immunogenetics Laboratory, Butantan Institute, São Paulo, Brazil.
7 - Department of Health Informatics, UNIFESP, São Paulo, Brazil.
8 - Medical Investigation Laboratory 60, School of Medicine, University of São Paulo, São Paulo, Brazil.
9. Escola Superior do Instituto Butantan, Butantan Institute, São Paulo, SP, Brazil.

## Abstract:

A scientific study begins with a central question, and search engines like PubMed are the first tools for retrieving knowledge and understanding the current state of the art. Large Language Models (LLMs) have been used in research, promising acceleration and deeper results. However, besides caution, they demand rigorous validation. Assessing complex biological relationships remains challenging for SQL-based tools and LLM models. Here, we introduce the Digital Pathway Curation (DPC) pipeline to evaluate the reproducibility and accuracy of the Gemini models against PubMed search and human expert curation. Using two omics experiments, we created a large dataset (Ensemble) based on determining pathway-disease associations. With the Ensemble dataset, we demonstrate that Gemini achieves high run-to-run reproducibility of approximately 99% and inter-model reproducibility of around 75%. Next, we calculate the crowdsourced consensus using a smaller dataset. The CSC allows us to calculate accuracies, and the Gemini multi-model consensus reached a significant accuracy of about 87%. Our findings demonstrate that LLMs are reproducible, reliable, and valuable tools for navigating complex biomedical knowledge.

## Keywords:

Artificial Intelligence, Large Language Model, PubMed search, Google, Gemini, Reproducibility, Accuracy, Curation, Biomedical Pathways, Bioinformatics, Semantic Queries

## Introduction:

Scientists frequently use computational tools such as PubMed[1], Web of Science[2], and Google Scholar[3] to search biomedical literature. These tools are essential for retrieving scientific knowledge, with PubMed being a particularly valuable resource within the biomedical domain, integrating NCBI's MEDLINE, PMC, and Bookshelf databases.

PubMed is the reference search tool in the National Library of Medicine (NLM), providing free access to more than 37 million scientific articles (as of November 2024). PubMed offers a website interface (https://pubmed.ncbi.nlm.nih.gov/), an API[4] and web services for querying biomedical questions. Functionally, it retrieves references based on a Structured Query Language (SQL) model, where queries use ordered terms and logical operators following Boolean algebra. It also employs Medical Subject Headings (MeSH) terms for classification and offers complementary databases like Genome, Nucleotide, and Protein for specialised searches.

The process of scientific discovery itself, encompassing steps from observation to hypothesis testing and communication, relies on evidence and reproducibility, with literature review being a crucial first step. Although the peer-review standard from most PubMed literature implies adherence to the scientific method and evidence-based reporting, this inference is questioned by some[5–9]. Furthermore, despite being reproducible and yielding identical responses for identical queries, PubMed's accuracy concerning false positives and negatives is not always as high as expected, which is an issue that is often overlooked.

As a potential solution to such limitations, Large Language Models (LLMs) like Gemini (Google[10]) and ChatGPT (Open AI[11]) offer a more flexible approach, allowing semantic constructs for effective searches within their trained embedding hyperspace[12]. The advent of LLMs offers new paradigms[13,14], popularising the ability to search scientific topics in natural language, engage in dialogue, and summarise texts. Some initiatives explore LLMs for biomedical research, including training models *ab initio* or refining them with specific corpora like PubMed to improve outcomes[15–21].

Despite LLMs' potential, significant concerns remain regarding their reliability, reproducibility, and the distinction between evidence-based assumptions and potential hallucinations. This uncertainty fuels ongoing investigation into whether LLMs might eventually replace established resources like PubMed or be incorporated to augment search capabilities and effectively carry out data curation[22]. Specifically, there is a need to evaluate how well broadly trained LLMs, without specific biomedical fine-tuning, perform in terms of accuracy and reproducibility against established methods like PubMed searches.

We developed the Digital Pathway Curation (DPC) pipeline to fill this gap by evaluating the capability of LLMs like Gemini to determine whether selected pathways are involved in modulation due to a disease. Tiwari et al. from the EMBL-EBI developed a similar approach to test the Reactome database completion using ChatGPT[23].

In the first part of this study, we build a large dataset called Ensemble to measure LLM reproducibility. Next, we compare the Ensemble consensuses to PubMed findings. In this analysis, we discuss PubMed's potential benefits and limitations.

In the second part of this study, we build a smaller dataset called the Multi-Source Dataset (MSD), which includes the Gemini consensus, PubMed findings, and Human Reviewers consensus. The MSD is used to calculate the Crowdsourced Consensus (CSC)[24] and assess the accuracy of each source.

Finally, in the third part, we analyse different confusion matrices. By analysing the Enriched Pathways Confusion Matrix, one can achieve one of the goals of DPC: to uncover FN pathways. Gemini multi-model consensus (MMC) demonstrated the highest accuracy and was used as a reference to uncover False Positives (FP) and False Negatives (FN) while also confirming True Positives (TP) and True Negatives (TN) among the enriched pathways identified through Gene Set Enrichment Analysis (GSEA).

## Methods:

All pipeline code was developed with Python version 3.11.9 and Gemini versions 1.5-pro and 1.5-flash.

We limited our study to PubMed and Gemini as search engines and Reactome as the pathway database. PubMed is an open-access biomedical reference tool, while Gemini is Google's generative AI tool based on many different LLM models. Reactome is the European Pathway Database. Our primary focus was on Gemini since it is an open-access AI tool due to its Web Service, API programmatic tool, and the availability of multiple models, including Gemini-1.0-pro, Gemini-1.5-pro, Gemini-1.5-flash, and others (as of November 2024). DPC is prepared for more than two models or other LLM tools, but some customisations must be applied.

We utilised data from the COVID-19 proteomic study performed at Instituto Butantan. This study included eight cases categorised by gender, age, and severity, where g2a stands for mild, g2b moderate, and g3 severe COVID-19. Additionally, we included a Medulloblastoma (MB) transcriptomic study conducted by the Human Genetics Lab, USP-RP, comprising the WNT and G4 subtypes. Both studies will be published in the future, pending the completion of the current study.

**Gemini datasets**

The Digital Pathway Curation (DPC) pipeline consists of two datasets: the "Ensemble" and the Multi-Source Dataset (MSD).

The Ensemble dataset is created using three selected groups of pathways from the GSEA table, and one additional group obtained outside the calculated table for each disease case/subtype, which is referred to as the 4 pathway groups. The method to select the four pathway groups (see Figure 1) can be found in the supplementary information (SI).

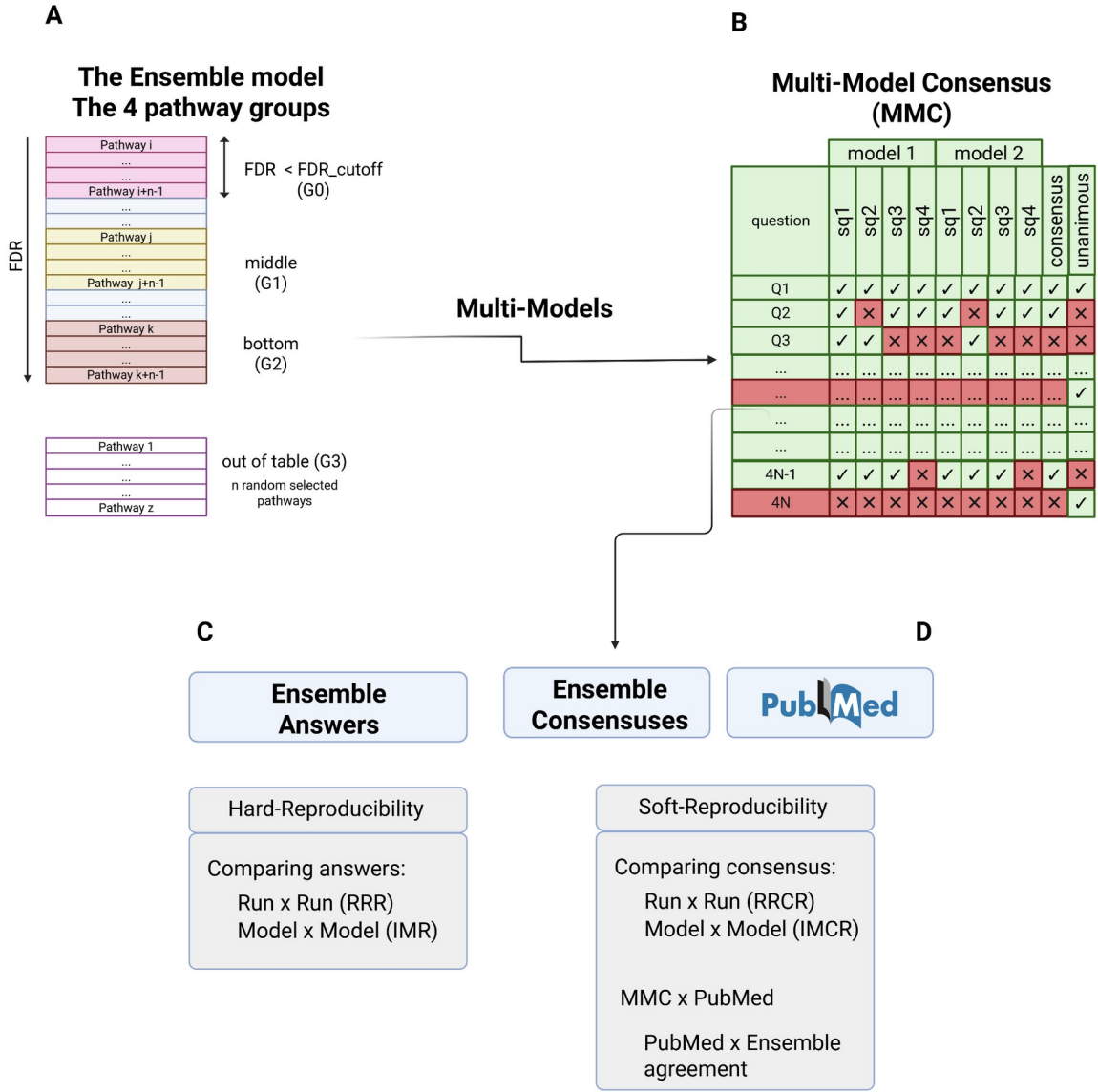

Figure 1 - In A, one can visualise the enriched pathway table from which we gather the selected pathways to build the Ensemble dataset. Pathways were calculated according to GSEA and pathway groups are divided into 4 groups: 1) the default enriched pathways and additional pathways having n pathways (G0), 2) n pathways in the middle of the table (G1), 3) n pathways at the end of the table (G2), and 4) n randomly selected pathways (G3) obtained using the complementary Reactome table without the calculated GSEA pathways. In B, one can visualise the Multi-Model Consensus (MMC) table, and in C, the hard-reproducibility, followed by, in D, the soft-reproducibility, the last, also containing the PubMed x Ensemble agreement.

The second dataset, the MSD, is based on a list of "two cases of randomly selected pathways" (2CRSP) containing 30 randomly selected pathways for each disease case or subtype. It comprises 3 sources (Gemini, PubMed, and Human) that evaluate if each selected pathway is modulated due to the disease. The MSD is a smaller dataset (Figure 2).

One of the DPC goals is to evaluate Gemini's semantic capabilities. To address this goal, we generated four different but semantically similar questions to perform this task. These questions will broadly ("a pathway related to") or specifically ("has a strong relationship to") ask whether a particular pathway is involved in modulating a disease case. Additionally, the second variation focuses on whether the relevant findings can be found on PubMed by adding "<according to PubMed>".

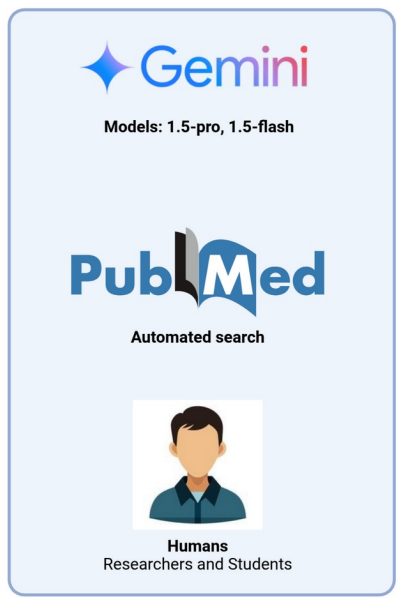

The resulting answers are categorised as 'Yes,' 'Possible,' 'Low evidence,' or 'No,' being counted and summarised in runs, models, cases, and pathway groups.

Figure 2 - The MSD includes the Gemini MMC, PubMed findings, and Human consensus based on the 2CRSP list of pathways. Besides Gemini and PubMed, we include human answers as an additional source required to calculate the CSC.

To summarise, given the Ensemble or the MSD dataset, we formulated the 4DSSQ for each pathway by merging the pathway name and the disease case or subtype to be part of the query template. This task is performed for each disease case and LLM model at least twice (runs >= 2).

Below, we present the basic semantic query format:

"Answer in the first line: Yes, Possible, Low evidence, or No, and explain: Is **<pathway> <related or have a strong relationship>** to **<disease> <severity> <age> <gender> <according to PubMed>**? Context: **<contextualise the disease> + <contextualise the pathway>.**"

The Ensemble allows for evaluating the "hard reproducibility" (focusing on each pathway answer) and the "soft reproducibility" (focusing on each pathway consensus) by comparing results across multiple runs or Gemini models. We also compare each pathway consensus to each PubMed finding using the Gemini MMC.

The first summary table is the One-Model Consensus (OMC), which determines the consensus for each pathway by identifying the most frequently voted answer among the 4DSSQ using a single Gemini model and a single run. Next, we combine many OMCs from different Gemini models, creating a more robust consensus table called the Multi-Model Consensus (MMC).

In addition to Gemini and PubMed results, we introduced human evaluation, which consists of sending the same list of pathways to researchers and students. The MSD allows for calculating the Crowdsourced Consensus (CSC) and assessing the accuracy of Gemini, PubMed, and human evaluations.

**PubMed**

The second source is PubMed, which can be challenging when querying Reactome's pathway names and diseases, often leading to many FNs. To circumvent this SQL-based query limitation, we developed a table that maps the Reactome's pathway names into simplified terms (see SI).

**Human reviewers**

The third source is the Human Reviewer evaluations. These tables, which we collected by sending spreadsheets to human reviewers, are based on pathways from the 2CRSP. This method was created to prevent humans from being overwhelmed with hundreds of questions.

**Crowdsourced consensus**

We established our gold-standard consensus table as the crowdsourced consensus (CSC), which is calculated using the consensus of MSD: Gemini MMC, PubMed findings, and Human consensus (see Figure 3 and SI). Therefore, the crowdsourced consensus can be understood as the "consensus of consensuses".

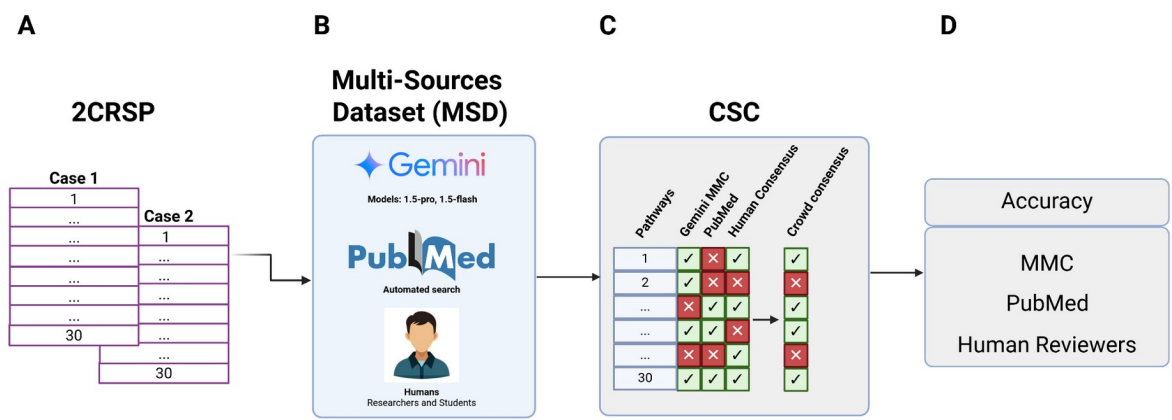

Figure 3 - In A, one can visualise the 2 selected cases and respective pathway lists (2CRSP). The pathways, as queries, were submitted to the Gemini and PubMed web services and shared with researchers and students. Data was collected and combined into the MSD (B). We used the calculated consensus for Gemini and Humans, while for PubMed, a single answer was obtained for each pathway ('Yes' if at least one PMID was found; otherwise, 'No'). In C, merging the multi-source results, we build the crowdsourced consensus table and calculate the CSC values for each pathway. Finally, the three accuracies (D) are calculated using the CSC values.

Having calculated the CSC table, one can calculate all three accuracies by comparing each pathway consensus to the respective CSC value. Gemini accuracy represents the percentage of agreements between the Gemini MMC and the CSC values. The same methodology was applied to calculate the accuracies for PubMed and Reviewers.

**Confusion matrix**

The confusion matrix, generated from the positive and negative control data, calculates accuracy, sensitivity, and specificity. Using the MMC results, we uncover FP and FN pathways. There are 4 confusion matrices (see Figure 4): three for the Ensemble model and another for the Enriched Pathways identified by GSEA. We cannot calculate a confusion matrix using the MSD because it was developed by sampling pathways based on the 1.5-flash Gemini model, which could lead to circular findings.

The Ensemble model allows us to assess the omic experiment effectively by comparing G3 (negative control) and G1 or G2 (intermediary negative controls, the cloudy zone) to G0 (positive control). In this context, the number of TPs and TNs must be confirmed by comparing G3 to G0, since one expects a large number of TPs and TNs in this comparison and an increased number of FPs and FNs by comparing the cloudy zone to G0 (see Figure 4)

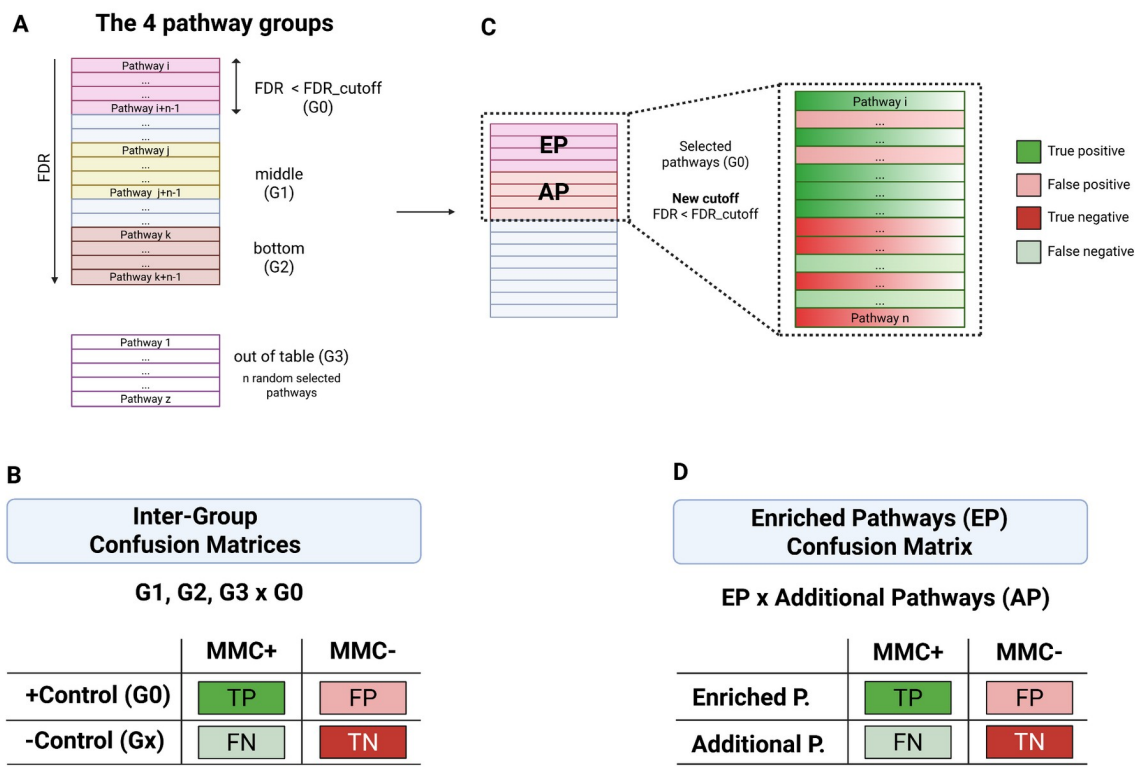

Figure 4 - In A, one can visualise the 4 pathway groups. These groups are used in B to calculate the inter-group confusion matrices. The primary comparison is G3 x G0, representing the true negative and positive controls. However, G1 x G0 and G2 x G0 analyse how the 'cloudy zone' behaves. The default enriched pathways (EP) and additional pathways (AP) can be seen in C. Additional pathways are obtained by flexibilising the cutoff parameters. Below, in D, is the Enriched Pathway Confusion Matrix. MMC uncovers FP and FN pathways for all confusion matrices since it is the closest method to the CSC. It also confirms the TP and TN pathways.

In contrast, the "Enriched Pathway Confusion Matrix" is defined based on the "Enriched Pathways" result, based on the default cutoff parameters, as the positive control, and "Additional Pathways," obtained from relaxed cutoff parameters, as the negative control. Therefore, TPs and TNs can be confirmed, FPs ask if possible LLM mistakes are factual, and FNs reveal possible new pathways to be validated.

## Results:

All data are presented by a value (mean, percentage, or other), followed by the standard deviation in parentheses. Bar errors are calculated according to the t-student distribution confidence interval, which has a 95% confidence level.

### Ensemble

The Ensemble dataset comprises four pathway groups (G0, G1, G2, and G3) containing tens or hundreds of pathways. It incorporates two Gemini models (1.5-pro and 1.5-flash) and four distinct semantic forms of questions for each disease case or subtype posed at different times, called runs (see Methods and SI). This dataset allows for evaluating the run-to-run, inter-model, and semantic reproducibility and comparing the Gemini MMC to the PubMed findings.

**Counts per group**

The curation count plot illustrates the number of 'Yes' responses for each case, categorised by pathway groups G0, G1, G2, and G3. For each case-group combination, it features four-point counts from "four different but semantically similar questions" (4DSSQ). The first solid segment represents model 1 (Gemini 1.5-pro), while the dashed segment corresponds to model 2 (Gemini 1.5-flash). We expect a decreasing number of 'Yes' curation responses as we progress from G0 to G1 and from semantic question 0 (sq0) to the semantic question 3 (sq3). Any discrepancies in the data may indicate inaccuracies in the GSEA pathway ranks, Gemini errors, or differences between the Gemini models.

Below is one example related to COVID-19, the first four cases, plotting the curation count percentages (Figure 1). For a complete overview and detailed explanations, please consult the SI.

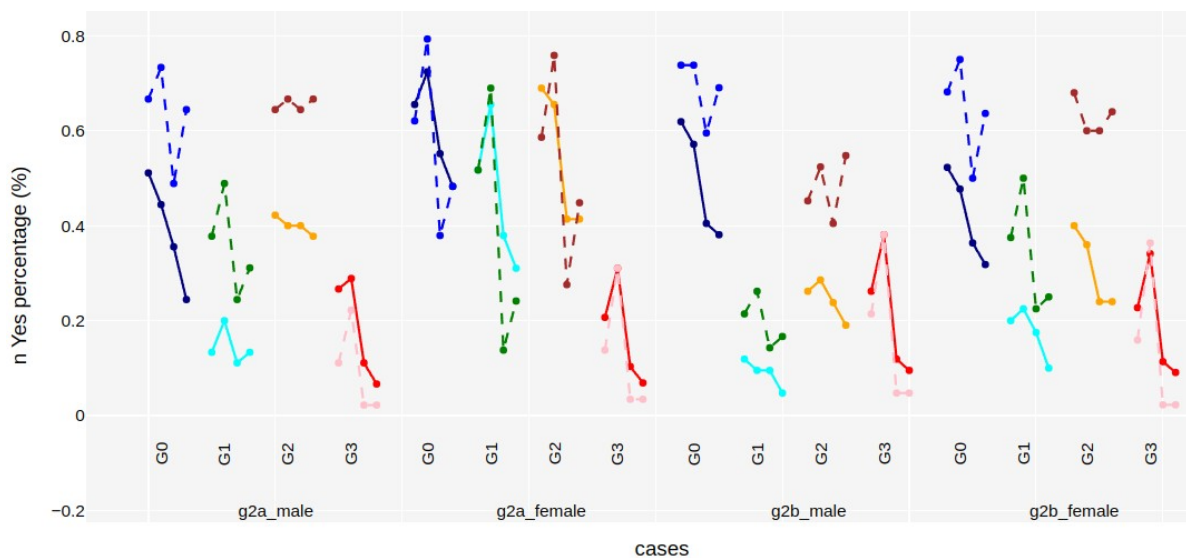

Figure 5 - The curation count plot illustrates the percentage of 'Yes' responses for each pathway group and disease case. Cases are categorised into pathway groups, each with 4 data points related to the four questions of the 4DSSQ. Each set of points on the plot consists of two segments: the solid line represents model 1 (1.5-pro), while the dashed line corresponds to model 2 (1.5-flash). We expect a decreasing percentage of 'Yes' responses as the pathway group increases and the number of semantic questions (sq) increases. Additionally, we expect the solid and dashed lines to overlap if both models provide similar answers.

**Gemini tables: OMC and MMC**

To generate the Gemini responses table for a single run and model, we access the Gemini web service, which provides a table similar to Table 1. The possible responses from the web service are: "Yes," "Possible," "Low Evidence," or "No", seen in the 'curation' column. These results are then consolidated into a one-model consensus (OMC) table by merging the 4DSSQ and calculating the consensus and unanimous columns (see section 'OMC' in the methods and SI). Following this, we can merge multiple OMCs from all selected Gemini models, such as the 1.5-pro and 1.5-flash models. This merging process produces the multi-model consensus table (MMC), which reflects the most robust consensus calculated using all 4DSSQ answers for each model. With the MMC table in hand, one can compare it against PubMed findings and evaluations from human reviewers.

Table 1 shows 5 lines of the MMC table related to the disease COVID-19 case g2a male, from run 'run02', merging model 1 (Gemini 1.5-pro) and model 3 (Gemini 1.5-flash).

|  |  | model 1 – 1.5-pro | | | | model 3 – 1.5-flash | | | | consensus | n yes | n no | unanimous |
|---|---|---|---|---|---|---|---|---|---|---|---|---|---|
| pathway id | pathway name | sq0 | sq1 | sq2 | sq3 | sq0 | sq1 | sq2 | sq3 | | | | |
| R-HSA-109582 | Hemostasis | Yes | Yes | Yes | Yes | Yes | Yes | Yes | Yes | Yes | 8 | 0 | True |
| R-HSA-114608 | Platelet Degranul… | Yes | Yes, | Yes | Low | Yes | Yes | Yes | Yes | Yes | 7 | 1 | False |
| R-HSA-216083 | Integrin Cell Surface … | Low | Yes, | Low | Low | Yes | Yes | No | Yes | Doubt | 4 | 4 | False |
| R-HSA-375276 | Peptide Ligand-Bind… | Yes | Yes | Yes | Yes | Yes | Yes | Low | Low | Yes | 6 | 2 | False |
| R-HSA-381426 | Regulation Of IGF Transp … | Low | Low | Low | Low | No | Low | No | Low | No | 0 | 8 | True |

Table 1 - The Gemini Multi-Model Consensus (MMC) table consolidates all semantic question (sq) responses by merging the selected Gemini models, given a disease like COVID-19, case g2a, from run 'run02'. The rows represent the Reactome IDs and their corresponding names, while the columns display multiple 4DSSQ answers (semantic questions: sq0, sq1, sq2, and sq3), a set of 4 answers for each model. Additionally, a "consensus" column is the most voted answer among all the answers; in the case of a tie, it is marked with a 'Doubt.' The "unanimous" column is marked True if all responses across all models are the same ('Yes' or 'No') and False otherwise.

**Reproducibility**

The Ensemble dataset calculates the hard reproducibility by pairing each pathway for different runs or Gemini models, using a single run. The Run-to-Run Reproducibility (RRR) metric compares all the answers from two distinct runs. Our findings showed an RRR around 99.3% for COVID-19 (see STable 13) and 99.9% for MB (see STable 14), by comparing 'run01' to 'run02'. This high level of reproducibility was not evident when comparing two models. Specifically, we found an Inter-Model Reproducibility (IMR) of about 73.1% for COVID-19 (see STable 19) and 78.4% for MB (see STable 20), using 'run01'.

To calculate the soft reproducibility, one must compare consensuses. The Run-to-Run Consensus Reproducibility (RRCR) metric evaluates all consensuses by pairing all models and pathways from two different runs. Our findings showed an RRCR around 97.5% for COVID-19 (n = 1262) and 99.8% for MB (n = 556) (see STable 21).

This high level of reproducibility was not observed when comparing different models. Using the Inter-Model Consensus Reproducibility (IMCR) metric, we calculated approximately 72.8% for COVID-19 and 62.9% for MB, based on the run equals 'run01' (see STable 32).

**Unanimous Reproducibility (UR)**

The Unanimous Reproducibility (UR) compares whether all 4DSSQs' answers are the same. The results are shown in Table 2.

| run | case | n | n yes | n no | cons yes | cons yes std | cons no | cons no std | unan | unan std |
|---|---|---|---|---|---|---|---|---|---|---|
| run01 | COVID-19 | 1262 | 502 | 760 | 39.8% | 49.0% | 60.2% | 49.0% | 55.5% | 49.7% |
| run02 | | 1262 | 475 | 787 | 37.6% | 48.5% | 62.4% | 48.5% | 59.9% | 49.0% |
| run01 | MB | 556 | 222 | 334 | 39.9% | 49.0% | 60.1% | 49.0% | 33.5% | 47.2% |
| run02 | | 556 | 223 | 333 | 40.1% | 49.1% | 59.9% | 49.1% | 33.5% | 47.2% |
| run03 | | 556 | 222 | 334 | 39.9% | 49.0% | 60.1% | 49.0% | 33.5% | 47.2% |

Table 2 - The UR table was calculated, counting only 'Yes' and 'No' responses. It presents the overall unanimous inner reproducibility and standard deviation for each study.

**Comparing Gemini to PubMed**

PubMed findings were compared to MMC, and the comparisons were categorised based on the "with_gender" variable only for COVID-19, which can be true or false. This distinction is crucial for COVID-19 because PubMed is sensitive to gender, age, and disease severity, while Gemini is not.

Below are the COVID-19 and MB agreements for run equals 'run01' (see Table 3).

| Disease | with gender | <agree> | std | n |
|---|---|---|---|---|
| COVID-19 | False | 72.7% | 6.7% | 1,262 |
| | True | 66.3% | 4.3% | 1,262 |
| MB | False | 66.8% | 5.6% | 556 |

Table 3 - The summary table comparing the PubMed and Gemini agreements for COVID-19 and MB has been filtered to include only entries where the run is 'run01'. For COVID-19, the data is categorised by "with_gender," while for MB, "with_gender" is set to False. Each row represents a study-gender pairing, and the columns display the mean agreement, standard deviation (std), and the number of compared consensuses (n). A more detailed table can be found in the SI file named gemini_x_PubMed_covid_MB.xlsx.

**Crowdsourced consensus (CSC) and Accuracies**

The crowdsourced consensus (CSC) is calculated using the MSD to identify the most voted answer for each pathway. Below are five rows of the CSC table where the filter 'with_gender' is set to False, and to agree means 'Yes' (see Table 4).

| pathway id | pathway | CSC | CSC n yes | CSC n no | agree gemini | agree pubmed | agree review |
|---|---|---|---|---|---|---|---|
| R-HSA-1280218 | Adaptive Immune System | Yes | 3 | 0 | True | True | True |
| R-HSA-389356 | CD28 Co-Stimulation | Yes | 3 | 0 | True | True | True |
| R-HSA-2262752 | Cellular Responses To Stress | Yes | 3 | 0 | True | True | True |
| R-HSA-983169 | Class I MHC Mediated Antigen Processing And Presentation | Yes | 3 | 0 | True | True | True |
| R-HSA-1442490 | Collagen Degradation | Yes | 3 | 0 | True | True | True |

Table 4: The CSC table presents pathways as rows and includes the following columns: 'CSC' as the calculated crowdsourced consensus ('Yes' or 'No'). Additionally, it shows the number of 'Yes' and 'No' responses obtained from the MSD (labelled as 'CSC n yes' and 'CSC n no'). For additional information, see "crowdsource_consensus.xlsx" (in SI).

Below is the accuracy table for COVID-19, for run 'run01' (Table 5).

| case | with gender | PubMed | | Gemini | | Reviewers | |
|---|---|---|---|---|---|---|---|
| | | percent | std | percent | std | percent | std |
| g3 male adult | False | 86.67% | 34.57% | 86.67% | 34.57% | 96.67% | 18.26% |
| g3 male adult | True | 53.33% | 50.74% | 100.00% | 0.00% | 83.33% | 37.90% |
| g3 female elder | False | 60.00% | 49.83% | 96.67% | 18.26% | 73.33% | 44.98% |
| g3 female elder | True | 50.00% | 50.85% | 93.33% | 25.37% | 76.67% | 43.02% |

Table 5 - The accuracy table for COVID-19 presents the accuracy of each source, categorised by cases using run equals 'run01'. The rows are the two selected cases defined in the MSD split by 'with_gender', while the columns include the three accuracies and their respective standard deviations: PubMed, Gemini, and Reviewers.

Next is the accuracy table for MB, for run 'run01' (Table 6).

|  |  | PubMed | | Gemini | | Reviewers | |
| --- | --- | --- | --- | --- | --- | --- | --- |
| case | with gender | percent | std | percent | std | percent | std |
| WNT | False | 80.00% | 40.68% | 86.67% | 34.57% | 86.67% | 34.57% |
| G4 | False | 86.67% | 34.57% | 66.67% | 47.95% | 86.67% | 34.57% |

Table 6 - The accuracy table for MB presents the accuracy of each source, categorised by cases (subtypes), using run equals 'run01'. The rows are the two selected subtypes defined in the MSD, while the columns include the three accuracies along with their respective standard deviations: PubMed, Gemini, and Reviewers. The filter "with_gender" is fixed and equals False.

## Confusion Matrices

### Inter-Group Confusion Matrix

The Inter-Group Confusion Matrix utilises the GSEA results grouping in four pathway groups. It compares the positive control group (G0), the enriched table with additional pathways, against the other three groups. The primary comparison is between the Positive Control Group (G0) and the Negative Control Group (G3).

Below is the G3 x G0 Confusion Matrix regarding statistical values for COVID-19, for run equal to 'run01' (Table 7)

| case | n | sens | spec | accu | prec | f1_score |
| --- | --- | --- | --- | --- | --- | --- |
| g2a_male | 90 | 84.8% | 70.2% | 75.6% | 62.2% | 71.8% |
| g2a_female | 58 | 82.6% | 71.4% | 75.9% | 65.5% | 73.1% |
| g2b_male | 84 | 76.3% | 71.7% | 73.8% | 69.0% | 72.5% |
| g2b_female | 88 | 77.1% | 67.9% | 71.6% | 61.4% | 68.4% |
| g3_male_adult | 48 | 73.7% | 65.5% | 68.8% | 58.3% | 65.1% |
| g3_male_elder | 120 | 75.0% | 62.5% | 66.7% | 50.0% | 60.0% |
| g3_female_adult | 78 | 81.3% | 71.7% | 75.6% | 66.7% | 73.2% |
| g3_female_elder | 78 | 86.2% | 71.4% | 76.9% | 64.1% | 73.5% |

Table 7 - The G3 x G0 Confusion Statistical table for COVID-19 is based on a single run equal to 'run01'. The rows of the table represent different cases, while the columns display the total number of pathways for each case (n = n(G0) + n(G3)) and the sensitivity (sens), specificity (spec), accuracy (accu), precision (prec), and F1-score.

Next is the G3 x G0 Confusion Matrix regarding statistical values for MB, for run equal to 'run01' (Table 8)

| case | n | sens | spec | accu | prec | f1_score |
|---|---|---|---|---|---|---|
| WNT | 136 | 71.7% | 67.1% | 69.1% | 63.2% | 67.2% |
| G4 | 142 | 72.0% | 62.0% | 65.5% | 50.7% | 59.5% |

Table 8 - The G3 x G0 Confusion Statistical table for MB is based on a run equal to 'run01'. The rows of the table represent different subtypes, while the columns display the total number of pathways for each case (n = n(G0) + n(G3)) and the sensitivity (sens), specificity (spec), accuracy (accu), precision (prec), and F1-score.

**Cloudy zone**

As mentioned, G0 is the positive control, while G3 is the negative control. It means that when we shift our comparisons from G1xG0 and G2xG0 to G3xG0, we can expect an increase in true values and a decrease in false values. In other words, comparing G1 and G2 (the cloudy zone) to G0 will likely yield fewer TPs and TNs and a higher number of FPs and FNs. In contrast, by comparing G3 to G0, we must achieve the maximum number of TPs and TNs while minimising FPs and FNs. Therefore, as we approach the comparison of G3xG0, both sensitivity and specificity are expected to trend toward 1.

Below, you can see the COVID-19 Sensitivity, Specificity, Accuracy, Precision, and F1-score by comparing G1, G2, and G3 to G0 (see Figure 6).

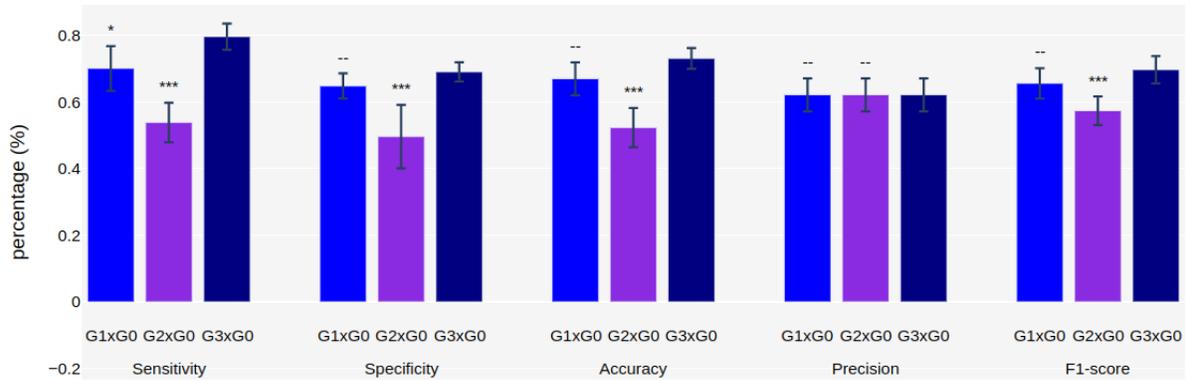

Figure 6 - Each statistical method (Sensitivity, Specificity, Accuracy, Precision, and F1-score) can be plotted for each comparison: G1xG0, G2xG0, and G3xG0, for COVID-19. The t-tests contrast G1xG0 and G2xG0 compared to G3xG0 (true negative control x positive control), for each statistical method. The bar errors represent the confidence interval at a 95% confidence level, and t-test p-values are represented as * for p-value < 0.05, ** p-value < 0.01, and *** p-value < 0.001.

Next, you can see the MB Sensitivity, Specificity, Accuracy, Precision, and F1 Score by comparing G1, G2, and G3 to G0 (see Figure 7).

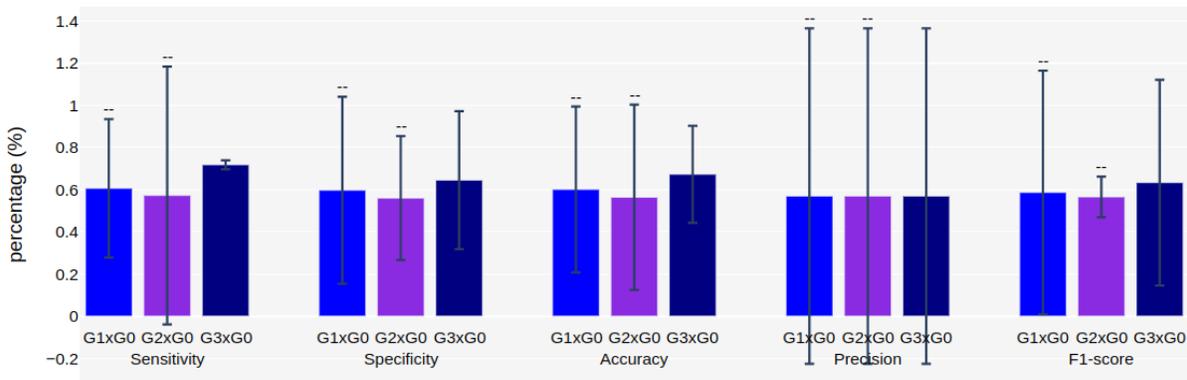

Figure 7 - Each statistical method (Sensitivity, Specificity, Accuracy, Precision, and F1-score) can be plotted for each comparison: G1xG0, G2xG0, and G3xG0, for MB. The t-tests contrast G1xG0 and G2xG0 compared to G3xG0 (true negative control x positive control), for each statistical method. The bar errors represent the confidence interval at a 95% confidence level, and t-test p-values are represented as * for p-value < 0.05, ** p-value < 0.01, and *** p-value < 0.001.

**Enriched Pathways Confusion Matrix**

The Enriched Pathway Confusion Matrix utilises the calculated enriched pathways based on the default cutoff parameters and the additional pathways obtained from relaxed cutoff parameters. The algorithm designates the enriched pathways as TPs and the additional pathways as TNs to initiate the construction of the confusion matrix. Following this, FPs and FNs are uncovered using the results from the MMC.

Below is the Enriched Pathway Confusion Matrix statistics for COVID-19, for run equals 'run01' (Table 9)

| case | n | sens | spec | accu | prec | f1_score |
|---|---|---|---|---|---|---|
| g2a_male | 45 | 0.0% | 100.0% | 37.8% | | |
| g2a_female | 29 | 0.0% | 100.0% | 34.5% | | |
| g2b_male | 42 | 0.0% | 100.0% | 31.0% | | |
| g2b_female | 44 | 88.9% | 29.4% | 65.9% | 66.7% | 76.2% |
| g3_male_adult | 24 | 21.4% | 100.0% | 54.2% | 100.0% | 35.3% |
| g3_male_elder | 60 | 26.7% | 90.0% | 58.3% | 72.7% | 39.0% |
| g3_female_adult | 39 | 96.2% | 0.0% | 64.1% | 65.8% | 78.1% |
| g3_female_elder | 39 | 88.0% | 0.0% | 56.4% | 61.1% | 72.1% |

Table 9 - The Enriched Pathway Confusion matrix statistics for COVID-19 were calculated using run equals 'run01'. The rows of the table represent different cases, while the columns display the total number of pathways for each case (n) and the sensitivity (sens), specificity (spec), accuracy (accu), precision (prec), and F1-score.

Next is the Enriched Pathway Confusion Matrix statistics for MB, for run equals 'run01' (Table 10)

| case | n | sens | spec | accu | prec | f1_score |
|---|---|---|---|---|---|---|
| WNT | 45 | 16.3% | 68.0% | 35.3% | 46.7% | 24.1% |
| G4 | 29 | 0.0% | 100.0% | 49.3% | | |

Table 10 - The Enriched Pathway Confusion matrix statistics for MB were calculated using run equals 'run01'. The rows of the table represent different cases, while the columns display the total number of pathways for each case (n) and the sensitivity (sens), specificity (spec), accuracy (accu), precision (prec), and F1-score.

**Discussion:**

Some experiments yield a low number of DEGs or DAPs, which can lead to an inability to predict pathway enrichment due to a perturbation or disease. Possible explanations include a low level of perturbation (such as a low dosage, mild disease, or reaching a recovery state) or improper sampling timing; many studies rely on a single sampling point. Therefore, researchers should assess how different gene expressions or protein productions change over time before designing an experiment. After collecting all data, if the number of DEGs or DAPs is still low and no enriched pathways are identified, researchers may need to abandon the experiment or adjust their cutoff parameters. One potential approach to continue the analysis is to lower the LFC cutoff and increase the FDR cutoffs.

For this purpose, we created the Digital Pathway Curation algorithm, which evaluates the significant modulation of the additional DEGs by uncovering extra pathways using AI, PubMed, and researchers' evaluations.

**Ensemble model**: First, we constructed a large dataset known as the Ensemble to measure the LLM reproducibility and to ensure robust statistical analysis. To construct the Ensemble, we cross all selected pathways (G0, G1, G2, and G3) to the 4DSSQ, adopting a standard query format to guide the LLM in focusing its answers on a specific region of the embedding hyperspace while testing its semantic capabilities (see methods and SI). We designed the queries to begin with a preamble, merging each pathway name and crossing to each disease case or subtype. The proposed query format is in the form: "<preamble> is the <pathway> related to <disease-case> <context>." The preamble is fixed - "Answer Yes or No in the first line" - ensuring that the model systematically responds with a "Yes" or "No" (or "Yes", "Possible", "Low evidence", or "No"). It ends with a short context about the pathway and disease. Our approach demonstrated high reproducibility when running the queries at different moments and with various Gemini models. We did not observe any generative hallucinations; additionally, we adjusted some parameters of the Gemini model, explicitly setting the model temperature to a low value equal to 0.1.

Regarding the Ensemble Reproducibility methods, hard reproducibility is an inflexible method for "comparing answers" of different runs or models using all questions. It is the most statistically powerful method implemented in the present study, as it evaluates responses across hundreds of pathways in various models, incorporating four different ways of posing questions, the 4DSSQ.

The Run-to-Run reproducibility (RRR) method compares all answers from two different runs, given all Gemini models and disease cases/subtypes. The RRR measures how consistently a generative language can respond without hallucinations, which are, in the best cases, random responses resulting in a percentage of around 50%. Gemini achieved a mean RRR of around 99.3% for the COVID-19 study and around 99.9% for the MB study. These results indicate a remarkable level of reproducibility for an LLM in its early development stages, particularly given that no refinements have been made (see also STable 13 and 14).

The Inter-model Reproducibility (IMR) compares two Gemini models based on a single run, making it a valuable measure for assessing the agreement of different models to address biomedical questions. Using run01 and comparing 1.5-pro and 1.5-flash models, IMR values were approximately 73.1% for the COVID-19 study and 78.4% for the MB study. These percentages, ranging from 73% to 78%, are far from 95% (p-value = 3.40e-193). Consequently, there is a discrepancy of approximately 25% between Gemini models in the context of hard reproducibility. Developing refined training models utilising biomedical texts will likely yield better IMR results. As mentioned, some initiatives in this area have already begun, but this is not the focus of our study. For more detailed data, please refer to STable 19 for COVID-19 and STable 20 for MB.

We then examined soft reproducibility, which evaluates Gemini's semantic capabilities by "comparing consensuses." There are two consensus measures: the One-Model Consensus (OMC) and the Multi-Model Consensus (MMC). The OMC calculates the consensus of the 4DSSQ answers using a single Gemini model in a single run. In contrast, the MMC merges multiple models, such as the 1.5-pro and 1.5-flash. The MMC is suggested as a more robust semantic measure because it aggregates the four different semantic responses from many different models.

The Run-to-Run consensus reproducibility (RRCR) uses all OMC results by comparing two runs. The RRCR measures how consistently an LLM can respond systematically. Gemini achieved an RRCR of about 97.5% for COVID-19 and 99.8% for MB. Like in the hard reproducibility methods, these results indicate a remarkable level of reproducibility for an LLM.

The Inter-model Consensus Reproducibility (IMCR) method compares two Gemini models based on a single run. The IMCR measures how consistently an LLM can systematically respond by comparing two models. The IMCR ranges from around 72.8% to 76.1% for COVID-19 and 62.9% to 63.5% for MB, given 'run01' and 'run02'. Given that COVID-19 was extensively studied during the pandemic, it can be inferred that the corpus used to train the LLM is more extensive for this topic than MB, resulting in higher IMCR for COVID-19 (see STable 32 and 33).

The Unanimous Reproducibility (UR), or inner-model response reproducibility, evaluates whether the 4DSSQ yields the same semantic answers merging all models using a single run. For each question, the algorithm evaluates whether the answers vary; if they do, they are considered semantically inconsistent, while equal answers are deemed semantically consistent. For COVID-19, the UR was around 55.5%, while for MB 33.5%, for run equals 'run01'. These results showed that reaching unanimous responses on MB was more challenging than COVID-19. These findings suggest the potential for improvement in Gemini's biomedical searches.

The last method, using the Ensemble dataset, compares the MMC results to PubMed findings. The agreement between PubMed and Gemini for COVID-19 was approximately 66.3% when gender was included in the PubMed queries and 72.7% when it was not. For MB, a gender independent study, the agreement was approximately 66.8%. It indicates that MMC, the most robust consensus, disagrees with PubMed by approximately 30%. The next step, the CSC calculus, will assess which of the two sources may be more accurate with the help of human reviewers' responses.

**MSD:** In the second part of the study, we focused on building a smaller dataset called the Multi-Source Dataset (MSD). As mentioned above, it is smaller since only 2 disease cases/subtypes with 30 randomly selected pathways were used to avoid overloading human reviewers. With the MSD, one can calculate the CSC, which is the most voted among the three sources: Gemini MMC, PubMed findings, and Human Reviewers' evaluations. Finally, we can calculate the three accuracies using the CSC as the gold standard.

Regarding COVID-19 research from PubMed, each pathway may generate a different CSC depending on whether the "with_gender" flag is True or False. For MB, "with_gender" is set to False since, as mentioned, MB subtypes are independent of gender. For COVID-19, setting "with_gender" to True leads PubMed to poorer responses (see Table 5) and biases the CSC towards 'No', resulting in different statistics compared to "with_gender" equals False.

The small size of the MSD may lead to potential biases: 1) When searching for COVID-19 pathways on PubMed, fewer pathways are identified when the gender is included, leading to an increase in FNs. This bias does not occur in MB, a disease unrelated to gender. 2) When examining the MB literature, one observes that there is significantly less material available in PubMed than in COVID-19 research. As of March 2025, there are approximately 12 thousand findings for MB, compared to more than 450 thousand studies for COVID-19. This limited corpus, at the time of the LLM training, makes it more challenging to locate articles when queries include cases/subtypes and pathway names. 3) In future studies, we will incorporate other LLMs and human reviewer groups to improve the CSC sources. The CSC with only 3 sources may lead to another bias: When human reviewers are more lenient in answering 'Yes', the CSC drifts further away from PubMed findings, while a more sceptical approach brings it closer to PubMed findings.

Additionally, the PubMed search tool does not handle multiple terms effectively. Successful retrieval of articles relies on authors clearly stating all necessary terms in the article's title, keywords, abstract, and text body. It is important to note that a free search on PubMed also looks at the reference section, which may lead to many FPs. The DPC PubMed algorithm restricts its searches to the title, abstract, and keywords to circumvent this problem, possibly increasing the FNs. Moreover, Gemini and PubMed may also produce FNs if the articles are not freely accessible or hidden within patent documentation.

The obtained accuracies using the two severe COVID-19 cases were as follows: 1) Gemini's accuracy was over 86% for both cases, independent of gender. These results demonstrate the ability of this semantic tool to retrieve consistent information. 2) PubMed's accuracy was around 87% for adults and 60% for elders, both with severe COVID-19. When filtering by gender, the accuracies dropped to approximately 53% for male adults and 51% for female elders. These results highlight PubMed's challenges in retrieving texts about severe COVID-19 when the gender is explicitly included in the query. 3) The reviewers' accuracy was close to that of Gemini. The reviewers' accuracies, having the CSC calculated without gender, were around 97% for male adults and 73% for female elders. However, by filtering the gender, the CSC changes, and the accuracies were around 83% for male adults and 77% for female elders.

The accuracies using the MB subtypes were: 1) Gemini's accuracy was around 87% for WNT and 67% for G4. As mentioned in the bias discussion, these results highlight the challenges Gemini faces in obtaining data for G4 compared to the WNT subtype. 2) PubMed's accuracy was around 80% for WNT and 87% for G4. As mentioned, the gender filter was not used for MB, and PubMed consensuses are prone to 'No' findings, which will probably result in CSC bias. 3) Reviewers' accuracies were around 87% for WNT and the same value for G4. These results demonstrate that, for the G4 subtype, Gemini's accuracy decreases and PubMed's accuracy increases, probably because there are fewer published studies compared to COVID-19.

**Confusion matrix:** In the third and last part of the study, we focused on building four confusion matrices. The Inter-Group Confusion Matrix compares G3 (negative control) and G1 and G2 (cloudy zone) to G0 (positive control) to assess GSEA effectiveness. Next, the Enriched Pathways Confusion Matrix aids in uncovering FP and FN pathways.

The Inter-Group Confusion Matrix provided intriguing insights. In addition to demonstrating high values for Accuracy and F1-Score, it revealed a significant number of FP and FN (see STable 50 for COVID-19 and STable 51 for Medulloblastoma). Regarding COVID-19, proteomics cannot find ten thousand different proteins, which is probably leading to low statistical power. Regarding the MB transcriptomics, the bias was introduced due to using a specific set of microarray probes dedicated to Medulloblastoma and including many long non-coding RNAs (lncRNAs) for a particular purpose.

The Enriched Pathways Confusion Matrix was the final goal for DPC. Identifying FPs and FNs among the enriched and additional pathways is a crucial outcome of bioinformatics research. Both disease studies show that there are zero or too few DEGs and DAPs in the following cases: g2a male and female, g2b male, and g3 adult male for COVID-19 (see STable 56), as well as both subtypes for MB (WNT and G4, see STable 57). Utilising the results from the MMC, the current method has identified numerous FPs and FNs that warrant further investigation in future studies.

> Box 1
>
> Some studies in omics experiments are powerful enough to distinguish between enriched pathways (the positive control group) and all other remaining pathways (the negative control group), others are not. However, it is essential to examine the cloudy zone (other pathways close to the cutoff region) and adjust the cutoff parameters to identify additional significant pathways.
>
> The Digital Pathway Curation pipeline (DPC) focuses on confirming whether the newly identified pathways have scientific evidence using LLMs and PubMed tools. Large Language Models (LLMs) are the primary tool supporting the DPC. They demonstrated about 99% reproducibility for both the run-to-run and run-to-run consensus reproducibility, RRR and RRCR, respectively. However, when comparing two models using the inter-model consensus reproducibility (IMCR), the reproducibility dropped by approximately 25%.
>
> The multi-model consensus (MMC) technique, combined with the Crowdsourced Consensus (CSC), showed that Gemini MMC is the most robust technique compared to PubMed findings and Human consensus. Therefore, MMC was used to support the calculation of the confusion matrices.
>
> Final calculus presents statistical analyses based on two distinct types of confusion matrices: 1) the inter-group and 2) the enriched pathway confusion matrix. The inter-group confusion matrix statistic assesses whether the cloudy zone (G1 and G2) is well separated from the enriched pathway (G0), supporting the adjusted cutoff parameters. However, the enriched pathway confusion matrix statistic supports uncovering the FP and FN pathways.

To summarise the main findings, the DPC pipeline was based on two data sources, each serving different purposes: 1) The Ensemble model involves a large dataset comprising selected pathways that allow for measuring the reproducibility of the LLM. The Ensemble dataset allows us to compare multiple runs and different LLM models using each pathway answer (hard reproducibility) or each consensus (soft reproducibility). The consensus reproducibility is a more flexible method, and the Gemini multi-model consensus (MMC) was used to compare Gemini to PubMed, thereby helping to evaluate possible limitations of the PubMed search. 2) The MSD was proposed to incorporate human responses to merge with Gemini MMC and PubMed findings. Due to the absence of a gold standard, we cannot ascertain whether there are relationships between pathways and diseases. Therefore, we proposed the CSC as the consensus among all values of the MSD (see also Box 1).

Our findings showed that both Gemini models responded reliably and reproducibly. Specifically, when questions are formulated correctly, following a previously outlined format and with the LLM temperature set to low values, this technique almost eliminates the likelihood of generating hallucinations and ensures reproducibility. However, when we assessed the reproducibility among Gemini models (IMCR), we found a concordance of about 75%, meaning that two trained Gemini models might produce different responses about 25% of the time. Our analysis showed that Gemini models are reproducible, but their performance depends on the specific model used.

DPC assessed the accuracy of Gemini, PubMed, and Humans by comparing them to the CSC values. The accuracy of the Gemini MMC was notably high (around 87%), showing that we can increasingly trust AI-generated responses. Compared to PubMed and human accuracies, we found that the MMC was the most accurate for inquiries about the relationship between pathways and diseases, specifically in COVID-19 and MB.

Limitations on PubMed search may be overcome by refining LLM and training with all available PubMed data or other techniques[25]. With AI support, one expects a significant decrease in false negatives.

Finally, we asked whether there is a method to demonstrate if "selected molecular biological pathways, calculated beyond the default cutoffs using GSEA, have relationships to a disease case". We proposed creating the Enriched Pathways Confusion Matrix to address this question, which can identify the FP and FN pathways using the MMC results. Discovering FPs and FNs was noteworthy, as no standard tools exist for their explicit detection. We presented these findings to several experts for review. They disagreed with some of the identified FPs, which we call the False-False Positives (FFP), while agreeing with many identified FNs. The FFPs may indicate potential errors in the Gemini responses, possibly due to a lack of supporting literature or a more complex training model. In contrast, the FN pathways were not detected by bioinformatics analysis due to restrictive default cutoff parameters, and uncovering them represents potential new hypotheses that can be explored in future studies.

The DPC pipeline was created to automatically curate biomedical literature, confirming TP and TN pathways and suggesting FP and FN ones. It confirmed that the Gemini multi-model consensus (MMC) effectively addressed biomolecular questions regarding the COVID-19 proteomics and the Medulloblastoma microarray studies.

**Data and Code Availability:**

The pipeline was developed in Python. Data, code, and Jupyter notebooks are publicly available at https://github.com/flalix/DPC, data at https://drive.google.com/drive/u/0/folders/1U6FBkKGE4SisHXUR9RhNiF6CyOQUa200, and respective documentation at https://mtdp.readthedocs.io/en/latest/index.html.

## Acknowledgements:

We thank the support and discussions with PhD Anatoly Yambartsev from the Mathematical Institute of USP (IME). We also thank the support of the Fundação Butantan and the Instituto Butantan. All drawings were created at [https://BioRender.com](https://BioRender.com).

## Author contributions:

Idea, development, testing, and writing (FL); testing, discussion, and writing (CEMT, DAS, LRL); initial ideas, discussion, and writing (VWSG, JPB, VMB, AMF, FALS, SAA); discussion and writing (OCMI, NS, LFO, AMCT).

## Competing interests:

The authors declare that they have no competing interests.

# Supplementary Information

## Introduction:

The scientific method - based on observation, hypothesis generation, feasible and reproducible experimental design, execution, data annotation, data analysis, hypothesis testing, and communication - relies on evidence and reproducibility. Moreover, the literature review is a crucial first step. Since most scientific literature stored in PubMed is published following peer review, one infers that all these articles follow the scientific method and are based on evidence, a statement questioned by some [8–12]. Since PubMed is a reproducible SQL-based engine, the responses are identical once queried using the same search terms in a short interval between queries. However, PubMed's accuracy is not as high as expected, particularly concerning false positives and negatives, which are often overlooked.

In the present study, Digital Pathway Curation confirms whether selected pathways are involved in modulating due to a disease. We propose using artificial intelligence, specifically the Gemini LLM models, to establish a reproducibility measure. The first step of our pipeline involves creating a large set of queries, called the Ensemble model. We can assess hard reproducibility by comparing the responses between runs or models. Additionally, we introduce a consensus method, which identifies the most frequently chosen answer among "four different but semantically similar questions" (4DSSQ). Using this consensus, we propose a measure of soft reproducibility, which involves comparing the results from different runs or models based on the consensus results.

Next, we propose a smaller dataset called the Multi-Source Dataset (MSD), based on a list of "two cases of randomly selected pathways" (2CRSP), containing randomly selected pathways for each case. This dataset includes a list of 30 randomly selected pathways related to each disease case or subtype. With MSD results, we can calculate the crowdsourced consensus score (CSS[25]) and assess the accuracy of each source.

The DPC pipeline evaluates the ability of LLMs to determine whether selected pathways are involved in modulating due to a disease, utilising GSEA results based on the Reactome database. Given an omic study, DPC selects calculated pathways stored in the GSEA pathway table using the "four pathway groups" method. These pathways are then used to perform the Ensemble dataset analyses.

The Ensemble dataset comprises the four pathway groups containing tens or hundreds of pathways. It incorporates at least two Gemini models and four distinct semantic forms of questions posed at different times, called runs. This dataset allows for evaluating the run-to-run, inter-model, and semantic reproducibilities. To analyse the semantic reproducibility, we introduced the concept of the "four different and semantically similar questions" (4DSSQ). Additionally, Tiwari et al. from the EMBL-EBI developed a similar approach to test the Reactome database completion using ChatGPT[24]. We also compared the Ensemble answers with those obtained from PubMed searches.

The CSC allows us to assess the accuracy of each source. Using the source with the best accuracy according to the CSC, we can uncover False Positives (FP) and False Negatives (FN) while also confirming True Positives (TP) and True Negatives (TN) among the enriched pathways calculated with GSEA.

In the upcoming sections, you will find a detailed description of each method and the results obtained.

## Methods:

All pipeline code was developed with Python version 3.11.9 and Gemini versions 1.5-pro and 1.5-flash. Private Python libraries, jupyter notebooks, and data can be found at https://github.com/flalix/DPC.

### Limiting the scope

In the present study, we limited our scope to PubMed and Gemini as search engines and Reactome as the pathway database. As mentioned above, PubMed is a scientific retrieval open-access reference tool, Gemini is Google's generative AI tool based on many different LLM models, supposedly trained from open-source texts on the open internet, plus other unknown sources and methods, and Reactome is the European Pathway database.

We focused on Gemini since Google is one of the first companies to develop an open-access AI tool due to the Web Service and API programmatic tools and the availability of multi-models such as Gemini-1.0-pro, Gemini-1.5-pro, Gemini-1.5-flash, and others.

### Data and studies

We utilised data from the COVID-19 study conducted at Hospital Municipal de Taubaté and proteomics analysis performed at Instituto Butantan, which will be published soon. This study included eight cases categorised by gender, age, and severity. Additionally, we included a transcriptomic study on Medulloblastoma (MB) conducted by the Genetics Lab, USP-RP, comprising the WNT and G4 subtypes, which will also be published in the future.

## DPC methods

### DPC datasets: Ensemble and DSP

The Digital Pathway Curation (DPC) pipeline consists of two datasets: the "Ensemble" and the Multi-Source Dataset (MSD). The ensemble dataset was created using three selected sets of pathways from the GSEA table, and one additional group was obtained outside the calculated table for each disease case/subtype, which is referred to as the 4 pathway groups. For each pathway, we formulate 4 different and semantically similar questions (4DSSQ) by merging the pathway name and the disease case or subtype. We used two accessible Gemini models to determine whether each pathway is modulated by or modulates the disease, performing multiple runs. The MSD comprises answers using the "two cases of randomly selected pathways" (2CRSP), containing 30 randomly selected pathways.

The Ensemble allows for evaluating "hard reproducibility" (comparing individual answers) and "soft reproducibility" (comparing the calculated consensuses) by examining results across multiple runs and Gemini models and comparing Gemini to PubMed.

In addition to having Gemini and PubMed answers, we introduced the human evaluation, which consists of sending the same 30 questions to researchers and students. The MSD allows for calculating the Crowdsourced Consensus (CSC) and assessing the accuracy of Gemini, PubMed, and human evaluations.

### The 4 pathway groups

The ensemble dataset aims to create sets of question-answers, selecting calculated pathways from the gene set enrichment analysis (GSEA) algorithm using Reactome pathways from omic experiments. First, one identifies statistically significant pathways calculated from the GSEA table using the default parameters (LFC cutoff=1 and FDR cutoff=0.05), called enriched pathways. However, we also explored whether additional pathways that follow along the table have biological meaning and can aid in studying each disease case. To do this, we introduce the first group (G0) of pathways, representing the positive control group consisting of the default enriched pathways and additional pathways selected by relaxing the cutoff parameters. Next, we select two cloudy groups and one negative control group, as detailed below. Each pathway allows for the building of queries arguing different Gemini models and PubMed, with the last using pathway simplified terms (see PubMed methods). Once we have compiled the question-answer ensemble, we can tally all "Yes" and "No" responses, calculate the consensus (the most frequently selected answers among the 4DSSQ, defined below), and conduct statistical tests.

Previously, we explained how to create G0; here, we explain how to create the following 3 pathway groups. The pathway groups are: 1) Group 0 (G0), the first set consisting of N top pathways calculated according to flexible cutoff parameters; 2) Group 1 (G1), the second set consisting of N pathways selected at the middle of the table; 3) Group 2 (G2), the third set consisting of N pathways selected at the very end of the table. Since we use the Reactome database, we can also find additional pathways that do not belong to the calculated GSEA table. Once Reactome has 2,610 annotated pathways (version 4.0.2, build 117, October 14, 2023) and the GSEA table has found about 1200 pathways, we randomly draw N pathways in the complementary pathway list, performing the last group, G3. Therefore, one can define G0 as the positive control group, G1 and G2 as cloudy zone groups, and G3 as the negative control group, as seen in SFigure 1. Related to the TP pathways, one expects that the number of TP(G3xG0) > TP(G2xG0) > TP(G1xG0), a similar tendency related to TNs, and an opposite tendency related to the FNs and FPs. Here, we impose N, an arbitrary number of selected pathways, but N comes from the flexibilisation of pathway cutoffs, a concept that is out of the scope of the present study.

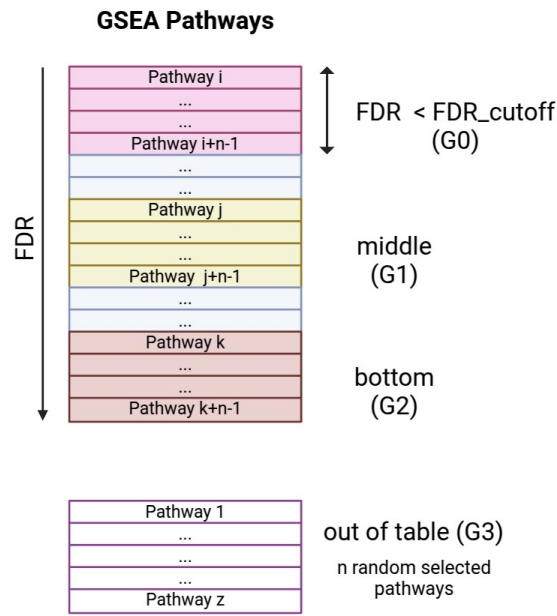

SFigure 1 - Above, one can visualise the GSEA table. Pathway groups are divided into 4 groups, the last, outside the table: 1) the enriched pathways and additional pathways having n pathways (G0), 2) n pathways in the middle of the table (G1), 3) n pathways at the end of the table (G2), and 4) n randomly selected pathways (G3) obtained using the complementary Reactome table without the calculated GSEA pathways.

**Controlled Query Methodology**

The main questions we started this study with were: How can one query Gemini in natural language and retrieve reproducible answers without hallucinations? How can one gather a positive or negative answer to infer if a pathway is related to a disease?

Below, one shows the basic query merging the disease case/subtype with a pathway:

"Is the <pathway> related to the <disease> for <case>?"
or,
"Is the <disease> related to the <pathway> for <case>?"

Concerning LLMs, a user asks a question in natural language, and LLM answers it in natural language. However, if the responses are ordered phrases with many paragraphs, how can we infer if the response has a positive mood (Yes) or a negative mood (No)? To summarise, we have two alternatives: a) calculate the sentiment analysis of each response (as a whole text or by splitting and weighting the paragraphs), or b) add a preamble to the question that guides the LLM to start answering with a 'Yes' or 'No.' The former method results in many uncertainties because the answer consists of many paragraphs, some with positive sentiments and others with negative ones (results not shown here). After many simulations, the process using a <preamble> yielded consistent results and demonstrated that, besides being a semantic tool, Gemini could reason and organise the answer the way we asked.

Given a pathway name, one can formulate the following query:

"**Is <pathway> related to <disease> <subtype> <severity> <patient classification>?**"

The machine will answer this question, write explanations, and a conclusion paragraph. Therefore, to gather a Yes or No, one must add a preamble (in red):

"**Answer in the first line, Yes or No;** is <pathway> related to <disease> <subtype> <severity> <patient classification>?"

This approach works fine according to many implemented tests, and we learned that one can try to introduce more qualitative classes than 'Yes' or 'No'. Observe the new terms in red:

"**Answer in the first line Yes, Possible, Low evidence, or No; is <pathway> related to <disease> <subtype> <severity> <patient classification>?**"

Depending on the selected Gemini model, the answer will return only Yes or No; other models may return Yes, Possible, Low evidence, or No. Moreover, by introducing this preamble, one loses the explanation if the machine decides only to answer shortly. Here, again, one can improve by adding "and explain" (in red):

"**Answer in the first line Yes, Possible, Low evidence, or No, and explain: Is <pathway> related to <disease> <subtype> <severity> <patient classification>?**"

Finally, one must add contexts to the query to help the LLM focus and improve reproducibility. Contexts direct the machine to a defined subspace in the high-dimensional embedding space. Therefore, the final proposed format is,

"**Answer in the first line Yes, Possible, Low evidence, or No, and explain: Is <pathway> related to <disease> <subtype> <severity> <patient classification>? Context: <contextualise the disease> + <contextualise the pathway>.**"

Above, in green, are the keywords that must be replaced for each pathway name and disease case/subtype name. The disease context must be a very short sentence (or sentences) regarding the studied disease and case/subtype. However, the pathway context can be retrieved from the chosen database. Since we are using Reactome, it has a small abstract for each annotated pathway, accessible on the website and downloadable through a Python API.

**The 4 different and semantically similar questions (4DSSQ)**

One of the main ideas of the present work is to test Gemini's semantic capability. Therefore, we propose to build the "4 different and semantically similar questions" (4DSSQ) for each pathway, allowing semantic variations.

The proposed method semantically varies the original query, presented in the last section, asking in a general and more restricted form whether a pathway is related/modulated due to a disease case. The second variation focuses on whether the achievements can be found on PubMed.

Here, a caveat must be enlightened. The two Gemini models, 1.5-pro and 1.5-flash, answer semantically using Yes, Possible, Low evidence, or No. Therefore, using the 4DSSQ, the qualitative answers enable us to quantify them probabilistically, transforming them into weights [1.0, 0.7, 0.35, 0.0] respectively, and measuring model differences.

**The semantic questions (sq)**

The original query is defined as:
"Answer in the first line Yes, Possible, Low evidence, or No, and explain: Is <pathway> related to <disease> <subtype> <severity> <patient classification>? Context: <contextualise the disease> + <contextualise the pathway>."

Below, one proposes 4 changes substituting "Is <pathway> related to <disease>" with new semantically similar terms,

1) Semantic Question 1 - general form without the term "PubMed":

    "Is <pathway> studied about the <disease>"

2) Semantic Question 2 - general form with "in PubMed":

    "Is <pathway> studied in PubMed, about the <disease>"

3) Semantic Question 3 - strong relationship without the term "PubMed":

    "has the <pathway> a strong relationship in studies related to the disease <disease>"

4) Semantic Question 4 - strong relationship with the term "in PubMed":

    "has the <pathway> a strong relationship in studies found in PubMed, related to the disease <disease>"

The four questions are different; however, they are semantically very close and named 4DSSQ. The first two are flexible about the disease, and the last 2 two ask for a strong relationship to the disease. It becomes the responsibility of the LLM to interpret these semantically similar nuances and answer in the same way or not. Finally, when one adds "found in PubMed," one expects that LLM restricts the search in the embedding space, focusing on vectors related to PubMed-related terms.

**Gemini response table**

DPC scans the four pathway groups, each containing 'n' pathways. For each pathway, it formulates queries based on the semantic questions and submits them to the corresponding Gemini web service model. If the response fails, DPC makes two additional attempts. Once a response is received, it begins with one of the following terms: 'Yes,' 'Possible,' 'Low evidence,' or 'No.' The algorithm retrieves this answer and stores it in the column labelled 'curation.' This method generates a table (see STable 3) for each run, Gemini model, semantic question, and disease case. The table includes Reactome IDs and names as rows, along with columns for 'curation,' 'response explanation' (which details the response), 'question' (the query), and other relevant columns.

**Run, Models, Consensus, and Reproducibility**

To evaluate the reproducibility of the ensemble model, one can compare two runs or two models based on their "answers" or "consensus." We use this variables as "hard reproducibility" when comparing answers and "soft reproducibility" when comparing consensuses. A run is a complete loop of queries covering all pathways for each disease case or subtype, using all selected Gemini models and the 4DSSQ. These queries, which can total hundreds or thousands, are typically completed within a few hours a day. The selected Gemini models refer to the specific models chosen from all trained and available by the Google IA team. It is important to note that these models can be discontinued and replaced with new ones at any time.

**OMC and Consensus definition**

The One-Model Consensus (OMC) determines the consensus for each query by identifying the most frequently voted answer among the 4DSSQ (as shown in SFigure 2) for a single Gemini model and a single run. For each pathway, four similar semantic questions (sq) are formulated. If three or four of the answers are "Yes" or "No," the consensus is classified as "Yes" or "No," respectively. An answer of "Possible" is treated as "Yes," while "Low evidence" is treated as "No." In cases with a tie, the consensus is labelled as "Doubt." Additionally, the unanimous flag is set to True if all four answers from the 4DSSQ are the same; otherwise, it is set to False.

**One model consensus**

| question | model 1 ||||  consensus | unanimous |
|---|---|---|---|---|---|---|
|  | sq1 | sq2 | sq3 | sq4 |  |  |
| Q1 | ✓ | ✓ | ✓ | ✓ | ✓ | ✓ |
| Q2 | ✓ | ✗ | ✓ | ✓ | ✓ | ✗ |
| Q3 | ✓ | ✓ | ✗ | ✗ | ✗ | ✗ |
| ... | ... | ... | ... | ... | ... | ... |
| ... | ... | ... | ... | ... | ... | ✓ |
| 4N-1 | ✓ | ✓ | ✓ | ✗ | ✓ | ✗ |
| 4N | ✗ | ✗ | ✗ | ✗ | ✗ | ✓ |

SFigure 2 - The One-Model Consensus (OMC) table consists of questions formulated from pathways as rows. The columns display the answers from the 4DSSQs (sq1, sq2, sq3, and sq4) and the consensus and unanimous flag columns. To calculate the consensus: a) if any permutation contains three 'Yes' answers or three 'No' answers, the consensus will be 'Yes' or 'No', respectively; b) if the answers consist of two 'Yes' and two 'No', the result will be 'Doubt;' c) if all four answers are 'Yes' or 'No', the consensus will be 'Yes' or 'No', respectively, and the unanimous flag will be True. In all other cases, the unanimous flag will be False.

The present study employed two Gemini models, 1.5-pro and 1.5-flash, available between October and December 2024. To build the Ensemble for COVID-19, we ran the two models twice; for MB, we ran the two models three times. We scanned all pathways and formulated the 4DSSQ for each one, for each run. These queries evaluate whether each pathway, belonging to the G0, G1, G2, or G3 groups, is modulated or modulates a given disease case or subtype.

**Hard and Soft reproducibilities**

We define "hard reproducibility" when comparing runs or models based on each individual "answer". In contrast, "soft reproducibility" is defined when comparing runs or models using the calculated "consensuses". Therefore, reproducibility is assessed by calculating the agreement between two different runs or models; when using "answers," one obtains the hard reproducibility, while the soft reproducibility is achieved using the "consensus" (see STable 1).

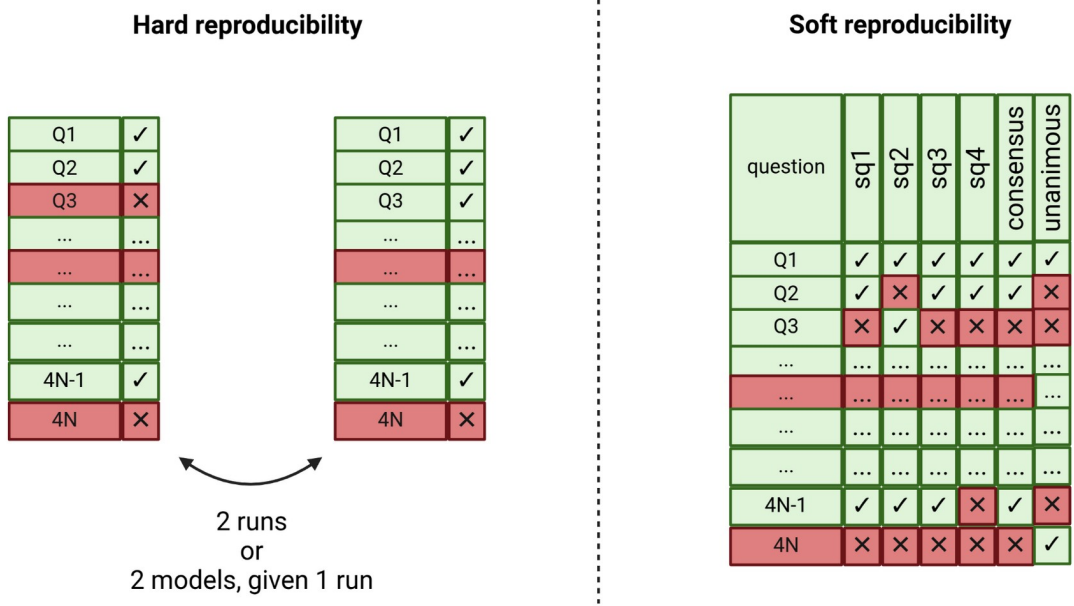

STable 1 - On the left, using two tables, one compares each answer to calculate the hard reproducibility. Therefore, the hard reproducibility can be calculated 1) by comparing two runs, the run-to-run reproducibility (RRR), or 2) by comparing two models, the inter-model reproducibility (IMR). On the right, one can see the OMC table. The soft reproducibility method counts how many equal consensuses can be found using two consensus tables: 1) by comparing two runs, the run-to-run consensus reproducibility (RRCR), and 2) by two models, the inter-model consensus reproducibility (IMCR). In the OMC table, rows are represented by all possible questions/pathways (four groups * N), and columns are the 4DSSQ, split into the 4 semantic questions (sq), the consensus column, fulfilled with a 'Yes', 'No', or 'Doubt', and the "unanimous" column, fulfilled with a True if all four questions are equal; otherwise, with a False. N is the number of enriched plus additional pathways.

The ensemble dataset allows us to compare runs and models. Here, we describe the hard and soft reproducibility methods:

1. Hard reproducibility uses all the answers:
    a. The run-to-run reproducibility (RRR) compares two runs using all answers
    b. The inter-model reproducibility (IMR) compares two models given one run

2. Soft-reproducibility uses all questions' consensus:
    a. The run-to-run consensus reproducibility (RRCR) compares consensuses by using two runs.

b. The inter-model consensus reproducibility (IMCR) compares whether two consensuses are equal, given two models and one run.
c. The multi-model Consensus reproducibility (MMCR) compares whether the consensuses are equal given two runs.
d. The unanimous reproducibility (UR) compares whether the 4DSSQs are the same.

Summarising,

1. Hard reproducibility is the percentage of equal **answers** obtained by comparing two runs (for all cases and models) or two models (for all cases, given a single run).
2. Soft reproducibility is the percentage of equal **consensus** obtained by comparing two runs (for all cases and models) or two models (for all cases, given a single run).

**The Multi-model Consensus (MMC)**

The Multi-model Consensus (MMC) is calculated based on the multi-model consensus table (MMCtab), which consists of two or more 4DSSQ, each corresponding to a different Gemini model (see SFigure 3). Unlike a typical voting process where the most-voted answer comes from a single model, the MMC reflects the collective answers from all selected models. We utilise the MMC to compare the answers generated by Gemini with those from PubMed and the human consensus, and we also use it to calculate the CSC.

**All models consensus**

| question | model 1 | | | | model 2 | | | | consensus | unanimous |
|---|---|---|---|---|---|---|---|---|---|---|
| | sq1 | sq2 | sq3 | sq4 | sq1 | sq2 | sq3 | sq4 | | |
| Q1 | ✓ | ✓ | ✓ | ✓ | ✓ | ✓ | ✓ | ✓ | ✓ | ✓ |
| Q2 | ✓ | ✗ | ✓ | ✓ | ✓ | ✗ | ✓ | ✓ | ✓ | ✗ |
| Q3 | ✓ | ✓ | ✗ | ✗ | ✗ | ✓ | ✗ | ✗ | ✗ | ✗ |
| ... | ... | ... | ... | ... | ... | ... | ... | ... | ... | ... |
| ... | ... | ... | ... | ... | ... | ... | ... | ... | ... | ✓ |
| ... | ... | ... | ... | ... | ... | ... | ... | ... | ... | ... |
| ... | ... | ... | ... | ... | ... | ... | ... | ... | ... | ... |
| 4N-1 | ✓ | ✓ | ✓ | ✗ | ✓ | ✓ | ✓ | ✗ | ✓ | ✗ |
| 4N | ✗ | ✗ | ✗ | ✗ | ✗ | ✗ | ✗ | ✗ | ✗ | ✓ |

SFigure 3 - The Multi-model Consensus Reproducibility (MMCR) uses two MMC tables by comparing two runs. It is the most robust soft reproducibility metric. In the MMCtab, rows are pathways (transformed into queries), columns are all possible questions (4 groups * N), and 'sq' denotes the semantic questions, with 4DSSQ for each model. Next, the consensus column is calculated as the most-voted answer using all models' sq, filled with a 'Yes', 'No', or 'Doubt'. Finally, "unanimous" behaves as a flag; it is a column filled with True if all 8 questions are equal; otherwise, it is False.

To achieve 100% hard reproducibility, all answers must be equal when comparing two runs or two models. However, soft reproducibility allows one or two discordances, being a more flexible measure.

Below is a macro view of DPC (see SFigure 4).

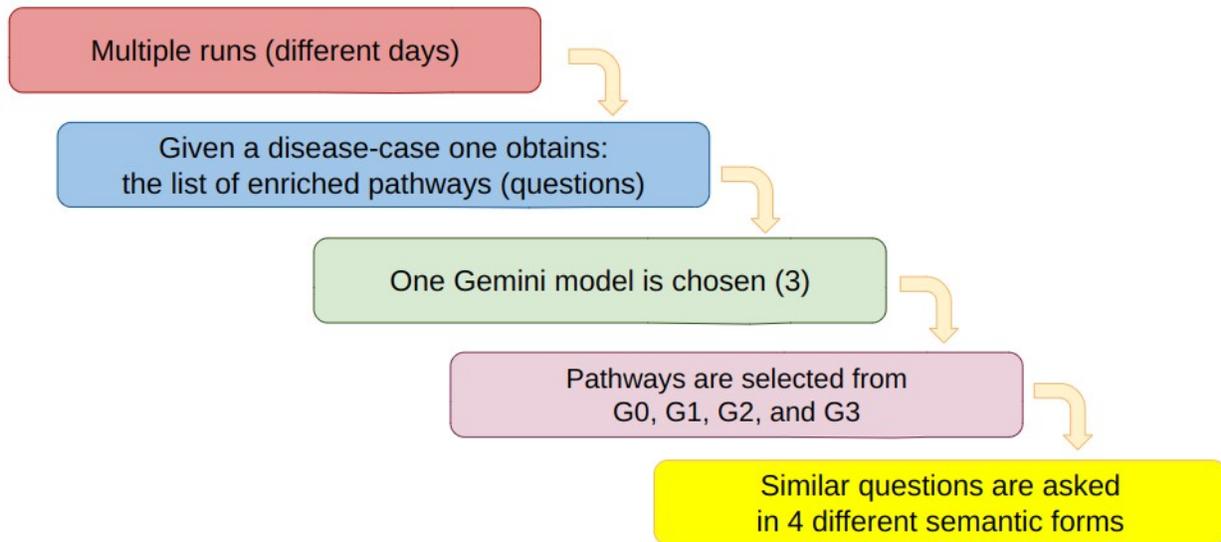

SFigure 4 - DPC pipeline: sequence of actions to create an ensemble of question-answers to measure Gemini consensus and reproducibility.

**Equations**

Below, one presents the hard reproducibility equations.

Equations 1a and 1b define the RRR method,

$$RRR\,list = ¿$$
$$RRR = mean(RRR\,list) - eq\,1\,b$$

Equations 2a and 2b define the IMR method,

$$IMR\,list = ¿$$
$$IMR = mean(IMR\,list) - eq\,2\,b$$

Where A denotes "answers", and N is the total number of questions found for each case. Moreover, N = n * 4 groups, where n is the number of enriched pathways added to the additional pathways.

Soft reproducibility methods compare each consensus,

Equations 3a and 3b define the RRCR method,

$$RRCR\,list = \lparen\ldots\rparen$$
$$RRCR = mean(RRCR\,list) - eq\,3b$$

Equations 4a and 4b define the IMCR method,

$$IMCR\,list = \lparen\ldots\rparen$$
$$IMCR = mean(IMCR\,list) - eq\,4b$$

Where C denotes Consensus and N is each case's total number of questions.

**PubMed**

The second source of information is PubMed, accessible through its web service. When querying PubMed to determine whether "a pathway is related to a disease case," it is essential to carefully define the terms that describe the pathway name, since PubMed is not a semantic machine and does not handle natural language queries. Instead, it is a well-structured SQL database where words can be retrieved using MESH and ordinary terms. These terms are primarily found in titles, abstracts, and keywords. However, if no restrictions are applied, PubMed also searches the body and reference sections. Therefore, it is critical to exercise caution, as a broad search may yield terms from the reference section, potentially resulting in many false positives.

Querying PubMed involves constructing logical phrases, where words are linked by Boolean operators (AND, OR, and NOT), and tags are auxiliary labels such as Author, Date, MESH, title, and abstract. However, these elements must be combined to create queries that traverse the selected pathways and disease cases. Therefore, we simplified all pathway names by replacing each with a few manually selected words to avoid the complexity leading to numerous FNs. The PubMed search is executed using this simplified set of words, besides replacing the original query with the disease name, severity, gender, and age. Finally, since PubMed is an SQL-based machine, it provides a single and precise answer for each query, ensuring reliability but not necessarily good accuracy. Therefore, multiple runs are unnecessary, and there is no PubMed consensus.

**Reactome term table**

The PubMed to Reactome terms table is designed to simplify PubMed queries by addressing the complexity of Reactome's pathway names. Initially, the algorithm automatically generates a table containing all pathway IDs and names, leaving the 'term' column empty. Users are then prompted to fill in this empty column manually. The DPC halts, and users should provide straightforward terms corresponding to the pathway names, steering clear of complex terminology when constructing PubMed SQL queries. Once all pathways have their corresponding terms filled in, the DPC can restart and call the PubMed web service algorithm. The algorithm replaces each pathway with the annotated terms and builds the query using the Boolean connector 'AND'.

**Human responses**

The third data source, human responses, was collected from researchers at the Laboratory of CDI (Instituto Butantan) and the Genetics Laboratory (University of São Paulo - Ribeirão Preto) and their students. For COVID-19, we recruited 4 researchers and 3 students; for the MB study, we also recruited 4 researchers and 3 students.

The team from Instituto Butantan was assigned to answer questions related to COVID-19, and the team from USP-RP was assigned to answer questions related to MB. All researchers and students were oriented to answer 'Yes' or 'No' to the 30 questions of the 2 cases/subtypes. The completed spreadsheets were sent back and processed using DPC algorithms. These spreadsheets and a letter advising participants not to use AI tools or communicate with each other are attached to the supplementary material.

**Multi-Source Dataset (MSD)**

To build the MSD, one merges the "3 Sources" results (see SFigure 5) - the Gemini MMC, PubMed answers, and Human consensus - we calculated the CSC followed by the three corresponding accuracies.

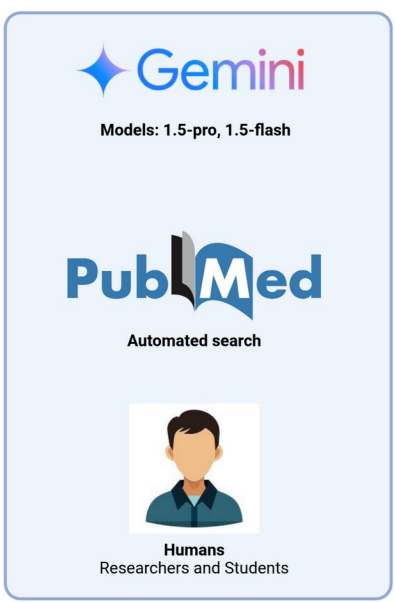

SFigure 5 - The Multi-Source Dataset (MSD) includes the Gemini MMC, PubMed answers, and Human consensus based on the 2CRSP list of pathways. Besides Gemini and PubMed, we include human answers as an additional source required to calculate the CSC.

The MSD was built by manually selecting cases and pathways randomly. We chose two cases for COVID-19: "severe adult male" and "severe old female," along with two subtypes for MB: WNT and G4. The algorithm randomly sorted 30 pathways using the Gemini 1.5-flash model, which included 15 'Yes' responses, indicating that Gemini recognised "the pathway is related to the disease," and 15 'No' responses for each case/subtype.

To compare Gemini to PubMed, we must use Gemini MMC and evaluate it against the corresponding answers from PubMed. This approach was performed using the Ensemble and MSD, the former as a measure and the latter as an assessment.

**Crowdsourcing consensus**

A gold-standard dataset is necessary to assess the accuracy of Gemini and PubMed. However, since no such repository exists, we propose a crowdsourced consensus (CSC) method[12,13], which we assume provides a more reliable measure. Given that Gemini and PubMed were found to be insufficient, we invited research experts and students to contribute their knowledge as additional sources. With the MSD, we could calculate the CSC, representing a consensus among the Gemini MMC, PubMed answers, and the Human consensus. Finally, the CSC allows us to assess the accuracy of Gemini, PubMed, and Humans.

**Definition**

To calculate the CSC, we utilised the 3 Sources evaluations. As mentioned, we did not include all the cases from the COVID-19 experiment in this model. Instead, the MSD model analysis focused on 2 manually selected cases for COVID-19 and 2 subtypes for MB. We randomly selected 30 pathways for each disease case; therefore, humans must answer all 60 pathways. One expects that this model does not yield robust statistics; thus, all calculated consensuses represent assessments. The MSD allowed us to compare the agreement between the Gemini consensus, the PubMed answers, and the Human consensus. Finally, we calculated the CSC, which represents the consensus among the 3 Sources, followed by the accuracy of each source.

**Method**

Using the MSD model (SFigure 6), Gemini MMC and PubMed answers were obtained by accessing the same web services and DPC techniques, allowing the calculation of both hard and soft reproducibilities.

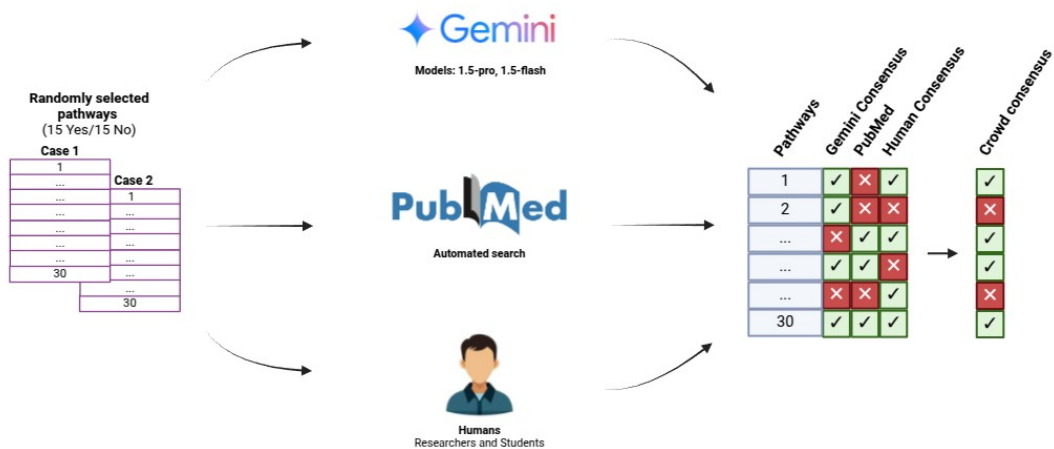

SFigure 6 - The 2 selected cases and respective pathway lists are displayed on the left (2CRSP). The pathways, as queries, were submitted to the Gemini and PubMed web services and shared with researchers and students. Data was collected and combined into the MSD. We used the calculated consensus for Gemini and Human data, while for PubMed, a single answer was obtained for each pathway ('Yes' if at least one PMID was found, otherwise 'No'). As a result, merging the "3 Sources" results, we build the MSD, enabling the crowdsourced consensus (CSC) calculation.

Different Methods Definitions:

1. The Gemini MMC table has pathways as rows, and the consensus column is calculated based on all models among the 4DSSQ, given one run.
2. The Human consensus is the consensus of all human answers, including researchers and students, according to the same set of 30 questions for each disease case/subtype.
3. PubMed has no reproducibility or consensus. Given a query (a logical and ordered set of terms), PubMed's web service answers with a list of references. If the result is an empty list, one says that PubMed did not find the pathway related to the disease; otherwise, it did. As mentioned, PubMed is an SQL-based engine with no semantic capabilities.
4. The Crowdsource consensus (CSC) is the consensus of Gemini MMC, PubMed answers, and the Human consensus. Each of these three sources holds equal weight in this composite measure. Metaphorically, Crowdsource consensus can be described as the consensus of the consensuses.

In summary, as seen in SFigure 7, all the selected queries were submitted to the Gemini web service (models 1.0-pro and 1.5-flash) and the PubMed web service. The same pathways were then provided to researchers and students in spreadsheet format. After three weeks, they returned the completed spreadsheets, which were subsequently imported into the DPC database. When gathering all this data, DPC can be calculated after merging the "3 Sources" evaluations and assessing the accuracy of each source.

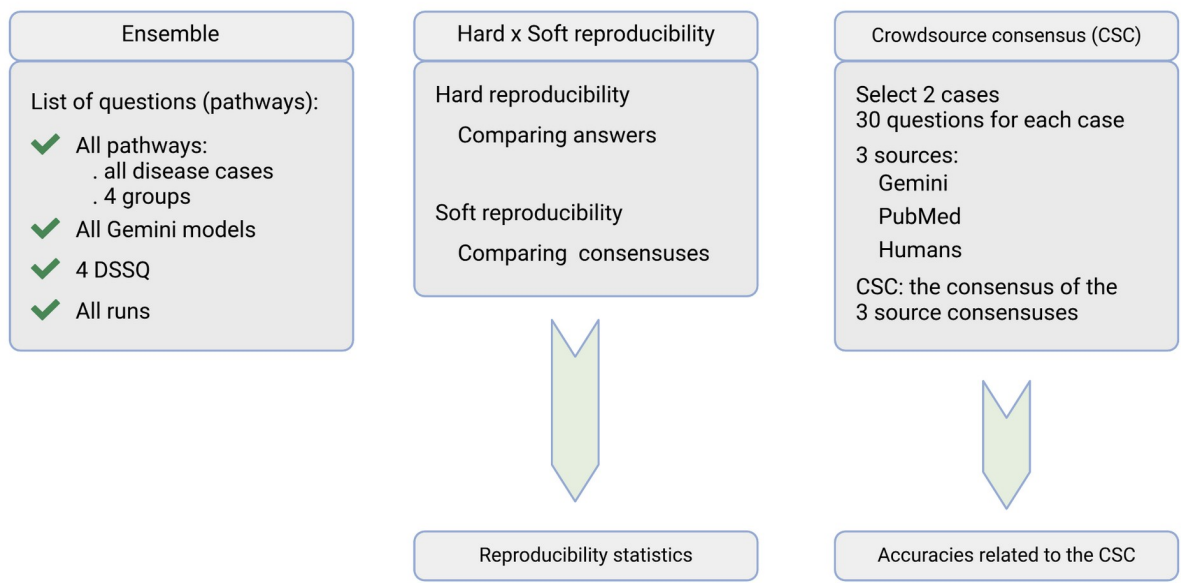

SFigure 7 - the ensemble, hard and soft reproducibility, and the crowdsource dataset structures.

**Accuracy**

Having calculated the CSC table, one can calculate all three accuracies: Gemini, PubMed, and Human accuracies, comparing each pathway consensus to the respective CSC value. Accuracy Definition: Gemini accuracy is the percentage of equal consensus between Gemini consensus and CSC. The same methodology can be applied to PubMed and Human accuracies.

We defined the CSC as our gold standard, which implies that the CSC is "more accurate" than the Gemini consensus, PubMed answers, and Human consensus. Therefore, with this methodology, one can:

1. Define the most accurate source
2. Compare the results of different sources
3. Evaluate all Gemini models
4. Propose an in silico method to uncover concealed FP and FN pathways.

**Confusion matrices**

The confusion matrix calculates accuracy, sensitivity, and specificity by merging two positive and negative control tables. Using the MMC results, we uncover FP and FN pathways.

There are many confusion matrices: three for the Ensemble model and another for the Enriched Pathways identified by GSEA. We cannot calculate a confusion matrix using the MSD because it was developed by sampling pathways based on the 1.5-flash Gemini model, which could lead to circular findings.

**Confusion matrix for Ensemble**

The confusion matrix for the Ensemble mode was built using the ensemble model. We compared G0, representing the extended enriched table or positive control, to G3, which was obtained from the out-of-GSEA table or negative control. In G0, MMC confirms the TP pathways while also uncovering the FPs. In contrast, MMC could validate the TN pathways in G3 and uncover the FNs. We utilised the MMC as the best source compared to the CSC.

The confusion matrix for the Ensemble model is defined as:

- Positive control: i_dfp==0 or G0
- Cloudy negative control 1: i_dfp==1 or G1
- Cloudy negative control 2: i_dfp==2 or G2
- Negative control: i_dfp==3 or G3

Using the Ensemble model, one expects that most of the pathways from G0 will fall under the TP pathways, with only a few FPs. In contrast, we expect that the pathways from G3 will correspond to the TN pathways, again with a few FNs. For G1 and G2, which are situated in the cloudy zone, we expect to see a gradual decrease in TPs and an increase in FNs. If this pattern is observed, it demonstrates that the ensemble model divided into pathway groups exhibits favourable behaviour regarding the false discovery rate (FDR).

**Confusion matrix for GSEA**

The confusion matrix for the Enriched Pathways is based exclusively on G0, the extended enriched table. As mentioned, one splits the pathways using the default cutoff parameters to identify the truly enriched pathways. The remaining pathways are classified as additional pathways. As a result, the enriched pathways fall into the True Positive (TP) group, while the additional pathways belong to the True Negative (TN) group. These classifications were determined using GSEA in conjunction with the Reactome database. Furthermore, MMC confirmed the TP pathways and uncovered False Positives (FPs), confronting GSEA findings. Next, analysing the additional pathways, MMC validated the TN pathways and uncovered FNs, disagreeing with GSEA-Reactome and suggesting that new modulated pathways may exist.

The confusion matrix for the enriched pathways is defined as:

- Positive control: enriched pathways using the default cutoff parameters
- Negative control: additional pathways using the flexible cutoff parameters

- Columns:
  - TP: true positives, based on the default cutoffs
  - FP: false positive, TP: that Gemini disagree
  - TN: true negatives, based on additional pathways
  - FN: false negatives, TN: that Gemini disagree
  - Sensitivity (sens): TP / (TP + FN)
  - Specificity (spec): TN / (TN + FP)

- Accuracy (accu): (TP + TN) / N
- Precision (prec): TP/(TP + FP)
- F1-score: (2*precision*sensitivity) / (precision+sensitivity)
- TN1, FN1: TN and FN from G1, cloudy zone
- TN2, FN2: TN and FN from G2, cloudy zone

Figure 8 - On the left, the selection of the four pathway groups belonging to the Ensemble model is represented. In the middle, the Enriched Pathways and Additional Pathways can be seen. The full enriched table and instructions for analysing TP, FP, FN, and FN are on the right.

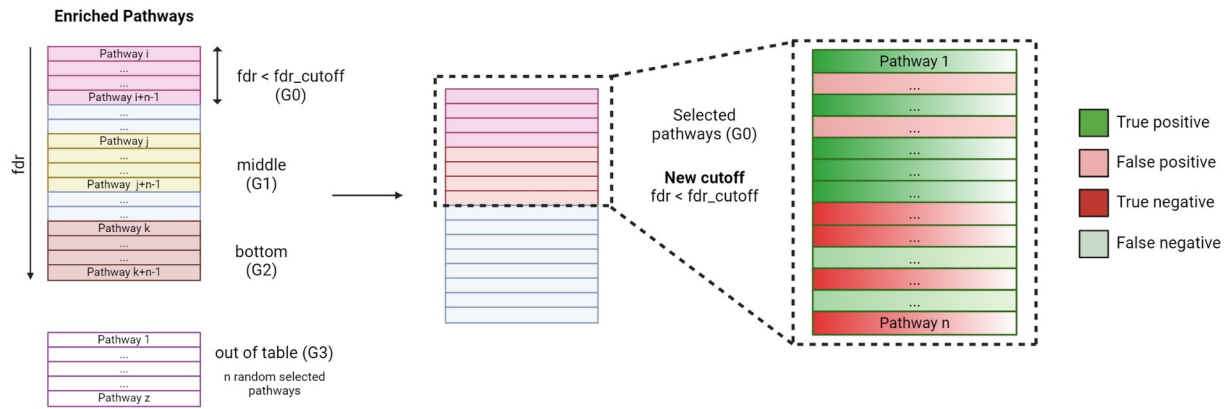

SFigure 8 – On the left, the figure illustrates the selection of the four pathway groups: G0, G1, G2, and G3. The enriched pathways calculated using the default cutoff parameters are highlighted in pink in the middle of the table. Below this, in orange, are additional pathways identified by flexibilising the cutoff parameters. On the right side, we examine whether the MMC agrees with the GSEA results. Here, pathways identified as TPs are green, while the uncovered FPs are pink. In addition, the confirmed TNs are represented in red, and the uncovered FNs are in light green.

## Statistical Analysis

The counting methods involve evaluating the number of 'Yes' and 'No' answers obtained from a single run within a specific Gemini model, categorised by pathway groups. The results consist of nominal categories with integer counts. Different aggregations allow the creation of n x m tables to perform statistical tests and compare runs, models, or groups. For such a table, the appropriate statistical test is the chi-squared test. However, if the table is 2 x 2 and any cells have a count of less than 5, we use the Fisher exact test (FET) instead.

One uses the t-test to compare agreements between runs or Gemini models for central values, like accuracy, sensitivity, and specificity. If multiple comparisons were performed, we corrected the p-values using Bonferroni or FDR.

## Results:

### Tables, consensus, and counts

### Gemini response table

Below is a brief example of the Gemini response table for run01, Gemini 1.5-pro, MB subtype WNT. STable 2 displays the columns pathway_id and pathway (name) based on the selected Reactome pathways. The curation column contains one of the possible responses: "Yes," "Possible," "Low evidence," or "No." Additionally, the "response explain" column provides a complete explanation: the Gemini answer. Finally, in STable 3, you can find the "question" column containing the formulated query.

| pathway id | pathway | curation | response explain |
|---|---|---|---|
| R-HSA-73856 | RNA Polymerase II Transcription Termination | Low | Low evidence. While RNA Polymerase II transcription termination is a fundamental process in gene expression, and dysregulation of gene expression is implicated in medulloblastoma development, there's no strong direct evidence linking *specific* defects in RNA Polymerase II termination pathways … |
| R-HSA-388841 | Costimulation By CD28 Family | Low | Low evidence. While the CD28 costimulation pathway is crucial for T-cell activation, and T-cells play a role in anti-tumor immunity, there's limited direct evidence strongly linking the CD28 family's costimulatory pathway specifically to the pathogenesis or progression of WNT-subtype medulloblastoma. …. |
| R-HSA-68949 | Orc1 Removal From Chromatin | Yes | Yes. While there isn't a wealth of direct studies explicitly linking Orc1 removal from chromatin *as the primary driver* of WNT-driven medulloblastoma, the strong relationship is inferred through several converging lines of evidence: Orc1's role in cell cycle regulation: Orc1's removal from chromatin is a crucial part of the cell cycle … |
| R-HSA-73884 | Base Excision Repair | Low | Low evidence. While base excision repair (BER) is a crucial pathway for maintaining genomic stability, and genomic instability is implicated in cancer development including medulloblastoma, there's limited direct evidence strongly linking *specific* BER deficiencies to the Wnt subtype of medulloblastoma. … |
| R-HSA-69017 | CDK-mediated Phosphorylation And Removal Of Cdc6 | Yes | Yes. While there isn't a wealth of research directly linking CDK-mediated Cdc6 phosphorylation and removal to Wnt-driven medulloblastoma specifically, the established role of Cdc6 in cell cycle regulation and the known involvement of cell cycle dysregulation in medulloblastoma strongly suggest a relationship… |

STable 2 - The Gemini response table has columns: "pathway_id" and pathway (name), and "curation" contains Gemini's decision: "Yes," "Possible," "Low evidence," or "No." Additionally, the "response explain" provides a complete explanation of Gemini's answer.

| pathway id | pathway | question |
|---|---|---|
| R-HSA-73856 | RNA Polymerase II Transcription Termination | Answer in the first line Yes, Possible, Low evidence, or No; and explain; has the pathway 'RNA Polymerase II Transcription Termination' a strong relationship in studies related to the disease medulloblastoma for type WNT? Context: This section includes the cleavage of both polyadenylated and non-polyadenylated transcripts. <p> In the former case polyadenylation has to precede transcript cleavage, while in the latter case there is no polyadenylation. Medulloblastoma is a rare and aggressive type of brain tumor that primarily affects children and young adults. It originates in the cerebellum, the part of the brain responsible for coordination, balance, and motor skills. |
| R-HSA-388841 | Costimulation By CD28 Family | Answer in the first line Yes, Possible, Low evidence, or No; and explain; has the pathway 'Costimulation by the CD28 family' a strong relationship in studies related to the disease medulloblastoma for type WNT? Context: Optimal activation of T-lymphocytes requires at least two signals. A primary one is delivered by the T-cell receptor (TCR) complex after antigen recognition and additional costimulatory signals are delivered by the engagement of costimulatory receptors such as CD28. The best-characterized costimulatory pathways are mediated by a set of cosignaling molecules belonging to the CD28 superfamily, including CD28, CTLA4, ICOS, PD1 and BTLA receptors. These proteins deliver both positive and negative second signals to T-cells by interacting with B7 family ligands expressed on antigen presenting cells. Different subsets of T-cells have very different requirements for costimulation. Medulloblastoma is a rare and aggressive type of brain tumor that primarily affects children and young adults. It originates in the cerebellum, the part of the brain responsible for coordination, balance, and motor skills. |
| R-HSA-68949 | Orc1 Removal From Chromatin | Answer in the first line Yes, Possible, Low evidence, or No; and explain; has the pathway 'Orc1 removal from chromatin' a strong relationship in studies related to the disease medulloblastoma for type WNT? Context: Mammalian Orc1 protein is phosphorylated and selectively released from chromatin and ubiquitinated during the S-to-M transition in the cell division cycle. Medulloblastoma is a rare and aggressive type of brain tumor that primarily affects children and young adults. It originates in the cerebellum, the part of the brain responsible for coordination, balance, and motor skills. |
| R-HSA-73884 | Base Excision Repair | Answer in the first line Yes, Possible, Low evidence, or No; and explain; has the pathway 'Base Excision Repair' a strong relationship in studies related to the disease medulloblastoma for type WNT? Context: Of the three major pathways involved in the repair of nucleotide damage in DNA, base excision repair (BER) involves the greatest number of individual enzymatic activities. This is the consequence of the numerous individual glycosylases, each of which recognizes and removes a specific modified base(s) from DNA. BER is responsible for the repair of the most prevalent types of DNA lesions, oxidatively damaged DNA bases, which arise as a consequence of reactive oxygen species generated by normal mitochondrial metabolism or by oxidative free radicals resulting from ionizing radiation, lipid peroxidation or activated phagocytic cells. BER is a two-step process initiated by one of the DNA glycosylases that recognizes a specific modified base(s) and removes that base through the catalytic cleavage of the glycosydic bond, leaving an abasic site without disruption of the phosphate-sugar DNA backbone. Subsequently, abasic sites are resolved by a series of enzymes that cleave the backbone, insert the replacement residue(s), and ligate the DNA strand. Medulloblastoma is a rare and aggressive type of brain tumor that primarily affects children and young adults. It originates in the cerebellum, the part of the brain responsible for coordination, balance, and motor skills. |
| R-HSA-69017 | CDK-mediated Phosphorylation And Removal Of Cdc6 | Answer in the first line Yes, Possible, Low evidence, or No; and explain; has the pathway 'CDK-mediated phosphorylation and removal of Cdc6' a strong relationship in studies related to the disease medulloblastoma for type WNT? Context: As cells enter S phase, HsCdc6p is phosphorylated by CDK promoting its export from the nucleus (see Bell and Dutta 2002). Medulloblastoma is a rare and aggressive type of brain tumor that primarily affects children and young adults. It originates in the cerebellum, the part of the brain responsible for coordination, balance, and motor skills. |

STable 3 - The Gemini response table. The last column is 'question', the formulated query merging pathway name, disease, semantic question type, and contexts.

**Quality control**

For the sake of quality control, we created a bar plot spanning each case/subtype and displaying bars for each two Gemini models, the first 2 in blue or green for 'Yes' responses, followed in red or orange for 'No' answers, and the third, in black, as counting the total number of pathways (n). The total number of cases in the ensemble model is n*4, where n represents the number of enriched pathways plus the additional pathways, and 4 represents the four groups from G0 to G4. However, in the MSD model, n is fixed and equals 30 randomly chosen pathways.

Below, we can see in SFigure 9 the bar plot for COVID-19 using the ensemble model.

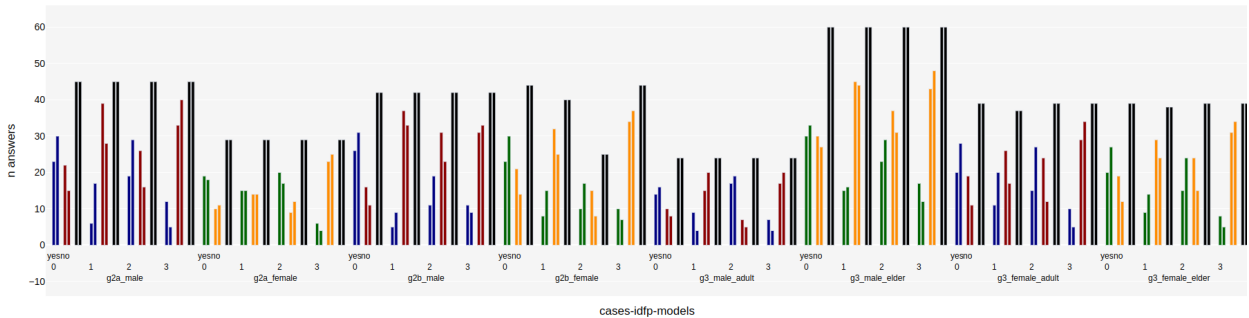

SFigure 9 - The bar plot displays the number of 'Yes' responses (shown in blue or green), 'No' responses (shown in red or orange), and the total number of responses (shown in black) for two consecutive models (1.5-pro and 1.5-flash) across all COVID-19 cases using the ensemble model. This data serves as a quality control measure that DPC employed to ensure all pathways are correctly accounted for in each case when building the ensemble.

Next, we can see in SFigure 10 the bar plot for MB using the ensemble model.

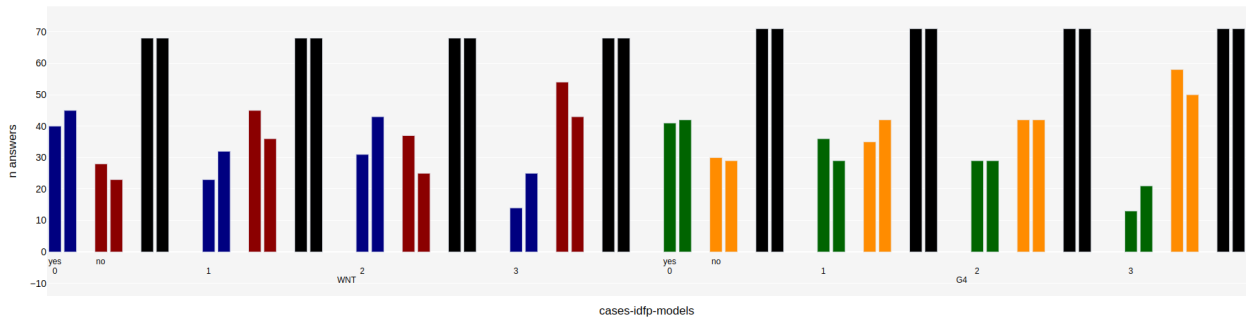

SFigure 10 - The bar plot displays the number of 'Yes' responses (shown in blue or green), 'No' responses (shown in red or orange), and the total number of responses (shown in black) for two consecutive models (1.5-pro and 1.5-flash) across all MB subtypes. This data serves as a quality control measure that DPC employed to ensure all pathways are correctly accounted for in each case when building the ensemble.

**Counting table**

The counting table stores the answers to each question collected after accessing Gemini's web service. The possible answer categories include: 'Yes,' 'Possible,' 'Low evidence,' and 'No.' For each case or subtype, the algorithm iterates over all selected Gemini models and runs, like 'run01' and 'run02'. As a result, the columns are labelled as follows: a) total yes (tot_yes), b) total possible (tot_pos), c) total low evidence (tot_low), and d) total no (tot_no), each followed by the suffixes 1 and 2, respectively, for each run.

Below is the counting table for the COVID-19 study (STable 4).

| case | tot yes1 | tot pos1 | tot low1 | tot no1 | tot yes2 | tot pos2 | tot low2 | tot no2 |
|---|---|---|---|---|---|---|---|---|
| g2a_male | 510 | 4 | 485 | 441 | 485 | 6 | 461 | 488 |
| g2a_female | 394 | 0 | 328 | 206 | 391 | 0 | 326 | 211 |
| g2b_male | 431 | 3 | 427 | 483 | 405 | 4 | 403 | 532 |
| g2b_female | 417 | 5 | 386 | 416 | 391 | 10 | 370 | 453 |
| g3_male_adult | 332 | 6 | 258 | 172 | 320 | 7 | 246 | 195 |
| g3_male_elder | 601 | 12 | 665 | 642 | 554 | 15 | 638 | 713 |
| g3_female_adult | 490 | 9 | 408 | 325 | 473 | 12 | 380 | 367 |
| g3_female_elder | 435 | 14 | 433 | 358 | 424 | 11 | 407 | 398 |

STable 4 - The counting table for COVID-19 has rows as cases and columns as different total answers. Here, total Yes (tot_yes), total Possible (tot_pos), total Low evidence (tot_low), and total No (tot_no) are followed by the suffix 1 for 'run01' and 2 for 'run02'.

Filename: run_run_hard_total_answers_per_case_for_COVID-19_runs_run01_x_run02_all_data.tsv

Next is the counting table for the MB study (STable 5).

| case | tot yes1 | tot pos1 | tot low1 | tot no1 | tot yes2 | tot pos2 | tot low2 | tot no2 |
|---|---|---|---|---|---|---|---|---|
| WNT | 929 | 0 | 725 | 522 | 930 | 0 | 725 | 521 |
| G4 | 744 | 0 | 1032 | 496 | 744 | 0 | 1034 | 494 |

STable 5 - The counting table for MB has rows as subtypes and columns as different total answers. Here, total Yes (tot_yes), total Possible (tot_pos), total Low evidence (tot_low), and total No (tot_no) are followed by the suffix 1 for 'run01' and 2 for 'run02'.

Filename: run_run_hard_total_answers_per_case_for_medulloblastoma_runs_run01_x_run02_all_data.tsv

**Comparing Counting Statistics**

The counting table statistic compares all answers from run01 versus run02. Observing the counting tables, they have 4 counting columns: total Yes (tot_yes), total Possible (tot_pos), total Low evidence (tot_low), and total No (tot_no). Therefore, for each run, we built an 8 x 4 matrix for COVID-19 (see STable 6) and a 2 x 4 matrix for MB (see STable 7). The chi2 test was applied, and the p-value cutoff is 0.05 divided by the number of cases, according to Bonferroni, since there are repeated tests.

Below is the counting statistical table for COVID-19 (STable 6).

| case | chi2 stats | pvalue | vals1 | vals2 |
|---|---|---|---|---|
| g2a_male | 3.941 | 2.68E-01 | [511 5 486 442] | [486 7 462 489] |
| g2a_female | 0.077 | 9.94E-01 | [395 1 329 207] | [392 1 327 212] |
| g2b_male | 3.971 | 2.65E-01 | [432 4 428 484] | [406 5 404 533] |
| g2b_female | 4.215 | 2.39E-01 | [418 6 387 417] | [392 11 371 454] |
| g3_male_adult | 2.005 | 5.71E-01 | [333 7 259 173] | [321 8 247 196] |
| g3_male_elder | 6.493 | 8.99E-02 | [602 13 666 643] | [555 16 639 714] |
| g3_female_adult | 4.225 | 2.38E-01 | [491 10 409 326] | [474 13 381 368] |
| g3_female_elder | 3.388 | 3.36E-01 | [436 15 434 359] | [425 12 408 399] |

STable 6 - The counting statistical table for COVID-19 compares all total counts from run01 versus run02. The rows represent the cases, while the columns include the chi2 statistics, p-value, and vals1 and vals2. The last columns, vals1 and vals2, are the two counting matrices for run01 and run02, respectively. The p-value cutoff is calculated by dividing 0.05 by the number of cases, according to Bonferroni. The number of cases equals 8 for COVID-19, and p-values below this cutoff are considered significant. Since no p-value is less than this threshold, there are no statistical differences between runs for all cases.

Filename: run_run_hard_total_answers_stats_per_case_for_COVID-19_runs_run01_x_run02_all_data.tsv

Next is the counting statistical table for MB (STable 7).

| case | chi2 stats | pvalue | vals1 | vals2 |
|---|---|---|---|---|
| WNT | 0.001 | 1.00E+00 | [930 1 726 523] | [931 1 726 522] |
| G4 | 0.006 | 1.00E+00 | [745 1 1033 497] | [745 1 1035 495] |

STable 7 - The counting statistical table for MB compares all total counts from run01 versus run02. The rows represent the cases, while the columns include the chi2 statistics, p-value, and vals1 and vals2. The last columns, vals1 and vals2, are the two counting matrices for run01 and run02, respectively. The p-value cutoff is calculated by dividing 0.05 by the number of cases, according to Bonferroni. The number of cases equals 2 for MB, and p-values below this cutoff are considered significant. Since no p-value is less than this threshold, there are no statistical differences between runs for all cases.

Filename: run_run_hard_total_answers_stats_per_case_for_medulloblastoma_runs_run01_x_run02_all_data.tsv

**Counts per group**

The consensus count plot displays the number of 'Yes' responses for each case categorised by the groups G0, G1, G2, and G3. Each case-group combination has 4-point counts associated with the four different but semantically similar questions (4DSSQ). We identified two segments: the first solid segment corresponds to model 1 (1.5-pro), while the dashed segment pertains to model 2 (1.5-flash). As detailed below, we expect a decreasing number of 'Yes' responses as we move from G0 to G1 and from question 0 (sq0) to question 3 (sq3). A case with more 'Yes' counts than another indicates that Geminy has a greater capacity to find pathways related to that case or subtype, which correlates with the number of enriched pathways identified. Consequently, we provide both the count plot and the normalised count plot.

The count of consensus algorithm counts the occurrences of consensus classes ('Yes' and 'No') for each case, group, and model, which is represented in a 4-point segment curve according to the 4DSSQ. The frequency table and corresponding plot provide valuable insights into how the Gemini LLM machine makes decisions. As illustrated in the figures below, we expect the number of 'Yes' responses to decrease from G0, the positive control, to G3, the negative control. As we move towards higher-rank pathways, we anticipate a lower likelihood that Gemini will agree that these pathways both modulate and are modulated by the disease case or subtype.

Another perspective involves 4DSSQ, corresponding to a 4-point segment within each case-group model. The first question, called sq1, addresses the number of 'Yes' responses to the query, "Is the <pathway> studied about the disease?" The second question, sq2, has a similar construct but we added a constrain "in PubMed', resulting in "Is the <pathway> studied in PubMed concerning the disease?" It is expected that this additional constraint will lead to fewer 'Yes' responses. The last two questions, sq3 and

sq4, maintain the same structure; however, they introduce the term "have a strong relationship." By imposing this further constraint, we anticipate that even fewer pathways will be identified, resulting in a diminished slope for the four-point segment.

Below, we can see the consensus count plots. SFigure 11 presents the first four cases of COVID-19, followed by SFigure 12, which presents the last four.

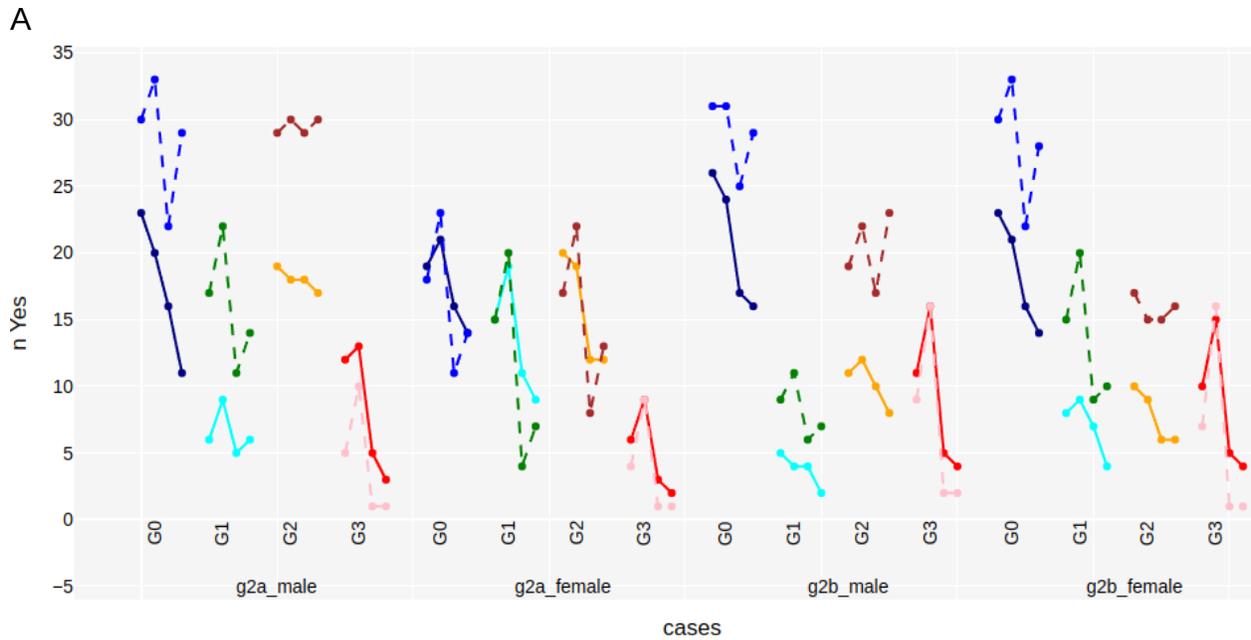

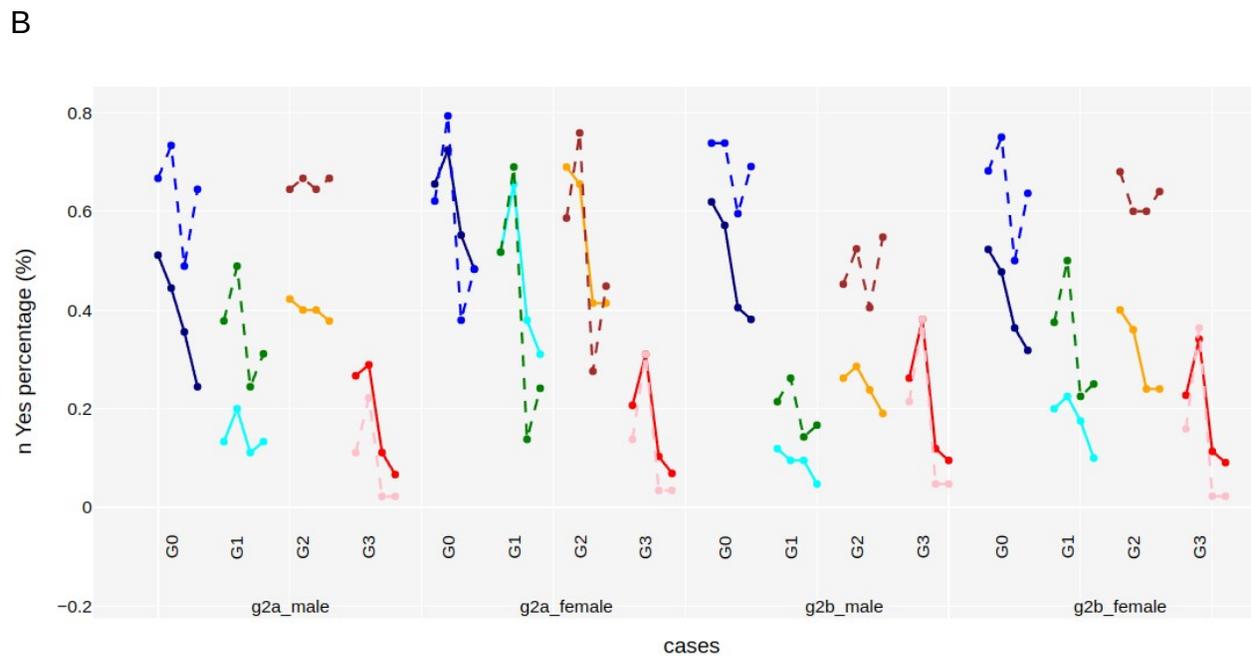

SFigure 11 - The consensus count plot displays each case's number of 'Yes' responses in A and B as percentages. The cases are divided into groups, each answering four questions about the 4DSSQ. Each data point consists of two segments: the first segment, represented by a solid line, corresponds to model 1 (1.5-pro), while the second segment, shown with a dashed line, corresponds to model 2 (1.5-flash). This data is related to COVID-19 cases, specifically groups g2a and g2b, including male and female participants. The cases in group g2a are associated with weak responses, whereas those in group g2b are related to mild responses.

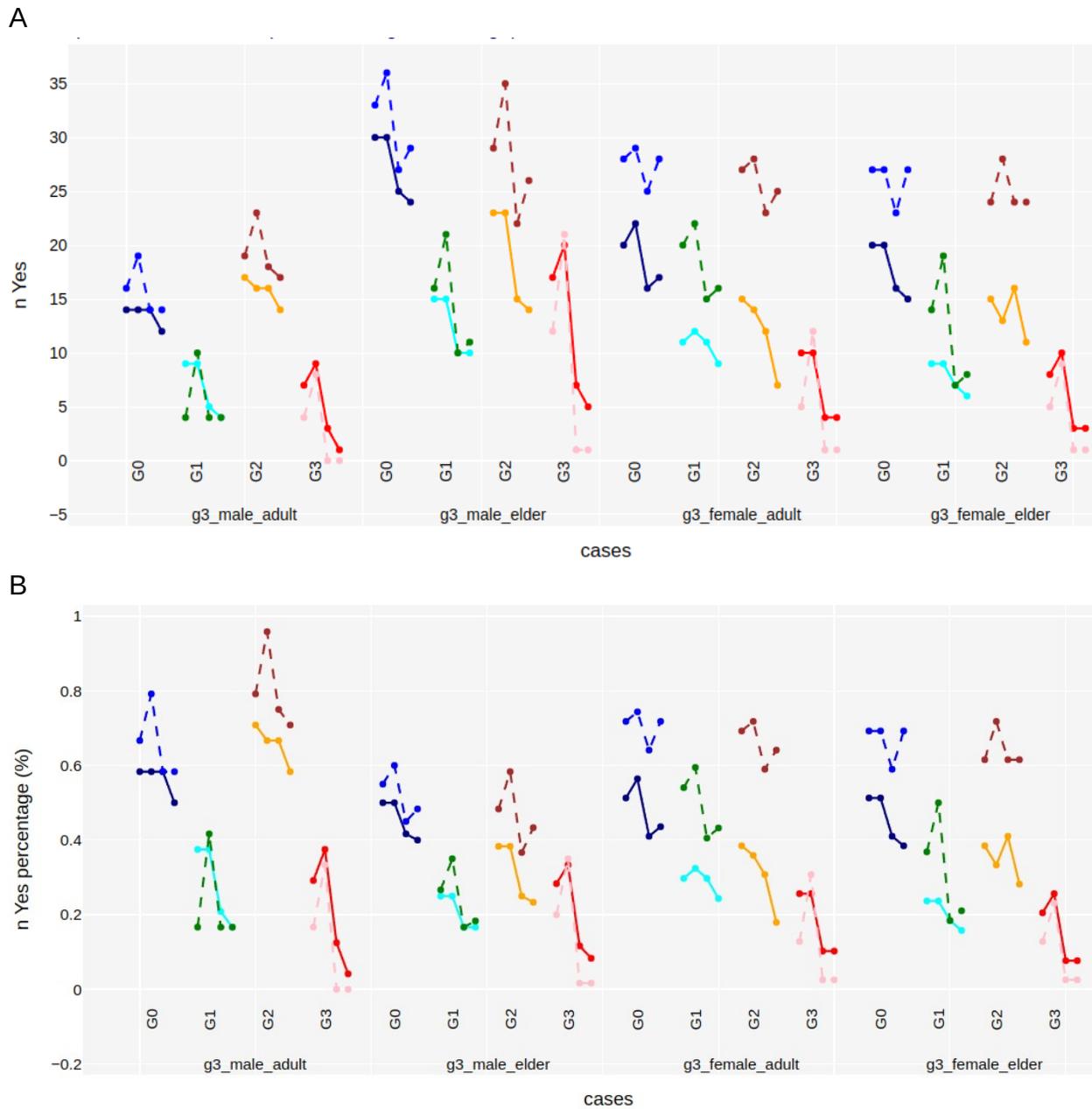

SFigure 12 - The consensus count plot displays each case's number of 'Yes' responses in A and B as percentages. The cases are divided into groups, each answering four questions about the 4DSSQ. Each data point consists of two segments: the first segment, represented by a solid line, corresponds to model 1 (1.5-pro), while the second segment, shown with a dashed line, corresponds to model 2 (1.5-flash). This data is related to COVID-19 cases, specifically groups g3, including male adults and elders and female adults and elderly participants. The cases in group g3 are associated with severe responses.

Next, we can see in SFigure 13 the consensus count plot for MB.

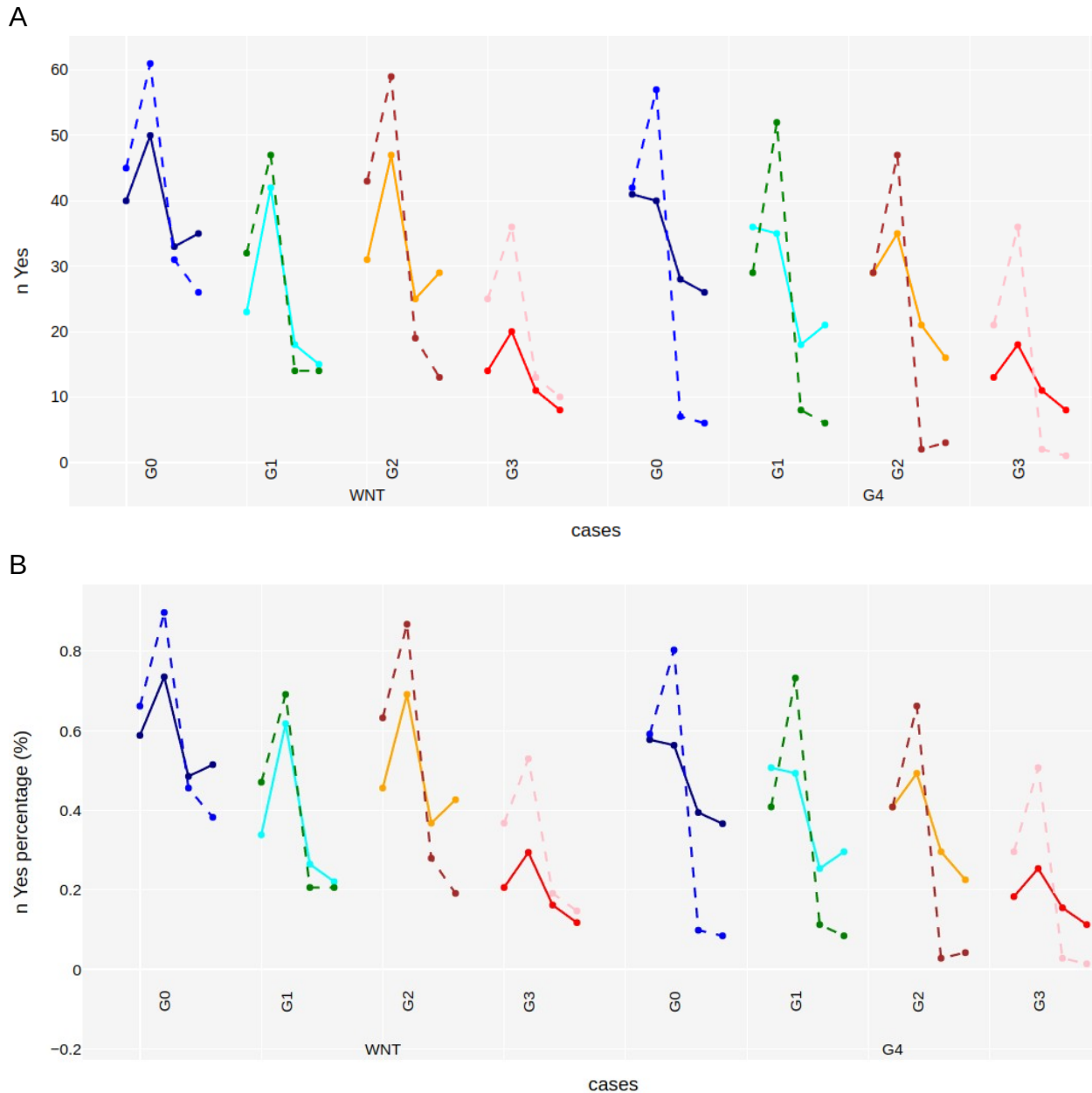

SFigure 13 - The consensus count plot displays each case's number of 'Yes' responses in A and B as percentages. The cases are divided into groups, each answering four questions about the 4DSSQ. Each data point consists of two segments: the first segment, represented by a solid line, corresponds to model 1 (1.5-pro), while the second segment, shown with a dashed line, corresponds to model 2 (1.5-flash). This data is related to MB subtypes, WNT and G4.

**OMC**

The Gemini One-Model Consensus (OMC) table combines Gemini's answers for a specific run, model, and case. To build the OMC table, the 4DSSQ answers are merged, and the consensus and unanimous columns are calculated.

Below, in STable 8, is a brief example related to the disease MB, subtype WNT, from run 'run02', using model 3 (Gemini 1.5-flash).

| pathway id | pathway | i dfp | consensus | unanimous | sq0 | sq1 | sq2 | sq3 |
|---|---|---|---|---|---|---|---|---|
| R-HSA-109582 | Hemostasis | 0 | No | False | No | Yes | Low | Low |
| R-HSA-111447 | Activation Of BAD And Translocation To Mitochondria | 1 | Yes | False | Yes | Yes | Low | Yes |
| R-HSA-111885 | Opioid Signaling | 1 | Doubt | False | Yes | Yes | Low | Low |
| R-HSA-111996 | Ca-dependent Events | 1 | Yes | True | Yes | Yes | Yes | Yes |
| R-HSA-112307 | Transmission across Electrical Synapses | 3 | No | False | Low | Yes | No | Low |
| R-HSA-112308 | Presynaptic Depolarization And Calcium Channel Opening | 0 | Yes | True | Yes | Yes | Yes | Yes |
| R-HSA-112310 | Neurotransmitter Release Cycle | 0 | Doubt | False | Yes | Yes | Low | Low |
| R-HSA-112314 | Neurotransmitter Receptors And Postsynaptic Signal Transmission | 0 | Yes | True | Yes | Yes | Yes | Yes |
| R-HSA-112315 | Transmission Across Chemical Synapses | 0 | Yes | True | Yes | Yes | Yes | Yes |
| R-HSA-112316 | Neuronal System | 0 | Yes | True | Yes | Yes | Yes | Yes |

STable 8 - The Gemini One-Model Consensus (OMC) table consolidates the four responses for a disease like medulloblastoma (MB), subtype WNT, from run 'run02' and model 3. The rows represent the Reactome IDs and their corresponding names, while the columns display the 4DSSQ answers (sq0, sq1, sq2, and sq3). Additionally, a "consensus" column is the most voted answer; in the case of a tie, it is marked with a 'Doubt.' The "unanimous" column is marked True if all responses are the same ('Yes' or 'No') and False otherwise.

Filename: gemini_OMC_for_medulloblastoma_run_run02_model_gemini-1.5-flash_all_data.tsv

**MMC**

The Multi-Model Consensus (MMC) table combines the results from multiple selected Gemini model answers for a specific run and case. To build the MMC, the 4DSSQ answers from different models are merged, and the consensus and unanimous columns among all answers are calculated.

Below, in STable 9, is a brief example of the MMC table related to the disease COVID-19 case g2a male, from run 'run02', merging model 1 (Gemini 1.5-pro) and model 3 (Gemini 1.5-flash).

| | | model 1 – 1.5-pro | | | | model 3 – 1.5-flash | | | | |
|---|---|---|---|---|---|---|---|---|---|---|
| pathway_id | pathway | sq0 | sq1 | sq2 | sq3 | sq0 | sq1 | sq2 | sq3 | consensus |
| R-HSA-109582 | Hemostasis | Yes | Yes | Yes, | Yes, | Yes | Yes | Yes | Yes | Yes |
| R-HSA-114608 | Platelet Degranulation | Yes | Yes, | Yes | Low | Yes | Yes | Yes | Yes | Yes |
| R-HSA-1280215 | Cytokine Signaling In Immune System | Yes | Yes | Yes | Yes | Yes | Yes | Yes | Yes | Yes |
| R-HSA-140877 | Formation Of Fibrin Clot (Clotting Cascade) | Yes | Yes | Yes | Yes, | Yes | Yes | Yes | Yes | Yes |
| R-HSA-1474244 | Extracellular Matrix Organization | Low | Yes | Low | Low | Yes | Yes | Yes | Yes | Yes |
| R-HSA-166658 | Complement Cascade | Yes | Yes | Yes | Yes | Yes | Yes | Yes | Yes | Yes |
| R-HSA-168249 | Innate Immune System | Yes | Yes | Yes | Yes | Yes | Yes | Yes | Yes | Yes |
| R-HSA-168256 | Immune System | Yes | Yes | Yes | Yes | Yes | Yes | Yes | Yes | Yes |
| R-HSA-216083 | Integrin Cell Surface Interactions | Low | Yes, | Low | Low | Yes | Yes | No | Yes | Doubt |
| R-HSA-375276 | Peptide Ligand-Binding Receptors | Yes | Yes | Yes | Yes | Yes | Yes | Low | Low | Yes |
| R-HSA-381426 | Regulation Of IGF Transport And Uptake By IGFBPs | Low | Low | Low | Low | No | Low | No | Low | No |

STable 9 - The Gemini Multi-Model Consensus (MMC) table consolidates all semantic questions (sq) responses by merging the selected Gemini models, given a disease like COVID-19, case g2a, from run 'run02'. The rows represent the Reactome IDs and their corresponding names, while the columns display multiple 4DSSQ answers (semantic questions: sq0, sq1, sq2, and sq3), a set of 4 answers for each model. Additionally, a "consensus" column is the most voted answer among all the answers; in the case of a tie, it is marked with a 'Doubt.' The "unanimous" column is marked True if all responses beyond all models are the same ('Yes' or 'No') and False otherwise.

Filename: gemini_semantic_consensus_for_COVID-19_all_models_[1,3]_run_run01_all_data.tsv

Next is a brief example of the MMC table related to the disease MB subtype WNT, from run 'run02', merging model 1 (Gemini 1.5-pro) and model 3 (Gemini 1.5-flash) - see STable 10.

|  |  | model 1 – 1.5-pro | | | | model 3 – 1.5-flash | | | | |
| --- | --- | --- | --- | --- | --- | --- | --- | --- | --- | --- |
| pathway_id | pathway | sq0 | sq1 | sq2 | sq3 | sq0 | sq1 | sq2 | sq3 | consensus |
| R-HSA-109582 | Hemostasis | No | Low | Low | Low | Yes | Yes | Low | Low | No |
| R-HSA-111885 | Opioid Signaling | Yes | Yes | Low | Low | Yes | Yes | No | Low | Doubt |
| R-HSA-112315 | Transmission Across Chemical Synapses | Yes | Yes | Yes | Yes | Yes | Yes | Low | Low | Yes |
| R-HSA-112316 | Neuronal System | Yes | Yes | Yes | Yes | Yes | Yes | Yes | Yes | Yes |
| R-HSA-114452 | Activation Of BH3-only Proteins | Yes | Yes | Yes | Yes | Yes | Yes | Low | Low | Yes |
| R-HSA-114508 | Effects Of PIP2 Hydrolysis | Yes | Yes | Low | Yes | Low | Yes | Low | Low | Doubt |
| R-HSA-1474244 | Extracellular Matrix Organization | Yes | Yes | Yes | Yes | Yes | Yes | Low | Low | Yes |
| R-HSA-1474290 | Collagen Formation | Yes | Yes | Yes | Low | Low | Yes | Low | Low | Doubt |
| R-HSA-1483257 | Phospholipid Metabolism | Yes | Yes | Yes | Yes | Yes | Yes | Low | Low | Yes |
| R-HSA-1489509 | DAG And IP3 Signaling | Yes | Yes | Low | Yes | Yes | Yes | Low | Low | Yes |
| R-HSA-162582 | Signal Transduction | Yes | Yes | Yes | Yes | Yes | Yes | Yes | Yes | Yes |

STable 10 - The Gemini Multi-Model Consensus (MMC) table consolidates all semantic question (sq) responses by merging the selected Gemini models, given a disease like medulloblastoma (MB), subtype WNT, from run 'run02'. The rows represent the Reactome IDs and their corresponding names, while the columns display multiple 4DSSQ answers (semantic questions: sq0, sq1, sq2, and sq3), a set of 4 answers for each model. Additionally, a "consensus" column is the most voted answer among all the answers; in the case of a tie, it is marked with a 'Doubt.' The "unanimous" column is marked True if all responses beyond all models are the same ('Yes' or 'No') and False otherwise.

Filename: gemini_semantic_consensus_for_MB_all_models_[1,3]_run_run01_all_data.tsv

**Multi-model Consensus and Unanimous summary counts**

The Multi-model Consensus and Unanimous Counts table is calculated based on the MMC results, summarising counts per run, case, and pathway groups (i_dfp). It is a support table to calculate the multi-model consensus and unanimous statistics.

Below is a brief overview of the Consensus and Unanimous Counts table for COVID-19 (STable 11)

| run | case | i_dfp | consensus | n | unanimous | not unanimous |
|---|---|---|---|---|---|---|
| run01 | g2a_male | 0 | Yes | 21 | 10 | 11 |
| run01 | g2a_male | 0 | No | 17 | 11 | 6 |
| run01 | g2a_male | 0 | Doubt | 7 | 0 | 7 |
| run01 | g2a_male | 1 | Yes | 5 | 3 | 2 |
| run01 | g2a_male | 1 | No | 35 | 18 | 17 |

STable 11 - The Consensus and Unanimous Counts table for COVID-19 is a comprehensive summary that displays all possible combinations of run, case, pathway group (i_dfp), and consensus. The results are categorised into 'Yes', 'No', and 'Doubt'. The column labelled 'n' represents the number of pathways identified on the MMC for each combination, the total counts of unanimous results, and their complement (non-unanimous results).

Filename: gemini_summary_yes_no_doubt_for_COVID-19_all_runs_models_[1, 3]_all_data.tsv

Next is a brief overview of the Consensus and Unanimous Counts table for MB (STable 12)

| run | case | i_dfp | consensus | n | unanimous | not unanimous |
|---|---|---|---|---|---|---|
| run01 | WNT | 0 | Yes | 39 | 20 | 19 |
| run01 | WNT | 0 | No | 25 | 7 | 18 |
| run01 | WNT | 0 | Doubt | 4 | 0 | 4 |
| run01 | WNT | 1 | Yes | 18 | 7 | 11 |
| run01 | WNT | 1 | No | 43 | 13 | 30 |

STable 12 - The Consensus and Unanimous Counts table for MB is a comprehensive summary that displays all possible combinations of run, case, pathway group (i_dfp), and consensus. The results are categorised into 'Yes', 'No', and 'Doubt'. The column labelled 'n' represents the number of pathways identified on the MMC for each combination, the total counts of unanimous results, and their complement (non-unanimous results).

Filename: gemini_summary_yes_no_doubt_for_medulloblastoma_all_runs_models_[1,3]_all_data.tsv

## Gemini Reproducibility

## Hard reproducibility

### Run-to-run reproducibility (RRR)

The run-to-run reproducibility (RRR) table compares each answer for two different runs.

Below is the RRR table summary for COVID-19 (see STable 13).

| run | hard_repro | mean | std | n | n_similar | n_different |
|---|---|---|---|---|---|---|
| | run-to-run | 99.31% | 6.74% | 7,520 | 7,471 | 49 |

STable 13 - The run-to-run reproducibility (RRR) table summarises the run-to-run comparisons. Here, one can see the comparisons between run01 and run02 for the COVID-19 proteomics study. The columns include the mean and standard deviation (std) of the RRR, based on n pathway responses labelled as 'Yes' and 'Possible,' which are considered equivalent, while 'No' and 'Low evidence' are also considered equivalent. The columns n_similar and n_different represent the number of similar and differing answers.

Next is the RRR table summary for MB (STable 14).

| run | hard repro | mean | std | n | n_similar | n_different |
|---|---|---|---|---|---|---|
| | run-to-run | 99.85% | 2.95% | 3,336 | 3,331 | 5 |

STable 14 - The run-to-run reproducibility (RRR) table summarises the run-to-run comparisons. Here, one can see the comparisons between run01 and run02 for the MB microarray study. The columns include the mean and standard deviation (std) of the RRR, based on n pathway responses labelled as 'Yes' and 'Possible,' which are considered equivalent, while 'No' and 'Low evidence' are also considered equivalent. The columns n_similar and n_different represent the number of similar and differing answers.

**Run-to-run reproducibility (RRR) per case**

The RRR per case table compares answers split by case for two runs.

Below is the RRR table split by case for COVID-19 (STable 15).

| case | mean | std | n | n_similar | n_different |
|---|---|---|---|---|---|
| g2a_male | 99.21% | 7.38% | 1,080 | 1,071 | 9 |
| g2a_female | 99.35% | 6.57% | 696 | 690 | 6 |
| g2b_male | 99.05% | 7.54% | 1,008 | 999 | 9 |
| g2b_female | 99.20% | 7.13% | 872 | 865 | 7 |
| g3_male_adult | 99.13% | 8.02% | 576 | 571 | 5 |
| g3_male_elder | 99.68% | 4.30% | 1,440 | 1,437 | 3 |
| g3_female_adult | 99.12% | 7.97% | 920 | 914 | 6 |
| g3_female_elder | 99.50% | 5.59% | 928 | 924 | 4 |

STable 15 - The RRR table split by cases summarises run-to-run comparisons. Here, one can see the comparisons between run01 and run02 for the COVID-19 proteomics study. The columns include the mean and standard deviation (std) of each RRR case, based on n pathway responses labelled 'Yes' and 'Possible,' considered equivalent, while 'No' and 'Low evidence' are also considered equivalent. The columns n_similar and n_different represent the number of similar and differing answers.

Next is the RRR table split by case for MB (STable 16).

| case | mu | std | n | n_similar | n_different |
|---|---|---|---|---|---|
| WNT | 99.82% | 3.26% | 1,632 | 1,629 | 3 |
| G4 | 99.88% | 2.61% | 1,704 | 1,702 | 2 |

STable 16 - The RRR table split by cases summarises run-to-run comparisons. Here, one can see the comparisons between run01 and run02 for the MB microarray study. The columns include the mean and standard deviation (std) of each RRR case, based on n pathway responses labelled 'Yes' and 'Possible,' considered equivalent, while 'No' and 'Low evidence' are also considered equivalent. The columns n_similar and n_different represent the number of similar and differing answers.

**Inter-model reproducibility (IMR) per case/subtype**

The inter-model reproducibility (IMR) compares each answer for two Gemini models in one run. It can be presented in a summarised form or per case/subtype. First, we present the per-case or subtype IMR table. Since Gemini's web service was accessed between October 2024 and December 2024, we selected the 1.5-pro and 1.5-flash Gemini models.

Below is the IMR table for COVID-19 (STable 17).

| case | mu | std | n | n_similar | n_different |
|---|---|---|---|---|---|
| g2a_male | 69.67% | 28.70% | 540 | 390 | 150 |
| g2a_female | 83.10% | 28.12% | 348 | 279 | 69 |
| g2b_male | 73.77% | 25.39% | 504 | 409 | 95 |
| g2b_female | 70.02% | 28.21% | 436 | 323 | 113 |
| g3_male_adult | 77.29% | 27.68% | 288 | 234 | 54 |
| g3_male_elder | 75.69% | 24.46% | 720 | 617 | 103 |
| g3_female_adult | 68.48% | 29.08% | 460 | 324 | 136 |
| g3_female_elder | 69.81% | 27.50% | 464 | 347 | 117 |

STable 17 - The inter-model reproducibility (IMR) table for the COVID-19 proteomics study presents the similarity between answers by comparing the 1.5-pro and 1.5-flash models using run01. The rows represent the cases, while the columns include the mean and standard deviation calculated from the answers' similarities, and n is the total number of queries, followed by the total number of similar answers (n_similar) and the total number of different answers (n_different). "Similar" means that the 'Yes' and 'Possible' answers are considered equal, and the 'No' and 'Low evidence' are also considered equal.

Filename: inter_model_hard_repro_per_case_for_COVID-19_models_[1,3]_run_run01_all_data.tsv

Below is the IMR table for MB (STable 18).

| case | mu | std | n | n_similar | n_different |
|---|---|---|---|---|---|
| WNT | 77.60% | 32.95% | 816 | 614 | 202 |
| G4 | 79.20% | 30.88% | 852 | 646 | 206 |

STable 18 - The inter-model reproducibility (IMR) table for the MB microarray study presents the similarity between answers by comparing the 1.5-pro and 1.5-flash models using run01. The rows represent the cases, while the columns include the mean and standard deviation calculated from the answers' similarities, and n is the total number of queries, followed by the total number of similar answers (n_similar) and the total number of different answers (n_different). "Similar" means that the 'Yes' and 'Possible' answers are considered equal, and the 'No' and 'Low evidence' are also considered equal.

Filename: inter_model_hard_repro_per_case_for_medulloblastoma_models_[1, 3]_run_run01_all_data.tsv

**Summary**

The RRR and IMR can be summarised in one RRR-IMR summary table to compare how the run-to-run and inter-model reproducibility behave. The first line shows the RRR results followed by two IMR results, one for each run.

Below is the RRR-IMR summary table for COVID-19 (STable 19).

| run | hard_repro | mu | std | n | n_similar | n_different |
|---|---|---|---|---|---|---|
| | run-to-run | 99.31% | 6.74% | 7,520 | 7,471 | 49 |
| run01 | inter-model | 73.11% | 27.52% | 3,760 | 2,923 | 837 |
| run02 | inter-model | 73.42% | 27.35% | 3,760 | 2,945 | 815 |

STable 19 - The RRR-IMR summary table for the COVID-19 proteomics study. The IMR compares 1.5-pro and 1.5-flash models. The first line shows the RRR results, followed by two IMR results, one for each run. Columns include the mean and standard deviation calculated from all answers' similarities, where n is the total number of all 8 cases' queries, followed by the total number of similar answers (n_similar) and the total number of different answers (n_different).

Filename: inter_model_hard_repro_summary_for_COVID-19_models_[1, 3]_run_run01_all_data.tsv

Below is the RRR-IMR summary table for MB (STable 20).

| run | hard_repro | mu | std | n | n_similar | n_different |
|---|---|---|---|---|---|---|
|       | run-to-run | 99.85% | 2.95%  | 3,336 | 3,331 | 5   |
| run01 | inter-model | 78.42% | 31.91% | 1,668 | 1,260 | 408 |
| run02 | inter-model | 78.51% | 31.86% | 1,668 | 1,261 | 407 |

STable 20 - The RRR-IMR summary table for the MB microarray study. The IMR compares 1.5-pro and 1.5-flash models. The first line shows the RRR results, followed by two IMR results, one for each run. Columns include the mean and standard deviation calculated from all answers' similarities, where n is the total number of all 2 cases' queries, followed by the total number of similar answers (n_similar) and the total number of different answers (n_different).

Filename: inter_model_hard_repro_summary_for_medulloblastoma_models_[1, 3]_run_run01_all_data.tsv

## Soft reproducibility

Soft reproducibility compares consensuses; therefore, it is expected to be more flexible and powerful than hard reproducibility, which compares each query answer.

### RRCR

The run-to-run consensus reproducibility (RRCR) table compares all consensuses for two runs, given a pathway and a model.

Below is the RRCR summary for COVID-19 and MB (STable 21).

| case     | mean repro | std repro | n     | fdr      | pvalue   | yes equal | yes diff | no equal | no diff |
|----------|------------|-----------|-------|----------|----------|-----------|----------|----------|---------|
| COVID-19 | 97.54%     | 15.48%    | 1,262 | 1.70E-08 | 1.88E-09 | 473       | 29       | 758      | 2       |
| MB       | 99.82%     | 4.24%     | 556   | 1.00E+00 | 1.00E+00 | 222       | 0        | 333      | 1       |

STable 21- The RRCR compares the number of consensus agreements by comparing two runs (run01 x run02). Consensuses were calculated using the multi-model approach; here, we used the Gemini models 1.5-pro and 1.5-flash. FDR > 0.05 means that agreements between runs are statistically similar. We used the Chi-squared test by comparing List 1 (run01) to List 2 (run02). The mean repro percentages and respective standard deviation are the mean percentages of agreement between run01 and run02.

Filename: run_run_soft_consensus_stats_for_<disease>_between_run01_x_run02_models_[1,3]_all_data.tsv

**RRCR per case**

Cases or subtypes split the RRCR table. This approach improves our understanding of reproducibility, as it is likely that some cases are more explicitly documented in the literature than others.

Below is the RRCR table categorised by cases for COVID-19 (STable 22).

| case | mean repro | std repro | n | fdr | pvalue | yes equal | yes diff | no equal | no diff |
|---|---|---|---|---|---|---|---|---|---|
| g2a male | 98.33% | 12.80% | 180 | 2.00E-01 | 1.11E-01 | 67 | 3 | 110 | 0 |
| g2a female | 98.28% | 13.02% | 116 | 1.00E+00 | 1.00E+00 | 52 | 1 | 62 | 1 |
| g2b male | 96.43% | 18.56% | 168 | 1.39E-02 | 4.64E-03 | 56 | 6 | 106 | 0 |
| g2b female | 96.73% | 17.78% | 153 | 4.48E-02 | 1.99E-02 | 56 | 5 | 92 | 0 |
| g3 male adult | 96.88% | 17.40% | 96 | 2.55E-01 | 1.99E-01 | 42 | 3 | 51 | 0 |
| g3 male elder | 97.50% | 15.61% | 240 | 6.19E-03 | 1.38E-03 | 70 | 6 | 164 | 0 |
| g3 female adult | 98.05% | 13.82% | 154 | 2.87E-01 | 1.91E-01 | 68 | 3 | 83 | 0 |
| g3 female elder | 98.06% | 13.78% | 155 | 8.52E-01 | 7.57E-01 | 62 | 2 | 90 | 1 |

STable 22- The RRCR split by cases compares the number of consensus agreements by comparing two runs (run01 x run02). Consensuses were calculated using the multi-model approach; here, we used the Gemini models 1.5-pro and 1.5-flash. FDR > 0.05 means that agreements between runs are statistically similar. We used the Chi-squared test by comparing List 1 (run01) to List 2 (run02). The mean repro percentages and respective standard deviation are the mean percentages of agreement between run01 and run02.

Filename: run_run_soft_consensus_stats_for_COVID-19_between_run01_x_run02_models_[1,3]_all_data.tsv

Next is the RRCR table categorised by cases for MB (STable 23).

| case | mean repro | std repro | n | fdr | pvalue | yes equal | yes diff | no equal | no diff |
|---|---|---|---|---|---|---|---|---|---|
| WNT | 100.00% | 0.00% | 272 | 1.00E+00 | 1.00E+00 | 124 | 0 | 148 | 0 |
| G4 | 99.65% | 5.92% | 284 | 1.00E+00 | 1.00E+00 | 98 | 0 | 185 | 1 |

STable 23- The RRCR split by cases compares the number of consensus agreements by comparing two runs (run01 x run02). Consensuses were calculated using the multi-model approach; here, we used the Gemini models 1.5-pro and 1.5-flash. FDR > 0.05 means that agreements between runs are statistically similar. We used the Chi-squared test by comparing List 1 (run01) to List 2 (run02). The mean repro percentages and respective standard deviation are the mean percentages of agreement between run01 and run02.

Filename: run_run_soft_consensus_stats_for_<disease>_between_run01_x_run02_models_[1,3]_all_data.tsv

**RRCR per case and group**

We investigate whether run01 yields a similar number of Yes-consensus counts compared to run02, specifically grouped by i_dfp, which means grouping by G0, G1, G2, and G3. Initially, for COVID-19, we compared whether the Yes-consensus distributions between run01 and run02 for G0 are statistically similar (STable 24). In this context, List 1 (run01) and List 2 (run02) represent the Yes counts for Gemini models 1.5-pro (model 1) and 1.5-flash (model 3).

Below is the COVID-19 Yes-consensus table for G0 between run01 and run02 (STable 24).

| case | n | p-value | repro yes mean perc1 | repro yes std perc1 | repro yes mean perc2 | repro yes std perc2 | repro yes List 1 | repro yes List 2 |
|---|---|---|---|---|---|---|---|---|
| g2a_male | 45 | 1.00E+00 | 47.78% | 14.44% | 47.78% | 14.44% | [15, 28] | [15, 28] |
| g2a_female | 29 | 1.00E+00 | 51.72% | 3.45% | 51.72% | 3.45% | [16, 14] | [16, 14] |
| g2b_male | 42 | 1.00E+00 | 54.76% | 11.90% | 54.76% | 11.90% | [18, 28] | [18, 28] |
| g2b_female | 44 | 1.00E+00 | 48.86% | 12.50% | 48.86% | 12.50% | [16, 27] | [16, 27] |
| g3_male_adult | 24 | 1.00E+00 | 54.17% | 4.17% | 54.17% | 4.17% | [12, 14] | [12, 14] |
| g3_male_elder | 60 | 1.00E+00 | 45.83% | 2.50% | 46.67% | 3.33% | [26, 29] | [26, 30] |
| g3_female_adult | 39 | 1.00E+00 | 56.41% | 12.82% | 56.41% | 12.82% | [17, 27] | [17, 27] |
| g3_female_elder | 39 | 1.00E+00 | 51.28% | 12.82% | 51.28% | 12.82% | [15, 25] | [15, 25] |

STable 24 - The Yes-consensus table compares the "Yes" consensus distribution between run01 and run02, given G0 (i_dfp=0, the positive control group). The rows are the COVID-19 cases, while columns include: "n" as the number of pathways for each case, "repo yes mean perc1 and perc2", which are the percentage of Yes for run01 and run02 according to n pathways, and "repo yes std perc1 and perc2" which present the respective standard deviation. List 1 and List 2 provide the counts of Yes consensuses for Gemini models 1 and 3. A p-value greater than 0.05 indicates that the distributions of "Yes" consensuses given the two runs are statistically similar, as tested by the Chi-squared test or FET.

Filename: run_run_soft_consensus_repro_per_idfp_for_medulloblastoma_between_run01_x_run02_models_[1, 3]_all_data.tsv

Next is the COVID-19 Yes-consensus distribution table for G1 between run01 and run02 (STable 25).

| case | n | p-value | repro yes mean perc1 | repro yes std perc1 | repro yes mean perc2 | repro yes std perc2 | repro yes List 1 | repro yes List 2 |
|---|---|---|---|---|---|---|---|---|
| g2a_male | 45 | 1.00E+00 | 16.67% | 10.00% | 16.67% | 10.00% | [3, 12] | [3, 12] |
| g2a_female | 29 | 1.00E+00 | 27.59% | 6.90% | 25.86% | 5.17% | [10, 6] | [9, 6] |
| g2b_male | 42 | 1.00E+00 | 11.90% | 4.76% | 11.90% | 4.76% | [3, 7] | [3, 7] |
| g2b_female | 44 | 1.00E+00 | 20.00% | 5.00% | 20.00% | 5.00% | [6, 10] | [6, 10] |
| g3_male_adult | 24 | 1.00E+00 | 20.83% | 4.17% | 20.83% | 4.17% | [6, 4] | [6, 4] |
| g3_male_elder | 60 | 1.00E+00 | 16.67% | 0.00% | 16.67% | 0.00% | [10, 10] | [10, 10] |
| g3_female_adult | 39 | 1.00E+00 | 35.14% | 8.11% | 35.14% | 8.11% | [10, 16] | [10, 16] |
| g3_female_elder | 39 | 1.00E+00 | 19.74% | 3.95% | 19.74% | 3.95% | [6, 9] | [6, 9] |

STable 25 - The Yes-consensus table compares the "Yes" consensus distribution between run01 and run02, given G1 (i_dfp=1, the cloudy group 1). The rows are the COVID-19 cases, while columns include: "n" as the number of pathways for each case, "repo yes mean perc1 and perc2", which are the percentage of Yes for run01 and run02 according to n pathways, and "repo yes std perc1 and perc2" which present the respective standard deviation. List 1 and List 2 provide the counts of Yes consensuses for Gemini models 1 and 3. A p-value greater than 0.05 indicates that the distributions of "Yes" consensuses given the two runs are statistically similar, as tested by the Chi-squared test or FET.

Next is the COVID-19 Yes-consensus distribution table for G2 between run01 and run02 (STable 26).

| case | n | p-value | repro yes mean perc1 | repro yes std perc1 | repro yes mean perc2 | repro yes std perc2 | repro yes List 1 | repro yes List 2 |
|---|---|---|---|---|---|---|---|---|
| g2a_male | 45 | 1.00E+00 | 47.78% | 12.22% | 47.78% | 12.22% | [16, 27] | [16, 27] |
| g2a_female | 29 | 1.00E+00 | 43.10% | 1.72% | 44.83% | 0.00% | [12, 13] | [13, 13] |
| g2b_male | 42 | 1.00E+00 | 32.14% | 13.10% | 32.14% | 13.10% | [8, 19] | [8, 19] |
| g2b_female | 44 | 1.00E+00 | 40.00% | 16.00% | 40.00% | 16.00% | [6, 14] | [6, 14] |
| g3_male_adult | 24 | 6.27E-01 | 66.67% | 8.33% | 68.75% | 10.42% | [14, 18] | [14, 19] |
| g3_male_elder | 60 | 1.00E+00 | 30.83% | 7.50% | 30.83% | 7.50% | [14, 23] | [14, 23] |
| g3_female_adult | 39 | 1.00E+00 | 41.03% | 20.51% | 46.15% | 15.38% | [8, 24] | [12, 24] |
| g3_female_elder | 39 | 1.00E+00 | 44.87% | 16.67% | 46.15% | 15.38% | [11, 24] | [12, 24] |

STable 26 - The Yes-consensus table compares the "Yes" consensus distribution between run01 and run02, given G2 (i_dfp=2, the cloudy group 2). The rows are the COVID-19 cases, while columns include: "n" as the number of pathways for each case, "repo yes mean perc1 and perc2", which are the percentage of Yes for run01 and run02 according to n pathways, and "repo yes std perc1 and perc2" which present the respective standard deviation. List 1 and List 2 provide the counts of Yes consensuses for Gemini models 1 and 3. A p-value greater than 0.05 indicates that the distributions of "Yes" consensuses given the two runs are statistically similar, as tested by the Chi-squared test or FET.

Next is the COVID-19 Yes-consensus distribution table for G3 between run01 and run02 (STable 27).

| case | n | p-value | repro yes mean perc1 | repro yes std perc1 | repro yes mean perc2 | repro yes std perc2 | repro yes List 1 | repro yes List 2 |
|---|---|---|---|---|---|---|---|---|
| g2a_male | 45 | 3.91E-01 | 6.67% | 4.44% | 5.56% | 1.11% | [5, 1] | [2, 3] |
| g2a_female | 29 | 1.00E+00 | 6.90% | 3.45% | 6.90% | 3.45% | [3, 1] | [3, 1] |
| g2b_male | 42 | 5.89E-01 | 9.52% | 2.38% | 7.14% | 2.38% | [5, 3] | [2, 4] |
| g2b_female | 44 | 5.40E-01 | 6.82% | 4.55% | 6.82% | 0.00% | [5, 1] | [3, 3] |
| g3_male_adult | 24 | 3.86E-01 | 6.25% | 6.25% | 6.25% | 2.08% | [3, 0] | [1, 2] |
| g3_male_elder | 60 | 2.35E-01 | 6.67% | 5.00% | 4.17% | 0.83% | [7, 1] | [2, 3] |
| g3_female_adult | 39 | 5.19E-01 | 6.41% | 3.85% | 6.41% | 1.28% | [4, 1] | [2, 3] |
| g3_female_elder | 39 | 4.80E-01 | 5.13% | 2.56% | 5.13% | 2.56% | [3, 1] | [1, 3] |

STable 27 - The Yes-consensus table compares the "Yes" consensus distribution between run01 and run02, given G3 (i_dfp=3, the negative control group). The rows are the COVID-19 cases, while columns include: "n" as the number of pathways for each case, "repo yes mean perc1 and perc2", which are the percentage of Yes for run01 and run02 according to n pathways, and "repo yes std perc1 and perc2" which present the respective standard deviation. List 1 and List 2 provide the counts of Yes consensuses for Gemini models 1 and 3. A p-value greater than 0.05 indicates that the distributions of "Yes" consensuses given the two runs are statistically similar, as tested by the Chi-squared test or FET.

Below is the MB Yes-consensus table for G0 between run01 and run02 (STable 28).

| case | n | p-value | repro yes mean perc1 | repro yes std perc1 | repro yes mean perc2 | repro yes std perc2 | repro yes List 1 | repro yes List 2 |
|---|---|---|---|---|---|---|---|---|
| WNT | 68 | 1.00E+00 | 50.00% | 2.94% | 50.00% | 2.94% | [36, 32] | [36, 32] |
| G4 | 71 | 1.00E+00 | 26.76% | 15.49% | 26.76% | 15.49% | [30, 8] | [30, 8] |

STable 28 - The Yes-consensus table compares the "Yes" consensus distribution between run01 and run02, given G0 (i_dfp=0, the positive control group). The rows are the MB cases, while columns include: "n" as the number of pathways for each case, "repo yes mean perc1 and perc2", which are the percentage of Yes for run01 and run02 according to n pathways, and "repo yes std perc1 and perc2" which present the respective standard deviation. List 1 and List 2 provide the counts of Yes consensuses for Gemini models 1 and 3. A p-value greater than 0.05 indicates that the distributions of "Yes" consensuses given the two runs are statistically similar, as tested by the Chi-squared test or FET.

Filename: run_run_soft_consensus_repro_per_idfp_for_medulloblastoma_between_run01_x_run02_models_[1,3]_all_data.tsv

Next is the MB Yes-consensus distribution table for G1 between run01 and run02 (STable 29).

| case | n | p-value | repro yes mean perc1 | repro yes std perc1 | repro yes mean perc2 | repro yes std perc2 | repro yes List 1 | repro yes List 2 |
|---|---|---|---|---|---|---|---|---|
| WNT | 68 | 1.00E+00 | 24.26% | 0.74% | 24.26% | 0.74% | [17, 16] | [17, 16] |
| G4 | 71 | 1.00E+00 | 21.13% | 8.45% | 21.13% | 8.45% | [21, 9] | [21, 9] |

STable 29 - The Yes-consensus table compares the "Yes" consensus distribution between run01 and run02, given G1 (i_dfp=1, the cloudy group 1). The rows are the MB cases, while columns include: "n" as the number of pathways for each case, "repo yes mean perc1 and perc2", which are the percentage of Yes for run01 and run02 according to n pathways, and "repo yes std perc1 and perc2" which present the respective standard deviation. List 1 and List 2 provide the counts of Yes consensuses for Gemini models 1 and 3. A p-value greater than 0.05 indicates that the distributions of "Yes" consensuses given the two runs are statistically similar, as tested by the Chi-squared test or FET.

Next is the MB Yes-consensus distribution table for G2 between run01 and run02 (STable 30).

| case | n | p-value | repro yes mean perc1 | repro yes std perc1 | repro yes mean perc2 | repro yes std perc2 | repro yes List 1 | repro yes List 2 |
|---|---|---|---|---|---|---|---|---|
| WNT | 68 | 1.00E+00 | 36.76% | 8.82% | 37.50% | 9.56% | [31, 19] | [32, 19] |
| G4 | 71 | 1.00E+00 | 15.49% | 11.27% | 15.49% | 11.27% | [19, 3] | [19, 3] |

STable 30 - The Yes-consensus table compares the "Yes" consensus distribution between run01 and run02, given G2 (i_dfp=2, the cloudy group 2). The rows are the MB cases, while columns include: "n" as the number of pathways for each case, "repo yes mean perc1 and perc2", which are the percentage of Yes for run01 and run02 according to n pathways, and "repo yes std perc1 and perc2" which present the respective standard deviation. List 1 and List 2 provide the counts of Yes consensuses for Gemini models 1 and 3. A p-value greater than 0.05 indicates that the distributions of "Yes" consensuses given the two runs are statistically similar, as tested by the Chi-squared test or FET.

Next is the MB Yes-consensus distribution table for G3 between run01 and run02 (STable 31).

| case | n | p-value | repro yes mean perc1 | repro yes std perc1 | repro yes mean perc2 | repro yes std perc2 | repro yes List 1 | repro yes List 2 |
|---|---|---|---|---|---|---|---|---|
| WNT | 68 | 1.00E+00 | 18.38% | 2.21% | 18.38% | 2.21% | [11, 14] | [11, 14] |
| G4 | 71 | 1.00E+00 | 9.15% | 4.93% | 9.15% | 4.93% | [10, 3] | [10, 3] |

STable 31 - The Yes-consensus table compares the "Yes" consensus distribution between run01 and run02, given G3 (i_dfp=3, the negative control group). The rows are the MB cases, while columns include: "n" as the number of pathways for each case, "repo yes mean perc1 and perc2", which are the percentage of Yes for run01 and run02 according to n pathways, and "repo yes std perc1 and perc2" which present the respective standard deviation. List 1 and List 2 provide the counts of Yes consensuses for Gemini models 1 and 3. A p-value greater than 0.05 indicates that the distributions of "Yes" consensuses given the two runs are statistically similar, as tested by the Chi-squared test or FET.

**IMCR**

The Inter-model Consensus Reproducibility (IMCR) table compares two Gemini models given one run. Below is the IMCR (see STable 32), comparing Gemini 1.5-pro and 1.5-flash for all possible runs.

| study | run | IMCR | std |
|---|---|---|---|
| COVID-19 | run01 | 72.8% | 4.8% |
| | run02 | 76.1% | 4.3% |
| Medulloblastoma | run01 | 62.9% | 0.1% |
| | run02 | 63.5% | 0.4% |
| | run03 | 62.9% | 0.1% |

STable 32 - The IMCR for COVID-19 and Medulloblastoma studies. Rows are disease and runs; columns are the mean IMCR and its standard deviation.

Below is the IMCR split by cases for COVID-19 (STable 33), comparing the Gemini 1.5-pro and 1.5-flash models.

| case | mean similarity | std similarity | mean sim yes | std sim yes | n | n similar | n similar yes |
|---|---|---|---|---|---|---|---|
| g2a_male | 68.33% | 46.65% | 56.11% | 49.76% | 180 | 123 | 101 |
| g2a_female | 70.69% | 45.72% | 59.48% | 49.31% | 116 | 82 | 69 |
| g2b_male | 80.36% | 39.85% | 64.88% | 47.88% | 168 | 135 | 109 |
| g2b_female | 71.24% | 45.41% | 59.48% | 49.25% | 153 | 109 | 91 |
| g3_male_adult | 72.92% | 44.67% | 58.33% | 49.56% | 96 | 70 | 56 |
| g3_male_elder | 79.17% | 40.70% | 66.67% | 47.24% | 240 | 190 | 160 |
| g3_female_adult | 66.88% | 47.22% | 55.19% | 49.89% | 154 | 103 | 85 |
| g3_female_elder | 72.90% | 44.59% | 60.00% | 49.15% | 155 | 113 | 93 |

STable 33 - The IMCR table is split by cases for COVID-19 by comparing two Gemini Models, 1.5-pro and 1.5-flash, for run01. Rows are disease; columns are the mean IMCR and its standard deviation when comparing consensus, respective mean/std comparing consensus and equal number of Yes, n as the total number of enriched pathways times 4 (number of groups), and "n similar" and "similar n yes" as the respective counts.

Next is the IMCR split by cases for MB (STable 34), comparing the Gemini 1.5-pro and 1.5-flash models.

| case | mean similarity | std similarity | mean sim yes | std sim yes | n | n similar | n similar yes |
|---|---|---|---|---|---|---|---|
| WNT | 62.87% | 48.40% | 44.49% | 49.79% | 272 | 171 | 121 |
| G4 | 63.03% | 48.36% | 41.55% | 49.37% | 284 | 179 | 118 |

STable 34 - The IMCR table is split by cases for MB by comparing two Gemini Models, 1.5-pro and 1.5-flash, for run01. Rows are disease; columns are the mean IMCR and its standard deviation when comparing consensus, respective mean/std comparing consensus and equal number of Yes, n as the total number of enriched pathways times 4 (number of groups), and "n similar" and "similar n yes" as the respective counts.

**IMCR Venn Diagram**

The Venn Diagram illustrates inter-model consensus for the two models, filtering equal levels of consensus and highlighting the agreements for a disease case and group. Three distinct diagrams can be seen in SFigure 14: in A, we see the common 'Yes' pathway consensuses for COVID-19 case "g2a male"; in B, the common 'No' pathway consensuses; and in C, the common 'Doubt' pathway consensuses. The OMC consensus table column includes entries for 'Yes,' 'No,' and 'Doubt' and compares two Gemini models based on a single run equal to 'run01.'

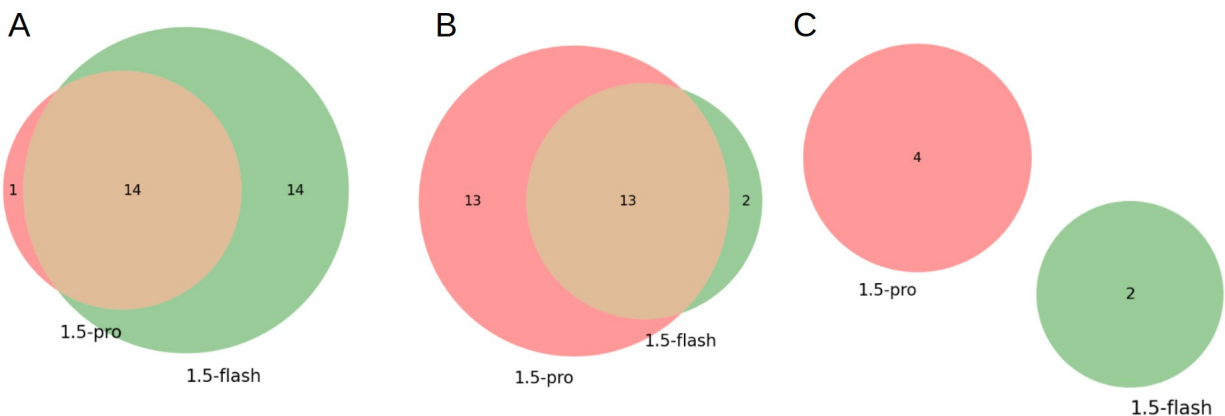

SFigure 14 - This Venn diagram illustrates inter-model comparisons between models 1.5-pro and 1.5-flash for the COVID-19 case 'g2a male' within group G0 (enriched pathways), based on run01. In A, the 'Yes' diagram, the two models agree on 14 pathways, constituting 48.28% of the 29 pathways identified as 'Yes' out of a possible 45 enriched pathways. The first model identified 15 pathways as 'Yes,' while the second identified 28. In B, the 'No' diagram, the two models agree on 13 pathways, constituting 46.43% of the 28 pathways identified as 'No.' The first model identified 26 pathways as 'No,' while the second identified 15. In C, the 'Doubt' diagram, the two models agree on 0 of the 6 pathways identified as 'Doubt.' The first model identified 4 pathways as 'Doubt,' while the second identified 2.

**UR**

The Unanimous Reproducibility (UR) is calculated over the 4DSSQs. "Unanimous" is defined when all 4DSSQs' answers for each Gemini model are equal 'Yes' or 'No.' Consequently, the UR indicates the semantic consistency of Gemini models.

Below is the UR table for COVID-19, for run equals 'run01' (STable 36)

| case | i_dfp | consensus | n | unanimous | not unanimous | unan percent |
|---|---|---|---|---|---|---|
| g2a_male | 0 | Yes | 21 | 10 | 11 | 47.6% |
| g2a_male | 0 | No | 17 | 11 | 6 | 64.7% |
| g2a_male | 0 | Doubt | 7 | 0 | 7 | 0.0% |
| g2a_male | 1 | Yes | 5 | 3 | 2 | 60.0% |
| g2a_male | 1 | No | 35 | 18 | 17 | 51.4% |
| g2a_male | 1 | Doubt | 5 | 0 | 5 | 0.0% |
| g2a_male | 2 | Yes | 22 | 13 | 9 | 59.1% |
| g2a_male | 2 | No | 18 | 11 | 7 | 61.1% |
| g2a_male | 2 | Doubt | 5 | 0 | 5 | 0.0% |
| g2a_male | 3 | Yes | 3 | 1 | 2 | 33.3% |
| g2a_male | 3 | No | 40 | 29 | 11 | 72.5% |
| g2a_male | 3 | Doubt | 2 | 0 | 2 | 0.0% |
| g2a_female | 0 | Yes | 17 | 9 | 8 | 52.9% |
| g2a_female | 0 | No | 10 | 6 | 4 | 60.0% |
| g2a_female | 0 | Doubt | 2 | 0 | 2 | 0.0% |
| g2a_female | 1 | Yes | 12 | 3 | 9 | 25.0% |
| g2a_female | 1 | No | 15 | 7 | 8 | 46.7% |
| g2a_female | 1 | Doubt | 2 | 0 | 2 | 0.0% |
| g2a_female | 2 | Yes | 12 | 7 | 5 | 58.3% |
| g2a_female | 2 | No | 13 | 6 | 7 | 46.2% |
| g2a_female | 2 | Doubt | 4 | 0 | 4 | 0.0% |
| g2a_female | 3 | Yes | 2 | 1 | 1 | 50.0% |
| g2a_female | 3 | No | 25 | 17 | 8 | 68.0% |
| g2a_female | 3 | Doubt | 2 | 0 | 2 | 0.0% |

STable 35 - The Consensus and Unanimous Counts table for COVID-19 is a comprehensive summary that displays all possible combinations of run, case, pathway group (i_dfp), and consensus. The results are categorised into 'Yes', 'No', and 'Doubt'. The column labelled 'n' represents the number of pathways identified on the MMC for each combination, the total counts of unanimous results, and their complement (non-unanimous results).

Filename: gemini_summary_yes_no_doubt_for_COVID-19_all_runs_models_[1,3]_all_data.tsv

Next is the UR table for MB (STable 36)

| case | i_dfp | consensus | n | unanimous | not_unanimous | unan percent |
|---|---|---|---|---|---|---|
| WNT | 0 | Yes | 39 | 20 | 19 | 51.3% |
| WNT | 0 | No | 25 | 7 | 18 | 28.0% |
| WNT | 0 | Doubt | 4 | 0 | 4 | 0.0% |
| WNT | 1 | Yes | 18 | 7 | 11 | 38.9% |
| WNT | 1 | No | 43 | 13 | 30 | 30.2% |
| WNT | 1 | Doubt | 7 | 0 | 7 | 0.0% |
| WNT | 2 | Yes | 32 | 7 | 25 | 21.9% |
| WNT | 2 | No | 29 | 8 | 21 | 27.6% |
| WNT | 2 | Doubt | 7 | 0 | 7 | 0.0% |
| WNT | 3 | Yes | 13 | 4 | 9 | 30.8% |
| WNT | 3 | No | 51 | 29 | 22 | 56.9% |
| WNT | 3 | Doubt | 4 | 0 | 4 | 0.0% |
| G4 | 0 | Yes | 28 | 5 | 23 | 17.9% |
| G4 | 0 | No | 35 | 12 | 23 | 34.3% |
| G4 | 0 | Doubt | 8 | 0 | 8 | 0.0% |
| G4 | 1 | Yes | 22 | 5 | 17 | 22.7% |
| G4 | 1 | No | 45 | 17 | 28 | 37.8% |
| G4 | 1 | Doubt | 4 | 0 | 4 | 0.0% |
| G4 | 2 | Yes | 16 | 2 | 14 | 12.5% |
| G4 | 2 | No | 49 | 19 | 30 | 38.8% |
| G4 | 2 | Doubt | 6 | 0 | 6 | 0.0% |
| G4 | 3 | Yes | 10 | 0 | 10 | 0.0% |
| G4 | 3 | No | 57 | 31 | 26 | 54.4% |
| G4 | 3 | Doubt | 4 | 0 | 4 | 0.0% |

STable 36 - The Consensus and Unanimous Counts table for MB is a comprehensive summary that displays all possible combinations of run, case, pathway group (i_dfp), and consensus. The results are categorised into 'Yes', 'No', and 'Doubt'. The column labelled 'n' represents the number of pathways identified on the MMC for each combination, the total counts of unanimous results, and their complement (non-unanimous results).

Filename: gemini_summary_yes_no_doubt_for_medulloblastoma_all_runs_models_[1,3]_all_data.tsv

The Unanimous Reproducibility (UR), only for 'Yes' and 'No,' compares if all the 4DSSQ's answers are the same. The results can be seen in STable 37.

| case | UR | std |
|---|---|---|
| COVID-19 | 53.6% | 12.3% |
| MB | 31.5% | 15.2% |

STable 37 - The UR table was calculated, counting only 'Yes' and 'No' responses. The table presents the overall unanimous reproducibility and standard deviation for each study.

**PubMed search**

The PubMed search differs from the Gemini search in two main ways:

1. PubMed is an SQL-based tool, while Gemini is a semantic one. As a result, each PubMed query is run only once, whereas multiple searches are performed with Gemini for each pathway, utilising many Gemini models and the 4DSSQ.
2. We can use Reactome's pathway names to query the Gemini web service since Gemini "understands" natural language. However, we have implemented a translation where each pathway name is replaced by a set of a few terms to minimise false negatives.

**Reactome terms table**

The PubMed to Reactome terms table is a static table after the user has filled the term column. Below are the first lines (STable 38).

| pathway_id | pathway | term |
|---|---|---|
| R-HSA-114608 | Platelet Degranulation | Platelet Degranulation |
| R-HSA-76005 | Response To Elevated Platelet Cytosolic Ca2+ | Platelet Calcium |
| R-HSA-140877 | Formation Of Fibrin Clot (Clotting Cascade) | Fibrin Clot |
| R-HSA-76002 | Platelet Activation, Signaling And Aggregation | Platelet Activation |
| R-HSA-109582 | Hemostasis | Hemostasis |
| R-HSA-381426 | Regulation Of IGF Transport And Uptake By IGFBPs | IGF |
| R-HSA-140875 | Common Pathway Of Fibrin Clot Formation | Fibrin Clot Formation |
| R-HSA-372708 | p130Cas Linkage To MAPK Signaling For Integrins | MAPK Signaling Integrins |
| R-HSA-354194 | GRB2:SOS Provides Linkage To MAPK Signaling For Integrins | GRB2 SOS |
| R-HSA-216083 | Integrin Cell Surface Interactions | Integrin Cell Surface Interaction |
| R-HSA-354192 | Integrin Signaling | Integrin Signaling |

STable 38: The PubMed to Reactome Terms table. The first two columns are Reactome ID and pathway name, and the third column contains terms used to build a PubMed query. Additional specific words, prepositions, and other unnecessary words were removed; plurals should be converted to singular.

Filename: pubmed_to_reatome_terms_table.tsv in folder refseq.

**Query**

To query the PubMed web service, we used all selected pathways (from G0 to G3) previously used for Gemini. Therefore, after gathering all the results, one can compare PubMed searches against Gemini searches.

Next, one scans each selected pathway by crossing all pathways to each disease case/subtype, replacing each pathway name with a concatenated boolean query built with the terms.

For example, here, one demonstrates how to query using the second pathway ("Response To Elevated Platelet Cytosolic Ca2+):

For COVID-19:
        COVID-19 AND Platelet* AND Degranulation*

For MB:
        (MB OR medulloblastoma) AND Platelet* AND Degranulation*

The disease may have many aliases concatenated with an OR, and the terms are concatenated with ANDs followed by asterisks.

**PubMed tables**

The PubMed web service responds to each query with a set of PMIDs (the PubMed IDs) or an empty list if no references are found. There are three final tables:

1. "pubmed_search_no_symbol_…tsv": These tables cross all pathways (IDs and names) with PMID, pub_date (date of publication), title, keywords, abstract, and other columns.
2. "pubmed_search_summary_no_symbol_by_pathway_…tsv": These tables provide a summary for each pathway; therefore, each pathway corresponds to a column having concatenated PMIDs;
3. "pubmed_search_summary_no_symbol_by_pmid_…tsv": These tables summarize data by PMID; each pmid corresponds to a column concatenated pathway ID.

## Comparing Gemini to PubMed

We compared the results from Gemini and PubMed using the Ensemble dataset, which includes all pathways for each disease case and subtype. The tables below are categorised by "with_gender" because PubMed is sensitive to gender factors and other factors such as age and disease severity, whereas Gemini is not. It is important to note that the level of agreement increases when the "which_gender" factor is turned off (set to False). The Gemini consensuses were obtained from the MMC using the 1.5-pro and 1.5-flash models to compare with PubMed. The statistical analysis employed was a Chi-squared test, using a 2 by 2 matrix to compare the Gemini consensus counts against the findings from PubMed articles, divided into positive findings (Yes) and negative findings (No).

## Agreement between Gemini and PubMed

Below is the table PubMed versus Gemini agreement (STable 39), for run equal 'run01' and "with gender" equal False,

|  |  |  | with_gender = False | | | | | | | | | |
|---|---|---|---|---|---|---|---|---|---|---|---|---|
| case | i dfp | n | agree | std | gem yes | gem no | pub yes | pub no | both yn | only gyes | only pyes | FDR |
| g2a_male | 0 | 45 | 64.4% | 48.4% | 28 | 17 | 20 | 25 | 29 | 12 | 4 | 2.19E-01 |
| g2a_female | 0 | 29 | 69.0% | 47.1% | 19 | 10 | 16 | 13 | 20 | 6 | 3 | 7.18E-01 |
| g2b_male | 0 | 42 | 38.1% | 49.2% | 29 | 13 | 3 | 39 | 16 | 26 | 0 | 3.32E-07 |
| g2b_female | 0 | 44 | 40.9% | 49.7% | 27 | 17 | 1 | 43 | 18 | 26 | 0 | 3.00E-07 |
| g3_male_adult | 0 | 24 | 75.0% | 44.2% | 14 | 10 | 18 | 6 | 18 | 1 | 5 | 4.91E-01 |
| g3_male_elder | 0 | 60 | 70.0% | 46.2% | 30 | 30 | 18 | 42 | 42 | 15 | 3 | 8.84E-02 |
| g3_female_adult | 0 | 39 | 76.9% | 42.7% | 26 | 13 | 27 | 12 | 30 | 4 | 5 | 1.00E+00 |
| g3_female_elder | 0 | 39 | 69.2% | 46.8% | 25 | 14 | 17 | 22 | 27 | 10 | 2 | 1.90E-01 |

STable 39 - The PubMed versus Gemini agreement table for COVID-19 has been filtered to include only entries where run equals 'run01,' pathway group equals 0, and with_gender equals False. The rows represent cases, while the columns include the mean agreement and its standard deviation, alongside the counts for responses from Gemini if 'Yes' ('gem yes') and 'No' ('gem_no'); the counts for PubMed if 'Yes' ('pub_yes') and 'No' ('pub_no'), and total counts of agreements ('both_yn'), along with the counts of 'Yes' responses exclusively for Gemini ('only_gyes') and for PubMed ('only_pyes'). The final column is the Chi-squared test (or FET) test result conducted on a 2x2 table comparing the totals of 'Yes' and 'No' responses for Gemini versus PubMed.

Next is the table PubMed versus Gemini agreement (STable 40), for run equal 'run01' and "with gender" equal True,

| case | i dfp | n | agree | std | gem yes | gem no | pub yes | pub no | both yn | only gyes | only pyes | FDR |
|---|---|---|---|---|---|---|---|---|---|---|---|---|
| | | | | | with_gender = True | | | | | | | |
| g2a_male | 0 | 45 | 53.3% | 50.5% | 28 | 17 | 7 | 38 | 24 | 21 | 0 | 1.00E-04 |
| g2a_female | 0 | 29 | 51.7% | 50.9% | 19 | 10 | 7 | 22 | 15 | 13 | 1 | 1.50E-02 |
| g2b_male | 0 | 42 | 33.3% | 47.7% | 29 | 13 | 1 | 41 | 14 | 28 | 0 | 6.69E-08 |
| g2b_female | 0 | 44 | 40.9% | 49.7% | 27 | 17 | 1 | 43 | 18 | 26 | 0 | 3.00E-07 |
| g3_male_adult | 0 | 24 | 54.2% | 50.9% | 14 | 10 | 9 | 15 | 13 | 8 | 3 | 3.75E-01 |
| g3_male_elder | 0 | 60 | 55.0% | 50.2% | 30 | 30 | 7 | 53 | 33 | 25 | 2 | 1.00E-04 |
| g3_female_adult | 0 | 39 | 61.5% | 49.3% | 26 | 13 | 13 | 26 | 24 | 14 | 1 | 2.37E-02 |
| g3_female_elder | 0 | 39 | 48.7% | 50.6% | 25 | 14 | 5 | 34 | 19 | 20 | 0 | 1.09E-04 |

STable 40 - The PubMed versus Gemini agreement table for COVID-19 has been filtered to include only entries where run equals 'run01,' pathway group equals 0, and with_gender equals True. The rows represent cases, while the columns include the mean agreement and its standard deviation, alongside the counts for responses from Gemini if 'Yes' ('gem yes') and 'No' ('gem_no'); the counts for PubMed if 'Yes' ('pub_yes') and 'No' ('pub_no'), and total counts of agreements ('both_yn'), along with the counts of 'Yes' responses exclusively for Gemini ('only_gyes') and for PubMed ('only_pyes'). The final column is the chi-square (or FET) test result conducted on a 2x2 table comparing the totals of 'Yes' and 'No' responses for Gemini versus PubMed.

**Agreement Summary**

PubMed findings were compared to MMC, and the comparisons were categorised based on the "with_gender" variable only for COVID-19, which can be true or false. This distinction is crucial for COVID-19 because PubMed is sensitive to gender, age, and disease severity, while Gemini is not.

The PubMed versus Gemini agreement summary table presents the mean, standard deviation, and number of consensuses (n) per disease. Below are the COVID-19 and MB agreements for run equals 'run01' (see STable 41).

| Disease | with gender | <agree> | std | n |
|---|---|---|---|---|
| COVID-19 | False | 72.7% | 6.7% | 1,262 |
| COVID-19 | True | 66.3% | 4.3% | 1,262 |
| MB | False | 66.8% | 5.6% | 556 |

STable 41 - The summary table comparing the PubMed and Gemini agreements for COVID-19 and MB has been filtered to include only entries where the run is 'run01'. For COVID-19, the data is categorised by "with_gender," while for MB, "with_gender" is set to False. Each row represents a study-gender pairing, and the columns display the mean agreement, standard deviation (std), and the number of compared consensuses (n). A more detailed table can be found in the file named gemini_x_PubMed_covid_MB.xlsx.

Filenames:
1. stat_gemini_x_pubmed_aggre_for_COVID-19_all_data.tsv
2. stat_gemini_x_pubmed_aggre_for_MB_all_data.tsv

### Crowdsourced consensus (CSC)

To calculate the crowdsourced consensus (CSC), one first must merge the consensus data from 3 Sources into a single table. With this table, we can calculate the CSC and each source's accuracy. The table presents three key metrics: CSC accuracy, reviewers' comparisons, and Yes agreements. Below, we demonstrate all three metrics along with their respective tables.

**Merging the 3 Sources and Calculating the CSC**

To build the MSD, we merged 3 Sources evaluations for both selected cases, and filtered 'with_gender' to True or False, due to PubMed biased answers. For each pathway in the 2CRSP, we include columns for the Gemini consensus, PubMed answer, and Reviewers' consensus, ultimately calculating the CSC. Each consensus is linked to a 'Yes' or 'No' response; to remind, 50% of the pathways derived from the Gemini 1.5-flash model resulted in 'Yes' answers, while the remaining 50% resulted in 'No' answers. As a result, we can expect the 'Yes' answer statistics to be around 50%. The final key metric is related to the Reviewers; thus, it allows for comparing the Reviewers' consensus against the findings from Gemini and PubMed.

Here is a brief example of the first five lines of the 3 Source table related to COVID-19, with 'run' set to 'run01', 'case' set to 'g3_male_adult', and 'with_gender' set to False (STable 42).

| pathway id | pathway | CSC | CSC n yes | CSC n no | agree gemini | agree pubmed | agree review |
|---|---|---|---|---|---|---|---|
| R-HSA-1280218 | Adaptive Immune System | Yes | 3 | 0 | True | True | True |
| R-HSA-389356 | CD28 Co-Stimulation | Yes | 3 | 0 | True | True | True |
| R-HSA-2262752 | Cellular Responses To Stress | Yes | 3 | 0 | True | True | True |
| R-HSA-983169 | Class I MHC Mediated Antigen Processing And Presentation | Yes | 3 | 0 | True | True | True |
| R-HSA-1442490 | Collagen Degradation | Yes | 3 | 0 | True | True | True |

STable 42: The CSC table presents pathways as rows and includes the following columns: 'CSC' as the calculated crowsourced consensus ('Yes' or 'No'), the number of 'Yes' responses and 'No' responses derived from the three sources (labeled as 'CSC n yes' and 'CSC n no'), and the agreements for Gemini, PubMed, and Reviewers (denoted as 'agree gemini', 'agree pubMed', and 'agree review', respectively). Several other columns are hidden for the sake of readability.

Filename: crowd_all_for_taubate_covid19.tsv

**Accuracy**

The CSC involves scanning each pathway and evaluating the agreement among the Gemini consensus, the PubMed responses, and the Reviewers' consensus. For each source, the accuracy score is determined by calculating the mean value of the comparisons between their responses and the corresponding CSC consensus for each pathway.

Below is the accuracy table for COVID-19, for run equal 'run01' (STable 43).

| case | with gender | PubMed | | Gemini | | Reviewers | |
|---|---|---|---|---|---|---|---|
| | | percent | std | percent | std | percent | std |
| g3 male adult | False | 86.67% | 34.57% | 86.67% | 34.57% | 96.67% | 18.26% |
| g3 male adult | True | 53.33% | 50.74% | 100.00% | 0.00% | 83.33% | 37.90% |

| g3 female elder | False | 60.00% | 49.83% | 96.67% | 18.26% | 73.33% | 44.98% |
| g3 female elder | True  | 50.00% | 50.85% | 93.33% | 25.37% | 76.67% | 43.02% |

STable 43 - The accuracy table for COVID-19 presents the accuracy of each source, categorised by cases using run equals 'run01'. The rows are the two selected cases defined in the MSD table, while the columns include the three accuracies along with their respective standard deviations: PubMed, Gemini, and Reviewers. We calculate the accuracies, both with and without considering gender, to assess their impact.

Filename: crowd_summary_for_taubate_covid19.tsv

Below is the accuracy plot for COVID-19, for run equal 'run01' (SFigure 15).

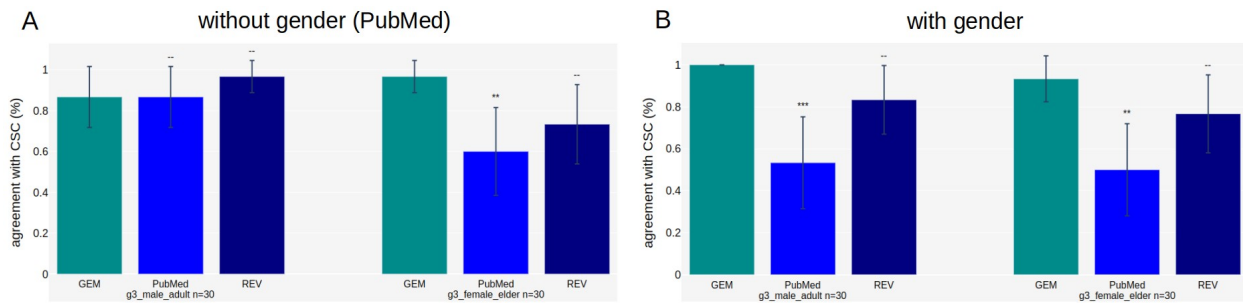

SFigure 15 - The accuracy table for MB presents the accuracy of each source, categorised by cases (subtypes) using run equals 'run01'. The rows are the two selected subtypes defined in the MSD table, while the columns include the three accuracies along with their respective standard deviations: PubMed, Gemini, and Reviewers. We calculate the accuracies, both with and without considering gender, to assess their impact.

Next is the accuracy table for MB, for run equal 'run01' (STable 44).

|      |             | PubMed  |        | Gemini  |        | Reviewers |        |
|------|-------------|---------|--------|---------|--------|-----------|--------|
| case | with gender | percent | std    | percent | std    | percent   | std    |
| WNT  | False       | 80.00%  | 40.68% | 86.67%  | 34.57% | 86.67%    | 34.57% |
| G4   | False       | 86.67%  | 34.57% | 66.67%  | 47.95% | 86.67%    | 34.57% |

STable 44 - The accuracy table for MB displays the accuracy of each source for 'run01'. The rows are the two subtypes, while the columns are the three accuracies along with their respective standard deviations: PubMed (accu_pubmed), Gemini (accu_gemini), and Reviewers (accu_review). Calculations are made both with and without the gender to evaluate the impact of gender.

Filename: crowd_summary_for_medulloblastoma.tsv

Below is the accuracy plot for MB, for run equal 'run01' (SFigure 16).

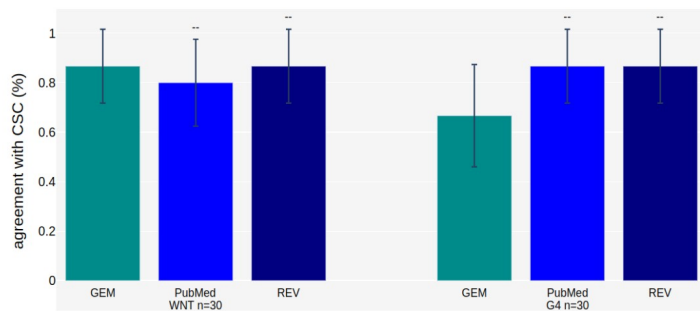

SFigure 16 - The accuracy plot for MB is not divided into the 'with_gender' factor, as MB is considered independent. The bars are grouped on the left 'WNT' and the right 'G4' subtypes. Each bar represents Gemini ('GEM'), PubMed, and Reviewers ('REV') accuracies.

**Yes agreements counts**

Here is a brief example of the first five lines of the 3 Source table focusing on the Yes count columns. These data are related to COVID-19, with 'run' set to 'run01', 'case' set to 'g3_male_adult', and 'with_gender' set to False (see STable 45).

| pathway id | pathway | gem cons | n yes gem | pub has pmid | n pmid | rev cons | rev yes | n yes rev |
|---|---|---|---|---|---|---|---|---|
| R-HSA-1280218 | Adaptive Immune System | Yes | 8 | Yes | 308 | Yes | True | 7 |
| R-HSA-389356 | CD28 Co-Stimulation | Yes | 5 | Yes | 3 | Yes | True | 6 |
| R-HSA-2262752 | Cellular Responses To Stress | Yes | 8 | Yes | 39 | Yes | True | 7 |
| R-HSA-983169 | Class I MHC Mediated Antigen Processing And Presentation | Yes | 8 | Yes | 40 | Yes | True | 7 |
| R-HSA-1442490 | Collagen Degradation | Doubt | 4 | Yes | 2 | Yes | True | 5 |

STable 45 - The 3 Source table focusing on the Yes counts presents pathways as rows and includes the following columns: the source's consensus ("gem_cons" and "rev_cons"), the PubMed answer ("pub_has_pmid"), and the number of "Yes" responses for Gemini ("n_yes_gem") and Reviewers ("n_yes_rev"). Additionally, it includes the counts of PMIDs found ("n_pmid").

**Yes agreements**

The table of 'Yes' agreements provides a statistical assessment based on 50% of the randomly selected pathways over the "possible modulated pathways," pathways with positive responses and the other 50% with negative responses. As a result, the "percentage of Yes agreements" indicates how close we are to matching the responses from Gemini 1.5-flash.

Below is the 'Yes' agreements table for COVID-19 (STable 46 and SFigure 17).

| case | with gender | CSC | | PubMed | | Gemini | | Reviewers | |
| --- | --- | --- | --- | --- | --- | --- | --- | --- | --- |
| | | percent | std | percent | std | percent | std | percent | std |
| g3 male adult | False | 80.00% | 40.68% | 80.00% | 40.68% | 66.67% | 47.95% | 83.33% | 37.90% |
| g3 male adult | True | 66.67% | 47.95% | 26.67% | 44.98% | 66.67% | 47.95% | 83.33% | 37.90% |
| g3 female elder | False | 63.33% | 49.01% | 23.33% | 43.02% | 66.67% | 47.95% | 83.33% | 37.90% |
| g3 female elder | True | 60.00% | 49.83% | 10.00% | 30.51% | 66.67% | 47.95% | 83.33% | 37.90% |

STable 46 - The table of 'Yes' agreements shows the percentages of 'Yes' (modulated pathways) in two selected cases for COVID-19. The rows in the table represent cases that include the 'with_gender' factor (set to True or False). The table also has repeated columns for CSC, PubMed, Gemini, and Reviewers, detailing the percentage of concordance and its standard deviation.

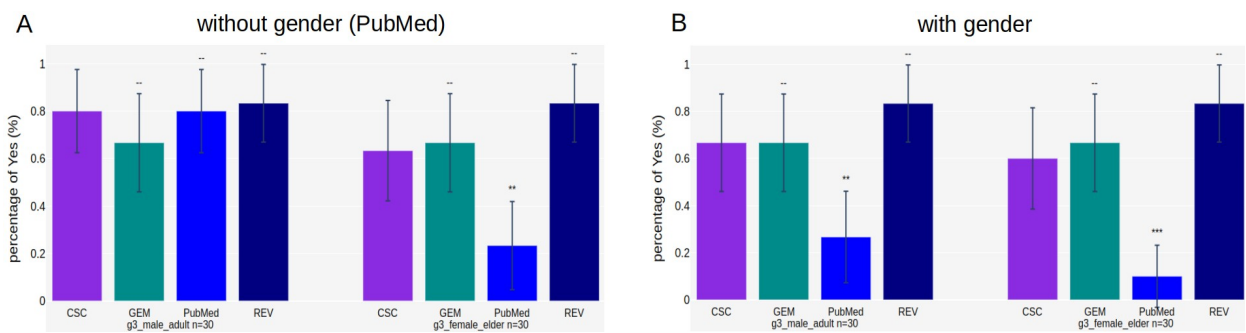

SFigure 17 - The 'Yes' agreements plot for COVID-19 is divided into 'without gender' on the left and 'with gender' on the right. For each 'with_gender' factor, there are 2 selected cases: on the left, we have 'g3_male_adult,' and on the right, 'g3_female_elder.' The bars represent the CSC, Gemini ('GEM'), PubMed, and Reviewers ('REV') agreements.

Next is the 'Yes' agreements table for COVID-19 (STable 47 and SFigure 18).

| case | with gender | CSC | | PubMed | | Gemini | | Reviewers | |
| --- | --- | --- | --- | --- | --- | --- | --- | --- | --- |
| | | percent | std | percent | std | percent | std | percent | std |
| WNT | False | 53.33% | 50.74% | 40.00% | 49.83% | 66.67% | 47.95% | 53.33% | 50.74% |
| G4 | False | 20.00% | 40.68% | 20.00% | 40.68% | 46.67% | 50.74% | 20.00% | 40.68% |

STable 47 - The table of 'Yes' agreements shows the percentages of 'Yes' (modulated pathways) in two selected subtypes for MB. The rows in the table represent cases that include the 'with_gender' factor (set to True or False). The table also has repeated columns for CSC, PubMed, Gemini, and Reviewers, detailing the percentage of concordance and its standard deviation.

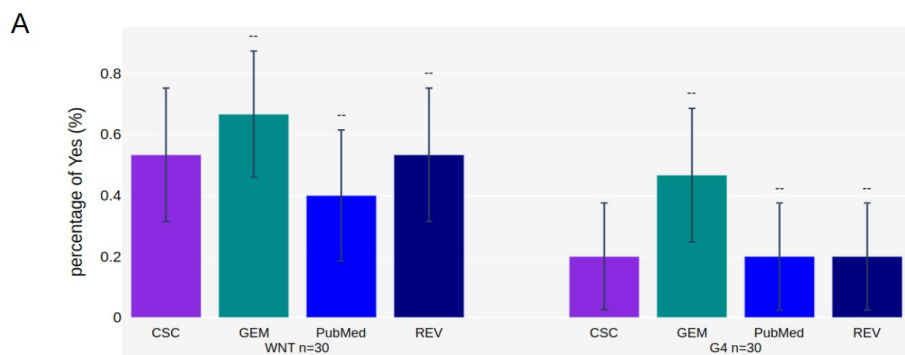

SFigure 18 - MB's 'Yes' agreements plot is not divided into the 'with_gender' factor, as MB is considered independent. The bars are grouped on the left 'WNT' and the right 'G4' subtypes. The bars represent the CSC, Gemini ('GEM'), PubMed, and Reviewers ('REV') agreements.

**Human reviewers' agreements**

The MSD table also includes columns related to the Human Reviewers. It allows us to calculate the agreement of their consensuses with PubMed answers and Gemini MMC consensuses, as each case/subtype shares 30 randomly selected pathways.

Below is the Reviewers' agreement table for COVID-19 (STable 48 and SFigure 19).

|  |  | reviewers x pubmed | | reviewers x gemini | |
|---|---|---|---|---|---|
| case | with_gender | agreement | std | agreement | std |
| g3 male adult | False | 83.33% | 37.90% | 83.33% | 37.90% |
| g3 male adult | True | 36.67% | 49.01% | 83.33% | 37.90% |
| g3 female elder | False | 33.33% | 47.95% | 70.00% | 46.61% |
| g3 female elder | True | 26.67% | 44.98% | 70.00% | 46.61% |

STable 48 - The Reviewers' Agreement table displays the level of agreement among reviewers regarding the PubMed answers and the Gemini MMC consensuses for COVID-19. The rows in the table represent cases that include the 'with_gender' factor, which may be set to either True or False. Additionally, the table features repeated columns for both PubMed and Gemini, providing the percentage of agreement along with its standard deviation.

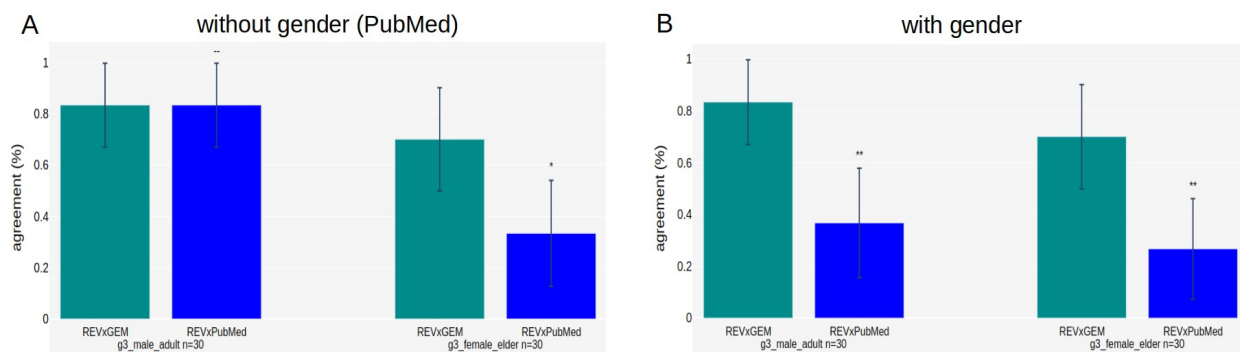

SFigure 19 - The Reviewers' agreement plot for COVID-19 is divided into 'without gender' on the left and 'with gender' on the right. For each 'with_gender' factor, there are 2 selected cases: on the left, we have 'g3_male_adult,' and on the right, 'g3_female_elder.' The bars represent the Reviewers' agreements against Gemini ('REVxGEM') and PubMed ('REVxPubMed').

Next is the Reviewers' agreement table for MB (STable 49 and SFigure 20).

|  |  | reviewers x pubmed | | reviewers x gemini | |
|---|---|---|---|---|---|
| case | with_gender | percent | std | percent | std |
| WNT | False | 66.67% | 47.95% | 73.33% | 44.98% |
| G4 | False | 73.33% | 44.98% | 53.33% | 50.74% |

STable 49 - The Reviewers' Agreement table displays the level of agreement among reviewers regarding the PubMed answers and the Gemini MMC consensuses for MB. The rows of the table represent subtypes, and since we are focusing on MB, we do not consider the 'with_gender' factor. Additionally, the table features repeated columns for both PubMed and Gemini, providing the percentage of agreement along with its standard deviation.

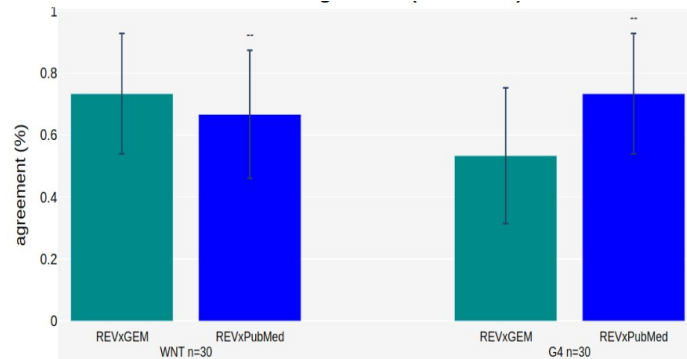

SFigure 20 - The Reviewers' agreement plot for MB is divided into the two subtypes, on the left 'WNT,' and on the right, 'G4.' The bars represent the Reviewers' agreements against Gemini ('REVxGEM') and PubMed ('REVxPubMed').

# Confusion Matrix

## Inter-Group Confusion Matrix

The Inter-Group Confusion Matrix utilises the GSEA results from the four pathway groups. It compares the positive control group (G0), the enriched table with additional pathways, against the other three groups.

### Counts

Below is the Inter-Group Confusion Matrix counts for COVID-19 comparing G0 to G3, for run equals 'run01' (STable 50)

| case | n | npos | nneg | TP | FP | TN | FN |
|---|---|---|---|---|---|---|---|
| g2a_male | 90 | 45 | 45 | 28 | 17 | 40 | 5 |
| g2a_female | 58 | 29 | 29 | 19 | 10 | 25 | 4 |
| g2b_male | 84 | 42 | 42 | 29 | 13 | 33 | 9 |
| g2b_female | 88 | 44 | 44 | 27 | 17 | 36 | 8 |
| g3_male_adult | 48 | 24 | 24 | 14 | 10 | 19 | 5 |
| g3_male_elder | 120 | 60 | 60 | 30 | 30 | 50 | 10 |
| g3_female_adult | 78 | 39 | 39 | 26 | 13 | 33 | 6 |
| g3_female_elder | 78 | 39 | 39 | 25 | 14 | 35 | 4 |

STable 50 - The Confusion matrix for COVID-19 displays counts obtained by comparing G0 against G3, using run equals 'run01'. The rows of the table represent different cases, while the columns indicate the total number of pathways for each case (n = n(G0) + n(G3)). These are further divided into positive G0 lengths (npos, 50% of n) and negative G3 lengths (nneg, 50% of n). Additionally, the table includes columns for True Positives (TP), False Positives (FP), True Negatives (TN), and False Negatives (FN).

Next is the confusion matrix for MB comparing G0 to G3, for run equal 'run01' (STable 51)

| case | n | n_pos | n_neg | TP | FP | TN | FN |
|---|---|---|---|---|---|---|---|
| WNT | 136 | 68 | 68 | 43 | 25 | 51 | 17 |
| G4 | 142 | 71 | 71 | 36 | 35 | 57 | 14 |

STable 51 - The Confusion matrix for MB displays counts and compares G0 against G3, using run equals 'run01'. The rows of the table represent different subtypes, while the columns indicate the total number of pathways for each case (n = n(G0) + n(G3)). These are further divided into positive G0 lengths (npos, 50% of n) and negative G3 lengths (nneg, 50% of n). Additionally, the table includes columns for True Positives (TP), False Positives (FP), True Negatives (TN), and False Negatives (FN).

**Statistics**

Below is the Inter-Group Confusion Matrix regarding statistical values for COVID-19 comparing G0 to G3, for run equals 'run01' (STable 52)

| case | n | sens | spec | accu | prec | f1_score |
|---|---|---|---|---|---|---|
| g2a_male | 90 | 84.8% | 70.2% | 75.6% | 62.2% | 71.8% |
| g2a_female | 58 | 82.6% | 71.4% | 75.9% | 65.5% | 73.1% |
| g2b_male | 84 | 76.3% | 71.7% | 73.8% | 69.0% | 72.5% |
| g2b_female | 88 | 77.1% | 67.9% | 71.6% | 61.4% | 68.4% |
| g3_male_adult | 48 | 73.7% | 65.5% | 68.8% | 58.3% | 65.1% |
| g3_male_elder | 120 | 75.0% | 62.5% | 66.7% | 50.0% | 60.0% |
| g3_female_adult | 78 | 81.3% | 71.7% | 75.6% | 66.7% | 73.2% |
| g3_female_elder | 78 | 86.2% | 71.4% | 76.9% | 64.1% | 73.5% |

STable 52 - The Confusion Statistical table for COVID-19 compares G0 against G3 based on a run equal to 'run01'. The rows of the table represent different cases, while the columns display the total number of pathways for each case (n = n(G0) + n(G3)) and the sensitivity (sens), specificity (spec), accuracy (accu), precision (prec), and F1-score.

Next is the Inter-Group Confusion Matrix regarding statistical values for MB comparing G0 to G3, for run equals 'run01' (STable 53)

| case | n | sens | spec | accu | prec | f1_score |
|---|---|---|---|---|---|---|
| WNT | 136 | 71.7% | 67.1% | 69.1% | 63.2% | 67.2% |
| G4 | 142 | 72.0% | 62.0% | 65.5% | 50.7% | 59.5% |

STable 53 - The Confusion Statistical table for MB compares G0 against G3 based on a run equal to 'run01'. The rows of the table represent different subtypes, while the columns display the total number of pathways for each case (n = n(G0) + n(G3)) and the sensitivity (sens), specificity (spec), accuracy (accu), precision (prec), and F1-score.

**Cloudy zone**

As mentioned, G0 is the positive control, while G3 is the negative control. It means that when we shift our comparisons from G1 and G2 against G0 to comparing G3 against G0, we can expect an increase in true values and a decrease in false values. In other words, comparing G1 and G2 (the cloudy zone) to G0 will likely yield fewer TPs and TNs and a higher number of FPs and FNs. In contrast, by comparing G3 to G0, we must achieve the maximum number of TPs and TNs while minimising FPs and FNs. Therefore, as we approach the comparison of G3xG0, both sensitivity and specificity are expected to trend toward 1.

See table: supp.mat. conf_table_covid_MB.xlsx

Below, you can see the Sensitivity, Specificity, Accuracy, Precision, and F1-score for COVID-19, comparing G1, G2, and G3 to G0 (see SFigure 21).

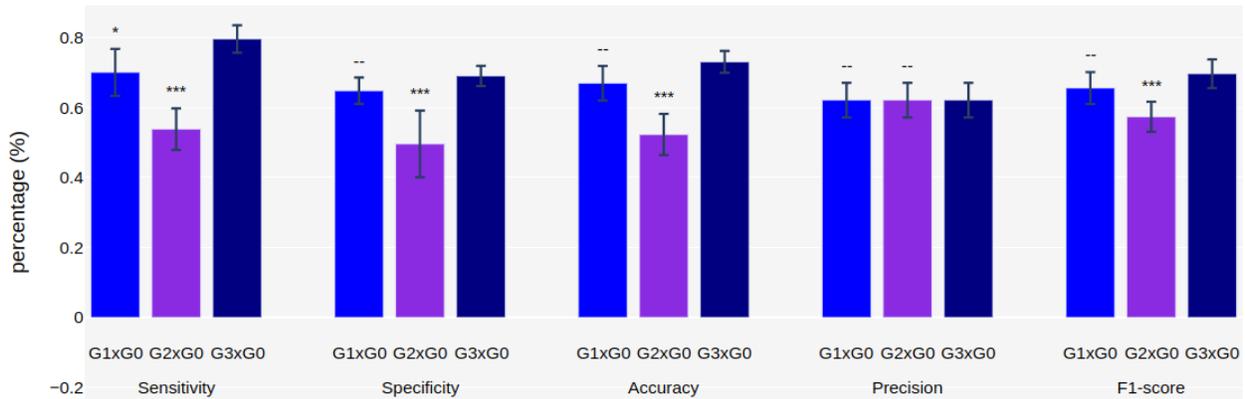

SFigure 21 - The confusion matrix statistical plot shows G1xG0, G2xG0, and G3xG0 Sensitivity, Specificity, Accuracy, Precision, and F1-score for COVID-19. The bar errors represent the confidence interval at a 95% confidence level. The comparisons involve t-tests contrasting each statistical method for each comparison.

Next, you can see the Sensitivity, Specificity, Accuracy, Precision, and F1-score for MB, comparing G1, G2, and G3 to G0 (see SFigure 22).

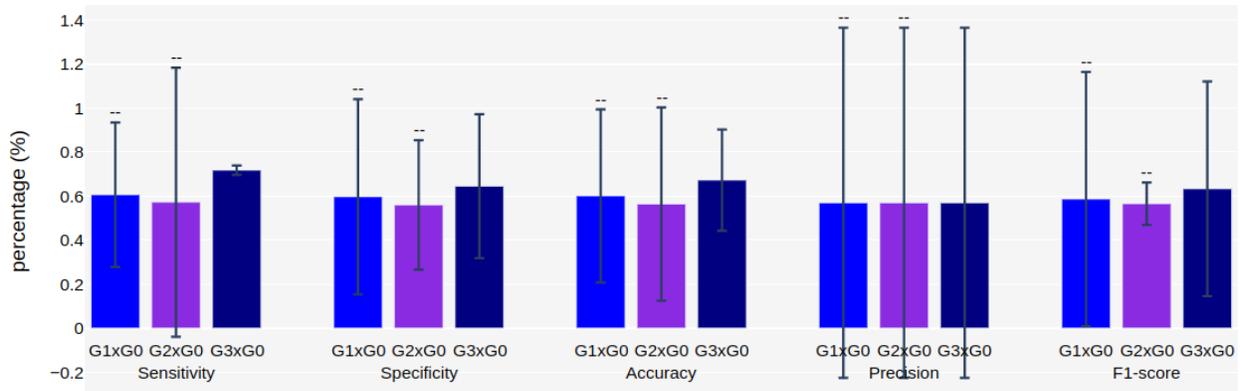

SFigure 22 - The confusion matrix statistical plot shows G1xG0, G2xG0, and G3xG0 Sensitivity, Specificity, Accuracy, Precision, and F1-score for MB. The bar errors represent the confidence interval at a 95% confidence level. The comparisons involve t-tests contrasting each statistical method for each comparison.

**Comparing runs**

The Confusion matrix compares the results of two runs, 'run01' and 'run02', using the t-test for COVID-19 (STable 54) and MB (STable 55). Each method is evaluated for different group comparisons. A t-test is conducted for each disease by comparing all case method evaluations.

The analysis shows no significant differences between the runs for both studies.

| which | method | mu_param0 | std_param0 | mu_param1 | std_param1 | pvalue |
|---|---|---|---|---|---|---|
| G0xG1 | Sensitivity | 70.1% | 8.0% | 70.3% | 7.6% | 9.55E-01 |
| G0xG1 | Specificity | 64.8% | 4.5% | 65.0% | 4.3% | 9.32E-01 |
| G0xG1 | Accuracy | 67.0% | 5.9% | 67.2% | 5.6% | 9.41E-01 |
| G0xG1 | Precision | 62.2% | 5.9% | 62.2% | 5.9% | 1.00E+00 |
| G0xG1 | F1-score | 65.6% | 5.4% | 65.7% | 5.3% | 9.64E-01 |
| G0xG2 | Sensitivity | 53.8% | 7.1% | 53.5% | 7.2% | 9.32E-01 |
| G0xG2 | Specificity | 49.6% | 11.4% | 49.1% | 11.3% | 9.34E-01 |
| G0xG2 | Accuracy | 52.3% | 7.0% | 51.9% | 7.0% | 9.16E-01 |
| G0xG2 | Precision | 62.2% | 5.9% | 62.2% | 5.9% | 1.00E+00 |
| G0xG2 | F1-score | 57.4% | 5.1% | 57.2% | 5.1% | 9.41E-01 |
| G0xG3 | Sensitivity | 79.6% | 4.7% | 89.3% | 3.4% | 3.21E-04 |
| G0xG3 | Specificity | 69.1% | 3.5% | 71.1% | 3.0% | 2.34E-01 |
| G0xG3 | Accuracy | 73.1% | 3.7% | 77.3% | 3.0% | 2.63E-02 |
| G0xG3 | Precision | 62.2% | 5.9% | 62.2% | 5.9% | 1.00E+00 |
| G0xG3 | F1-score | 69.7% | 4.9% | 73.1% | 4.5% | 1.67E-01 |

STable 54 - The Confusion Matrix Statistics Comparing Runs for COVID-19: rows represent group comparisons (G1xG0, G2xG0, and G3xG0) and the statistical methods used. The columns include mu_param and std_param, and the mean and standard deviation of the parameter methods. The indices 0 and 1 correspond to 'run01' and 'run02,' respectively. The p-value column represents the p-value obtained from the t-test.

| which | method | mu_param0 | std_param0 | mu_param1 | std_param1 | pvalue |
|---|---|---|---|---|---|---|
| G0xG1 | Sensitivity | 60.6% | 3.7% | 61.0% | 3.2% | 9.32E-01 |
| G0xG1 | Specificity | 59.7% | 4.9% | 60.1% | 4.4% | 9.46E-01 |
| G0xG1 | Accuracy | 60.1% | 4.4% | 60.5% | 3.9% | 9.40E-01 |
| G0xG1 | Precision | 57.0% | 8.9% | 57.7% | 7.9% | 9.41E-01 |
| G0xG1 | F1-score | 58.7% | 6.4% | 59.2% | 5.7% | 9.37E-01 |
| G0xG2 | Sensitivity | 57.3% | 6.8% | 57.6% | 7.3% | 9.68E-01 |
| G0xG2 | Specificity | 56.0% | 3.3% | 56.4% | 3.8% | 9.30E-01 |
| G0xG2 | Accuracy | 56.4% | 4.9% | 56.8% | 5.4% | 9.52E-01 |
| G0xG2 | Precision | 57.0% | 8.9% | 57.7% | 7.9% | 9.41E-01 |
| G0xG2 | F1-score | 56.6% | 1.1% | 57.1% | 0.3% | 5.54E-01 |
| G0xG3 | Sensitivity | 71.8% | 0.2% | 72.1% | 0.6% | 6.19E-01 |
| G0xG3 | Specificity | 64.5% | 3.6% | 64.9% | 3.2% | 9.30E-01 |
| G0xG3 | Accuracy | 67.3% | 2.6% | 67.7% | 2.1% | 8.94E-01 |
| G0xG3 | Precision | 57.0% | 8.9% | 57.7% | 7.9% | 9.41E-01 |
| G0xG3 | F1-score | 63.3% | 5.4% | 63.9% | 4.6% | 9.20E-01 |

STable 55 - The Confusion Matrix Statistics Comparing Runs for MB: rows represent group comparisons (G1xG0, G2xG0, and G3xG0) and the statistical methods used. The columns include mu_param and std_param, the mean and standard deviation of the parameter methods. The indices 0 and 1 correspond to 'run01' and 'run02,' respectively. The p-value column represents the p-value obtained from the t-test.

### Enriched Pathways Confusion Matrix

The Enriched Pathway Confusion Matrix utilises the calculated enriched pathways based on the default cutoff parameters and additional pathways obtained from relaxed cutoff ones. The algorithm designates the enriched pathways as TPs and the additional pathways as TNs to initiate the construction of the confusion matrix. Following this, FPs and FNs are uncovered using the results from the MMC.

### Counts

Below is the Enriched Pathway Confusion Matrix for COVID-19, for run equal 'run01' (STable 56)

| case | n | n_pos | n_neg | fdr | pvalue | TP | FP | TN | FN |
|---|---|---|---|---|---|---|---|---|---|
| g2a_male | 45 | 0 | 45 | 9.61E-05 | 2.40E-05 | 0 | 0 | 17 | 28 |
| g2a_female | 29 | 0 | 29 | 1.56E-03 | 5.84E-04 | 0 | 0 | 10 | 19 |
| g2b_male | 42 | 0 | 42 | 1.31E-05 | 1.64E-06 | 0 | 0 | 13 | 29 |
| g2b_female | 44 | 36 | 8 | 9.06E-02 | 9.06E-02 | 24 | 12 | 5 | 3 |
| g3_male_adult | 24 | 3 | 21 | 5.23E-02 | 4.57E-02 | 3 | 0 | 10 | 11 |
| g3_male_elder | 60 | 11 | 49 | 2.27E-03 | 1.14E-03 | 8 | 3 | 27 | 22 |
| g3_female_adult | 39 | 38 | 1 | 5.77E-02 | 4.33E-02 | 25 | 13 | 0 | 1 |
| g3_female_elder | 39 | 36 | 3 | 6.13E-03 | 3.83E-03 | 22 | 14 | 0 | 3 |

STable 56 - The Enriched Pathway Confusion Matrix for COVID-19 presents TP, FN, FP, and FN counts by comparing the Enriched Pathways and the additional pathways by confirming pathway modulation using the MMC, for one run equals 'run01'. The rows of the table represent different cases, while the columns indicate the total number of pathways for each case (n). These columns are divided into positive (npos) and negative (nneg). The table also includes columns for TP, FP, TN, and FN counts. Finally, a Chi-squared test compares the confusion matrix to one containing 90% TP and TN counts and 10% FP and FN counts represented by the columns p-value and FDR.

Next is the confusion matrix for MB comparing G0 to G3, for run equal 'run01' (STable 57)

| case | n | n_pos | n_neg | fdr | pvalue | TP | FP | TN | FN |
|---|---|---|---|---|---|---|---|---|---|
| WNT | 68 | 15 | 53 | 1.13E-09 | 5.67E-10 | 7 | 8 | 17 | 36 |
| G4 | 71 | 0 | 71 | 5.84E-06 | 5.84E-06 | 0 | 0 | 35 | 36 |

STable 57 - The Enriched Pathway Confusion Matrix for MB presents TP, FN, FP, and FN counts by comparing the Enriched Pathways and the additional pathways by confirming pathway modulation using the MMC, for one run equals 'run01'. The rows of the table represent different cases (subtypes), while the columns indicate the total number of pathways for each case (n). These columns are divided into positive (npos) and negative (nneg). The table also includes columns for TP, FP, TN, and FN counts. Finally, a Chi-squared test compares the confusion matrix to one containing 90% TP and TN counts and 10% FP and FN counts represented by the columns p-value and FDR.

**Statistics**

Below is the Enriched Pathway Confusion Matrix regarding statistical values for COVID-19, for run equals 'run01' (STable 58)

| case | n | sens | spec | accu | prec | f1_score |
|---|---|---|---|---|---|---|
| g2a_male | 45 | 0.0% | 100.0% | 37.8% | | |
| g2a_female | 29 | 0.0% | 100.0% | 34.5% | | |
| g2b_male | 42 | 0.0% | 100.0% | 31.0% | | |
| g2b_female | 44 | 88.9% | 29.4% | 65.9% | 66.7% | 76.2% |
| g3_male_adult | 24 | 21.4% | 100.0% | 54.2% | 100.0% | 35.3% |
| g3_male_elder | 60 | 26.7% | 90.0% | 58.3% | 72.7% | 39.0% |
| g3_female_adult | 39 | 96.2% | 0.0% | 64.1% | 65.8% | 78.1% |
| g3_female_elder | 39 | 88.0% | 0.0% | 56.4% | 61.1% | 72.1% |

STable 58 - The Enriched Pathway Confusion matrix regarding statistical values, for COVID-19, was calculated using run equals 'run01'. The rows of the table represent different cases, while the columns display the total number of pathways for each case (n) and the sensitivity (sens), specificity (spec), accuracy (accu), precision (prec), and F1-score.

Next is the Enriched Pathway Confusion Matrix regarding statistical values for MB, for run equals 'run01' (STable 59)

| case | n | sens | spec | accu | prec | f1_score |
|---|---|---|---|---|---|---|
| WNT | 45 | 16.3% | 68.0% | 35.3% | 46.7% | 24.1% |
| G4 | 29 | 0.0% | 100.0% | 49.3% | | |

STable 59 - The Enriched Pathway Confusion matrix regarding statistical values, for MB, was calculated using run equals 'run01'. The rows of the table represent different cases, while the columns display the total number of pathways for each case (n) and the sensitivity (sens), specificity (spec), accuracy (accu), precision (prec), and F1-score.

**Model discontinuation**

Regrettably, Google discontinued the 1.0-pro model during our research, and new models are forthcoming.

**Terms definitions - Glossary**

    RRR - run-to-run reproducibility

    IMR - inter-model reproducibility

    RRCR - run-to-run consensus reproducibility

    IMCR - inter-model consensus reproducibility

    MMC - multi-model consensus

    UR - unanimous reproducibility

    CSC - crowdsourced consensus

    AI - artificial intelligence

    LLM - large language model

    Temperature - LLM temperature parameter

    P - LLM P parameter

    Hallucination - generates false responses that are unrelated to the input

    Confabulation - generates false responses that appear correct

    FDR - false discovery rate

    LFC - log fold change

    ET - enriched table

    GSEA - geneset enrichment analysis

    SENS - sensitivity

    SPEC - specificity

ACCU - accuracy

PREC - precision